\documentclass[utf8]{frontiersFPHY} 
\setcitestyle{square} 
\usepackage{url,hyperref,lineno,microtype,subcaption}
\usepackage[onehalfspacing]{setspace}

\def\mbh{$M_{\rm BH}$\/}

\def\lledd{$L/L_{\rm Edd}$}

\def\rfe{$R_{\rm FeII}$}

\def\feiiq{{\rm Fe}{\sc ii}$\lambda$4570\/}
\def\msol{M$_\odot$\/}

\def\ltsima{$\; \buildrel < \over \sim \;$}
\def\ltsim{\lower.5ex\hbox{\ltsima}}  
\def\gtsima{$\; \buildrel > \over \sim \;$}

\def\gtsim{\lower.5ex\hbox{\gtsima}}

\def\civ{{\sc{Civ}}$\lambda$1549\/}

\def\cm3{cm$^{-3}$\/}
\def\hb{{\sc{H}}$\beta$\/}

\def\mgii{{Mg\sc{ii}}$\lambda$2800\/}

\def\ciii{{\sc{Ciii]}}$\lambda$1909\/}
\def\oiiiopt{{\sc{[Oiii]}}\-$\lambda\lambda$\-4959,\-5007\/}
\def\o4363{{\sc{[Oiii]}}$\lambda$4363\/}

\def\aliii{Al{\sc iii}$\lambda$1860\/}

\def\feii{{Fe\sc{ii}}\/}

\def\fe{{\sc{Fe}}\/}

\def\fe76087{{\sc [Fe vii]}$\lambda$6087\/}
\def\oiii{{\sc [Oiii]}$\lambda$5007}

\def\kms{km~s$^{-1}$}

\def\mbu{$M_\mathrm{bulge}$\/}

\def\keyFont{\fontsize{8}{11}\helveticabold }
\def\firstAuthorLast{D'Onofrio {et~al.}} 
\def\Authors{Mauro D'Onofrio\,$^{1,2}$, Paola Marziani\,$^{2}$ and Cesare Chiosi\,$^{1}$}
\def\Address{$^{1}$Department of Physics and Astronomy "G. Galilei", University of Padua, Italy \\
$^{2}$INAF-Osservatorio Astronomico di Padova, Italy}
\def\corrAuthor{D'Onofrio Mauro}
\def\corrEmail{mauro.donofrio@unipd.it}
\begin{document}
\onecolumn
\firstpage{1}

\title[Past, Present \& Future of the SRs of Galaxies and AGN]{Past, Present and Future of the Scaling Relations of Galaxies and Active Galactic Nuclei} 

\author[\firstAuthorLast ]{\Authors} 
\address{} 
\correspondance{} 

\extraAuth{}

\maketitle

\begin{abstract}

\section{}
We review the properties of  the established Scaling Relations (SRs) of galaxies and active galactic nuclei (AGN), focusing on their origin and expected evolution back in time, providing a short history of the most important progresses obtained up to now and  discussing  the possible future studies. We also try to connect the observed SRs with the physical mechanisms behind them, examining to what extent current models reproduce the observational data. 

The emerging picture clarifies the complexity intrinsic to the galaxy formation and evolution process as well as the basic uncertainties still affecting our knowledge of the AGN phenomenon. At the same time, however, it suggests that the detailed analysis of the SRs can profitably contribute to our understanding of galaxies and AGN.

\tiny
 \keyFont{ \section{Keywords:} Galaxies, structure, evolution, scale relations, simulations. AGN, structure, evolution, scale relations} 
\end{abstract}

\section{Introduction}
\label{intro}

With the term "Scaling Relations" (hereafter, SRs) astronomers indicate a series of correlations between the parameters describing the physical characteristics of galaxies. These can be radii, mean  velocities of stars and gas, stellar population proxies as colors or mass-to-light ratios, density and total amount of gas and dust, black-hole masses, etc. 

The study of SRs started when Edwin Hubble presented his famous tuning fork diagram for the morphological classification of galaxy types \cite{Hubble1936}. Very soon this beautiful scheme prompted the idea that the morphological sequence is driven by some physical parameters, such as mass, luminosity, color, angular momentum and gas content, that progressively change along the sequence determining the observed types. Attempts to build a "physical" classification of galaxies characterized the following years \cite[see e.g.][]{deVaucouleurs1962,Brosche1973,Benderetal1992,Cappellarietal2011a,KormendyBender2012}. 

The first questions arising from the morphological sequence concerned the different flattening observed among galaxies \cite{SandageFreemanStokes1970}. In this work the authors  tried to answer why some galaxies have a flat disk while others do not and, in connection with this, why the spheroidal components of all galaxies contain only old stars,  why S0's and early-type spirals have lost their spiral arms and why up to 50\% of galaxies are barred. 

The basic idea was that the Hubble sequence is essentially an angular momentum sequence \cite{Brosche1970,SandageFreemanStokes1970}, where star formation (SF) occurs at increasing gas density. The spread of color within the morphological types was attributed to the different star formation rates (SFR) inside galaxies \cite{Searle1973} and to the different stellar populations inside them \cite{King1971}.

Quite soon however, it was clear that the parameters describing the properties of galaxies can be considered a mathematical manifold \cite{Brosche1973}, because several correlations among them are in place. If we consider for example the galaxy luminosity ($L$), we observe that it correlates with: the effective radius ($R_e$; the radius enclosing half the total luminosity) \cite{Fish1964}, the central velocity dispersion of the stars ($\sigma$) (hereafter Faber-Jackson FJ relation \cite{faberjackson76}), the effective surface brightness ($I_e$; the mean surface brightness inside $R_e$) \cite{Kormendy1977, BinggeliSandageTarenghi1984}, color \cite{Sandage1972} and line-strength index ($Mg_2$) \cite{Terlevichetal1981}.

The great number of observed correlations promptly arose other fundamental questions. What are the most fundamental correlations? What parameters better describe their physics? How do the SRs evolve with time? 

In an attempt to answer these questions \citet{GuzmanLuceyBower1993} claimed that only three fundamental relations are necessary to describe all global SRs among the spheroidal systems, while \citet{Disneyetal2008} found a striking correlation among five basic parameters that govern the galactic dynamics ($R_{50}$, $R_{90}$, $M_{HI}$, $M_d$,  $L$: respectively the 90\%-light radius, the 50\%-light radius, the H I mass, the dynamical mass and the luminosity) and the color. The principal component analysis (PCA) further showed that the first eigenvector dominates the correlations among the parameters and can explain up to 83\% of the variance in the data. 

Unfortunately, the next investigations demonstrated that the SRs cannot be used as a basis for a theoretical understanding of galaxy formation and evolution. They can be used only 'a posteriori' to verify the ability of theories in reproducing the observed correlations. Galaxies are complex and evolving systems requiring much complex statistical tools than simple PCA \cite{Fraix-Burnetetal2019}.

In other words the Hubble classification  is only a qualitative scheme, influenced by subjective decisions and difficult to use for distant galaxies. The sequence rests only on the morphological parameters measured in the visual bands, while galaxies are complex systems that can be observed from X-rays to radio wavelengths. In addition a lot of information, such as chemical compositions, stellar populations, central black hole masses, kinematics of stars and gas, etc., can be obtained from the spectral analysis \cite{Sandage2005}.

Recently new support to the study of the SRs was gained  thanks to the data of the large sky surveys, such as the Sloan Digital Sky Survey  (SDSS \cite{Abazajianetal2003}), SAURON \cite{Baconetal2001}, WINGS \cite{Fasanoetal2006}, ATLAS3D \cite{Cappellarietal2011b}, CALIFA \cite{Sanchezetal2012}, SAMI \cite{Croometal2012}, MaNGA \cite{Bundy2015}, etc.
These surveys have provided data for thousands of galaxies permitting a more robust statistical analysis of the physical drivers behind their formation and evolution. Several SRs, such as the velocity-luminosity or Tully-Fisher relation (hereafter TF, \cite{TullyFisher1977, Courteauetal2007}), the Faber-Jackson (FJ) relation \cite{faberjackson76}, the $I_e - R_e$ (hereafter Kormendy relation KR \cite{Kormendy1977}), the fundamental plane of galaxies (hereafter FP, \cite{DjorgovskiDavis1987, Dressleretal1987, Benderetal1992, Bernardietal2003,Cappellarietal2006,LaBarberaetal2008}, the bulge mass - black hole (BH) mass relation \cite{Magorrianetal1998}, the mass-radius (MR) relation \cite{Chiosietal2020} are now robust for the galaxies of the nearby Universe and have now well constrained the physical laws governing the assembly of stellar systems.

{On the theoretical side, despite the recent progresses, galaxy formation models are still in difficulties with some basic  properties of galaxies. For instance  colors, radii \cite{Donofrioetal2020}, structural bimodalities \cite[see e.g.][]{DekelBirnboim2006,McDonaldetal2009}, angular momentum content \cite{FallRomanowsky2013,ObreschkowGlazebrook2014}, variations of the stellar initial mass function (IMF), mass-to-light ratios \cite{Duttonetal2011,Cappellarietal2012,Smith2014}, central versus satellite distributions \cite{Rodriguez-Pueblaetal2015}, and others cannot be satisfactorily matched by the models. Some fundamental dynamical tracers of galaxy structure (e.g. the circular velocity of galaxies and stellar-to-halo mass ratio) predicted by the models are  still discrepant with observations}. 

{Another remark to keep in mind is that the technical analysis of the SRs must be considered with due caution. The observed relations often depend on a number of factors}, last but not least the structural parameter definitions \cite{Courteau1996,Courteau1997}, the environment that could influence the general distribution of galaxies\cite{Moczetal2012}, the different fitting algorithms \cite{Courteauetal2007,Avila-Reeseetal2008,Halletal2012} that provide different coefficients, redshift and peculiar motions of the galaxies in the sample used \cite{Willicketal1997,FernandezLorenzoetal2011,Milleretal2011}, projection effects and bandpass \cite{Aaronsonetal1986,Donofrioetal2008,Halletal2012}, the morphology of galaxies in the sample \cite{Courteauetal2007,Tollerudetal2011}, the stellar population content \cite{Cappellarietal2006,Falcon-Barrosoetal2011,Cappellarietal2013}, the metallicity \cite{Wooetal2008} and the statistical properties of the dark matter (DM) halos \cite[see e.g.][]{Chiosietal2020}.

In general we want to stress that SRs are today universally considered convenient tools to estimate quantities such as distances and masses in an efficient way (when the data sample is large), but most importantly, they permit a much deeper understanding of galaxy structure, formation and evolution. {For example \citet{Kassinetal2012}, by examining the $V_{rot}/\sigma$ ratio across redshift, found that galaxies accrete baryons at different rates during evolution.}
{At the  same time, \citet{ObreschkowGlazebrook2014} pointed out the link between the FP and FJ relations with the angular momentum ($j$), the stellar mass ($M_s$), and the bulge fraction ($\beta$) of spiral galaxies \cite[see also][]{Peebles1969,Fall1983}. \citet{Lagosetal2017}, using cosmological simulations, confirmed the correlation between galaxy mass and specific angular momentum, and the evolution of the $M_s - j$ relation in passive and active galaxies, while \citet{Ferrareseetal2006} showed that the correlation of the mass of the BHs and the bulge mass is a key element in favor of the co-evolution of the AGN with their host galaxies. }
\citet{DesmondWechsler2017} used the FP to predict the amount of DM in the central regions of elliptical galaxies, while \citet{Ouelletteetal2017} found that the tilt of the FP correlates with the DM fraction of each galaxy and \citet{Chiosietal2020} demonstrated that the DM halo growth function is able to shape the mass-radius relation. We will see many other examples of the utility of SRs in this review.
 
The utility of SRs has been recognized not only for galaxies. They are also very important to understand the central BHs in galaxies and the nature of the active galactic nuclei (AGN). {The co-evolution of the central black holes and galaxies has been  known for  more than twenty years \cite[see e.g.][]{KormendyRichstone1995, HaringRix2004,FerrareseMerritt2000,Gebhardtetal2000,Grahametal2001, Ferrarese2002} }.  Even the active nuclei have shown to obey several SRs that are useful to clarify their structure and evolution.  We will therefore address in these pages several of these correlations involving the parameters that describe the properties of the central active nucleus in galaxies. This analysis will permit to conclude that, even in this context, SRs are fundamental tools to infer the physical mechanisms at work in galaxies and AGN.
 
In conclusion we can say that SRs are fundamental for any theory of galaxy formation and evolution. The current view is that the diversity of galaxies appears to increase rapidly with the instrumental improvements so that a good understanding of their physics requires sophisticated numerical simulations that reproduce realistic objects. The physical processes that operate together during galaxy evolution are numerous and imply that the morphological Hubble sequence is only the first approach to the complex problem of galaxy classification \cite{Fraix-Burnetetal2019}. The SRs are the network of properties that the modern statistical tools and theoretical simulations must be able to explain and reproduce. How are their properties interwined? { How do they evolve over time?} This is the challenge of future investigations.

{
In this work we will review some of the established SRs of galaxies and AGN, discussing our current understanding of their origin and evolution. The first six sections are dedicated to the SRs originating from the coupling of galaxies dynamics and stellar population properties. 
We start in Sec. \ref{FJ} with the FJ relation, addressing next the TF (Sec. \ref{TF}), the KR (Sec. \ref{KR}), the MR relation (Sec. \ref{MR} and Sec. \ref{MR_cosmo}), and the FP (Sec. \ref{FP}). We have analyzed the MR relation with more details because of its cosmological implication. 
We  go on with the color-magnitude (CM) relation (Sec. \ref{CM}),  the relation between the star formation (and star formation history) with the mass and initial halo density in galaxies of different morphological types (Sec. \ref{sf_mode}), the mass-metallicity relation (Sec. \ref{MZR}). They all  provide a useful insight of the stars and gas evolutionary properties. 
Then, we address the correlation among the DM halos and baryonic matter properties (Sec. \ref{stellar_to_halo}) and the angular momentum - mass relationship (Sec. \ref{AMM}). 
Finally, we enter into the AGN domain, starting with a discussion of the correlations of the black-hole masses with the galaxy host properties (Sec. \ref{BHM}) and the most popular correlations observed among AGN (Sec. \ref{AGNSRs} and \ref{virial}). Some conclusions are finally drawn in Sec. \ref{CONCLUSIONS}.
}

\section{The Faber-Jackson relation}
\label{FJ}

The FJ relation is by far the most misunderstood correlation between galaxies parameters. Discovered by \cite{faberjackson76} in 1976, it is a correlation between the total luminosity of early-type galaxies (ETGs) $L$ and the central velocity dispersion of their stars $\sigma$. The authors themselves did not attribute any physical significance to this relation,  considering the observed trend a byproduct of the virial theorem, i.e. a translation of the correlation between mass and velocity dispersion, induced by the strong link between mass and luminosity.

The first fit on a sample of 25  ETGs gave $L\propto\sigma^4$, while further investigations provided values of the FJ parameters (slope and scatter) that depend on the  magnitude range of the sample considered \cite{Nigoche-Netroetal2010} (as in the case of the FP \cite{Donofrioetal2008}). The slope varies from $\sim2$ to $\sim5$ and the scatter of the residuals ($\sim0.30$) correlates with the effective radius $R_e$ (in the sense that smaller than average objects have larger velocity dispersion) and with the mass-to-light ratio \cite{Donofrioetal2020,Cappellarietal2013b}. The correlation however extends over 8 dex in luminosity, from Globular Clusters to Galaxy Clusters. A small curvature seems to exist at $M_V\sim-21.5$ mag, separating bright and faint objects. {The bright galaxies have a slope around 4-5, while the faint ones have it much closer to 2-3 \cite{Choietal2014,Donofrioetal2021}}.

The FJ is not one of the orthogonal projections of the FP relation $\sigma \propto I^a_e R^b_e$ (with a scatter of $\sim0.09$ in $R_e$). In the FJ relation the variable $L$ include both $R_e$ and $I_e$. We can better say that it is a sort of 2D version of the FP\footnote{By the way the FP was discovered by studying the residuals of the FJ relation.}. The deep analysis of \citet{Nigoche-Netroetal2011b} concluded that the scatter of the FJ depends on the history of galaxies, i.e  on the number and nature of the transformations that have affected the galaxies along their life times (collapse, accretion, interaction and merging).
The investigations of ETGs from the ATLAS-3D survey have indeed shown that many of these galaxies possess high rotational velocities, while slow-rotating objects often present counter-rotating cores. There are multiple channels of formation, where secular processes, disk instability, mergers, and gas accretion are possible mechanisms. Star formation events are sometimes observed even in the brightest cluster galaxies (BCGs), today almost quenched,  down to low redshifts \cite[see e.g.][]{LiuMaoMeng2012,Oliva-Altamiranoetal2015}. 

Despite this complexity there is an ample consensus on the fact that ETGs are approximately virialized object from a dynamical point of view.  Since luminosity is in general a quite good tracer of stellar mass, the deviation from the expected virial slope of 2, was explained with a smooth transition of the zero-point of the relation, essentially due to a variation of the mean mass-to-light ratio. This is the same explanation given for the observed tilt of the FP (see Sec. \ref{FP}).

The existence of a  physical correlation between luminosity and velocity dispersion of stars has never been considered as a concrete possibility. Why should the global stars emission  be aware of the mean stars velocity in a galaxy? This appears as an unphysical possibility.
Recently however, \citet{Donofrioetal2020} has opened the door to this remote possibility. The idea is that the total luminosity of galaxies is essentially the result of the stars assembly, of the SF history (SFH) and the stellar evolution. Luminosity is a non monotonic function of star's evolution. In 1973 \citet{Brosche1973} first suggested a failing of the simple SF law of \citet{Schmidt1959}, based only on the gas density $\rho$, favoring a scenario in which the SF is a function $\sim f(\rho v^{\beta})$, where $v$ is the velocity of stars and $\beta\sim3.6$ for most of the galaxies. Stars born in large gas aggregates have a characteristic velocity that depends on the physical condition of the galaxy during the SF event (collapse, shock, merging, etc.). For this reason the global SF might keep memory of the velocity of this gas. The SFH could therefore preserve such information, leading to a "physical" connection between $L$ and $\sigma$.

{The proof that such a physical link exists between luminosity and stellar velocity dispersion is encrypted in the appearance of some SRs.} The way to demonstrate this is to write the FJ relation in this way:

\begin{equation}
L = L'_0 \sigma^{\beta},
\label{eq1}
\end{equation}

where $L'_0$ and $\beta$ are now fully variable parameters that depend on the complex channel of stars assembly inside galaxies (new SF, accretion and removal events, etc.), in other words on the complex SFH we mentioned above. The connection of this empirical law (that is valid for a single galaxy and should not be confused with the fit of the whole distribution of ETGs in the FJ relation in which $L_0$ and $\beta$ are constant) with the virial theorem is the key to understand the observed distribution of galaxies in the SRs. The MR relation, the KR relation and the FP relations are in fact perfectly reproduced when the parameters $\beta$ and $L'_0$ change. The data of the Illustris simulation \cite{Vogelsbergeretal2014} used by \citet{Donofrioetal2020} have shown that the values of $\beta$ have an ample spectrum, going from large positive values, typical of star forming objects, to large negative values, typical of passive and quenched objects (quite often the more massive old galaxies). The peak of the distribution is observed at $\beta\sim3$, i.e. exactly coincident with the slope of the fitted FJ relation  \cite{Donofrioetal2020}.

In this new framework the slope of the classical FJ relation is mainly driven by the channel governing the star assembly inside galaxies. The bottom-up scenario of hierarchical merging gives the imprint on the slope of the FJ.  Simulations indicate that this slope changes with time. The trend is from a slope equal to $\sim5$ at high redshift to $\sim3$ observed today. In other words there is a progressive convergence toward the value of 2 expected for the virial dynamical equilibrium. This is valid for all SRs involving mass, velocity and luminosity. 

\begin{figure}[h!]
	\begin{center}
		\includegraphics[width=15cm]{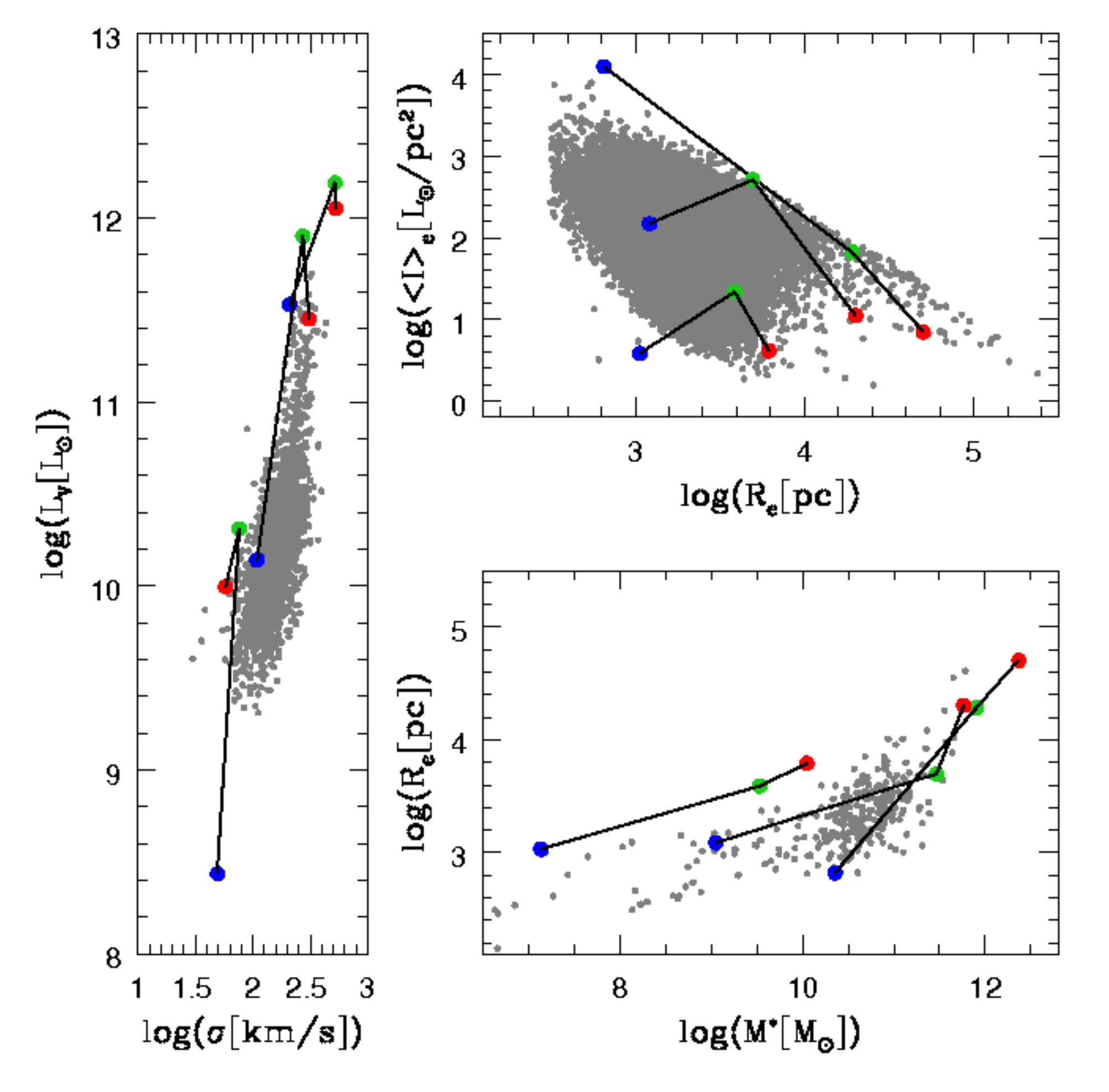}
	\end{center}
	\caption{Left panel: the $\log(\sigma)-\log(L)$ FJ plane. The gray dots mark the observational data extracted from the WINGS database. The coloured bigger points connected by lines are three objects extracted from the Illustris simulation respectively at redshift $z=4$ (blue dot), $z=1$ (green dot) and $z=0$ (red dot). The lines show the evolution of these objects across the cosmic epochs. Right upper panel: the $\log(R_e) - \log(I_e)$ KR plane. The WINGS galaxies are in gray and the coloured dots are the same objects of the left panel. Right lower panel: the $\log(M_s) - \log(R_e)$ MR plane. The symbols used are the same as before. The number of galaxies changes in each panel because masses and velocities are not available for the whole set of ETGs, in particular for the faint objects.}\label{fig:1}
\end{figure}

Figure \ref{fig:1} shows the FJ plane, the KR plane and the MR plane. The gray dots mark the observational data extracted from the WINGS database \cite{Fasanoetal2006,Morettietal2014}. Three artificial galaxies simulated by Illustris are shown with different colors marking their evolution in these planes from $z=4$ (blue dot) to $z=1$ (green dot) and $z=0$ (red dot). Note how the simulation is able to follow the FJ relation, keeping small the scatter of the relation despite the variations occurring in $\sigma$ and $L$. This happens because the relation is driven by mass. The classical FJ is essentially a relation between mass and velocity dispersion. Since galaxies are always close to the dynamical equilibrium, the variation expected in luminosity and velocity dispersion, due to SF or merging events, are never bigger than the scatter of the relation ($\sim0.3$ dex, that corresponds to a factor of 2). 
The DGs are the systems that are much distant to the virial equilibrium, probably for the strong feedback effects and SF activity still going on in many of them \cite{Chiosietal2020}.

Note how the distribution at $z=0$ in the KR and MR planes depends on the evolution of $\beta$. This parameter can be approximately estimated by looking at the direction of the lines connecting two redshift epochs ($\beta$ is the slope of the $\log(\sigma)-\log(L)$ relation). Negative values of $\beta$ in the FJ plane are those allowed only to quenched galaxies in passive evolution (where $L$ decreases at nearly constant $\sigma$).
As far as $\beta$ becomes progressively negative the distributions in the KR and MR planes converge toward the slopes expected for virialized objects ($-1$ in the KR plane and $1$ in the MR plane). In particular the tails observed in these two planes is that corresponding to the most massive and bright galaxies now in a quenched state of passive evolution \cite[see Table 4 in][for more details]{Donofrioetal2020}. This means that the full virialization in a galaxy can be achieved only when SF and feedback effects are stopped.

The emerging picture from the hierarchical model of galaxy assembly is that the KR and MR relations, that is the linear relations (in log units) visible when the samples contain only massive and bright ETGs, are formed by the tails of massive and large objects appearing after $z\sim1.5$. This is the location of the objects that today are almost quenched and passive. Their SF is over, the systems have reached a full virial configuration.

\section{The Tully-Fisher relation}
\label{TF}

As we have seen for the FJ relation, the complex process of galaxy assembly has produced some regular SRs, that ultimately suggest a tight connection between the stellar component and the hosting DM halos.
The Tully–Fisher relation (TF; \cite{TullyFisher1977}) is another example of a scaling law involving the luminosity of a galaxy (in this case of late type spiral galaxies) and the rotation velocities $V$ of stars. 
The dust-corrected TF relation has the form  $L\propto V^3$  in the optical band, with a slope that steepens toward redder passbands ($L\propto V^4$ in the near-infrared; \cite{Verheijen1997,Tullyetal1998}).
The variation of the slope with the passband indicate that there is a trend in color and in the stellar $M/L$ ratio with the galaxy mass. This change constrains galaxy formation and evolution models 
\cite[see e.g.][]{Coleetal2000,NavarroSteinmetz2000,vandenBosch2000}.

The TF is almost linear in log units for disk galaxies with well ordered rotation, while objects with disturbed morphology and compact galaxies do not follow the main relation, exhibiting lower rotations at a given stellar mass \cite{Kassinetal2007,Kassinetal2012}. The velocity fields are affected by major merging events or tidal disruptions \cite{Rampazzoetal2005,Kronbergeretal2007,Covingtonetal2010,DeRossietal2012}, by accretion of external angular momentum \cite{Brooksetal2009,ElmegreenBurkert2010} and/or by disruptive feedback events \cite{MacLowFerrara1999,Lehnertetal2009}. With spirals of the local universe the TF relation is tight \cite{Verheijen2001,Bekeraiteetal2016,Ponomarevaetal2017}. Galaxies with rising rotation curves and those with declining rotation curve are differently distributed in the TF relation \cite[e.g.][]{Persicetal1996}. 

The TF was used to measure the distance of spiral galaxies \cite[see e.g.][]{Giovanellietal1997} and  to test  cosmological models, arguing that its slope, zero-point and tightness are set by the cosmological evolution of the galactic DM halos
\cite{ColeSalamanca94, EisensteinLoeb1996, Moetal1998, Avila-Reeseetal1998, CourteauRix1999, NavarroSteinmetz2000}. 
The properties of these halos were often derived from the rotation curves of galaxies. However, the ignorance of the values of the stellar $M/L$ ratio (the gas contribution is typically well understood and relatively small \cite{Verheijen1997,SwatersMadoreTrewhella2000}) determines a degeneracy: many rotation curves can be equally well fitted by models in which the central part is dominated by stellar mass or by DM \cite{vanAlbadaetal1985,Swaters1999}. In order to resolve the degeneracy, some independent constraints on $M/L$ ratios are required.

{The TF relation is considered a product of the virial theorem and the almost constant mass-to-light ratio of spiral galaxies. Its origin has been discussed by \citet[e.g.][]{Silk1997, Moetal1998}. In their semi-analytical approach, \citet{Moetal1998} reproduced the TF relation assuming a constant mass-to-light ratio and an empirical profile for disks and halos. \citet{HeavensJimenez1999} used a similar approach, including an empirical star formation model, and successfully reproduced the TF relation in four pass-bands simultaneously. However, the exponential profile and the flat rotation curves of these galaxies were not constructed as the results of simulations, but assumed a priori. \citet{SteinmetzNavarro1999} provided the first numerical simulations within a cosmological context and explained the slope and scatter of the TF relation. They considered a volume much larger than the scale of galaxies, and some environmental effects (e.g., tidal field and infall/outflow of mass). \citet{KodaSofueWada2000} also reproduced the slope and scatter of the TF relation. In their approach the slope originate from the difference of mass among galaxies, while the scatter from the difference in the initial spin.}

A breakthrough was the discovery that the baryonic mass better correlates with rotational velocity than luminosity \cite{McGaughetal2000}. The baryonic TF relation (BTF) is remarkably tight \cite{BelldeJong2001, Verheijen2001, Pizagnoetal2005, KassindeJongWeiner2006, Courteauetal2007, MastersSpringobHuchra2008,Reyesetal2011}, but the exact slope still depends on the filters used 
\cite{Courteauetal2007, Ponomarevaetal2017, Schulz2017}.

The parametrization of the BTF gives important constraint for models of disk galaxy formation \cite{Moetal1998,SomervillePrimack1999,NavarroSteinmetz2000,Duttonetal2007}. Using a semi-analytic model, \citet{Dutton2012} predicted a minimum intrinsic scatter of $\sim0.15$ dex for the BTF while \citet{DiCintioLelli2016} had a scatter of $0.17$, using semi-empirical models that were able to reproduce the mass discrepancy acceleration, i.e. the ratio of total-to-baryonic mass at a given radius that anti-correlates with the acceleration due to baryons \cite{McGaughetal2004}. According to \citet{Bullocketal2001} most of the scatter comes from the mass–concentration relation of DM halos well-constrained by cosmological simulations. The scatter of the BTF is therefore a key test for the $\Lambda$CDM model. The scatter is minimum when the velocity is measured in the flat part of the rotation curve well beyond the optical extent of the galaxies \cite{Verheijen2001, NoordermeerVerheijen2007}, probably because such velocity is close to the virial velocity.

{As remarked before, one possible application of the BTF is to constrain the properties of the DM halos. \citet{Uebleretal2017} by investigating the stellar mass and BTF relations of massive star-forming disk galaxies at redshift $z\sim2.3$ and $z\sim0.9$ (using the data of the KMOS3D integral field spectroscopy survey), found that the contribution of DM to the dynamical mass increases toward lower redshift. Their comparison with the local relations reveals a negative evolution of the stellar and baryonic TF zero points from $z=0$ to $z\sim0.9$, no evolution of the stellar TF from $z\sim0.9$ to $z\sim2.3$, and a positive evolution of the BTF from $z\sim0.9$ to $z \sim2.3$.}

A useful progress came with the demonstration by \citet{Weineretal2006} and \citet{Kassinetal2007} that, accounting for disordered motions ($\sigma$) and ordered rotation  ($V$) in a new parameter $S_{0.5}=\sqrt{0.5V^2+\sigma^2}$, it is possible to get a tight $S_{0.5}–M_s$ relation \cite{Aquinio-Ortizetal2018}. This relation is independent of the morphology of galaxies and is coincident with the FJ relation of ETGs, when $\sigma$ dominates over $V$, and coincident with the TF when the opposite occurs. Numerical simulations seem to indicate that $S_{0.5}$ traces the potential well of the DM halos even in the case of merger events \cite{Covingtonetal2010}.  The inclusion in the TF of galaxies with disordered velocity components (often due to major mergers) has been addressed by several people \cite{Lemoine-BusserolleLamareille2010,Puechetal2010,Catinellaetal2012,Verganietal2012,Corteseetal2014, Wisnioskietal2015}. The scatter of the relation seems mainly due to merger events as we have seen for the FJ relation. 

Galaxy morphology is another possible source of scatter being a strong function of stellar mass and the less luminous systems quite often exhibit an irregular morphology, \cite[see e.g.][]{RobertsHaynes1994,BothwellKennicuttLee2009,Mahajanetal2015}. In general disturbed galaxies are increasingly more common at low masses in the early Universe \cite{Mortlocketal2013}.
The kinematic surveys are often biased against galaxies with disturbed {morphology}, because their aim is to study the DM content \cite{Bershadyetal2010}. Dwarfs galaxies (DGs) show rotational signatures in both their HI and stellar components \cite{Swatersetal2002,McConnachie2012} and when irregular galaxies, compact galaxies, and close pairs are analyzed in their kinematics the presence of peculiar velocity fields and thick disks are found \cite[see e.g.][]{Bartonetal2001,KannappanFabricantFranx2002,Vaduvescuetal2005, Corteseetal2014,Kirbyetal2014} together with high star-forming dwarfs \cite{vanZeeSkillmanSalzer1998, Cannonetal2004,LelliVerheijenFraternali2014}. However, only few studies have placed large samples of these disordered systems on the TF.

In the future, it will be interesting to study the TF relation in the same perspective of the FJ, distinguishing the relation valid for a set of galaxies, which is a translation of the virial theorem (once the variations in the stellar population are taken into account), and the relation valid for single galaxies, where the luminosity and the rotational velocity are the result of the mass assembly history and of the stellar evolution. The work of \citet{Donofrioetal2020} has demonstrated that it is important to look at the variations of the positions of each galaxy in the different SRs if we want to understand the origin of the observed distributions.

\section{The Kormendy relation}
\label{KR}

The $I_e -R_e$ relation of ETGs (often known as KR \cite{Kormendy1977}) is a projection of the FP. In this case the variables are the effective radius and the mean surface brightness inside it. It is the most easily accessible correlation of galaxies parameters even at high redshift. First discovered by Kormendy in 1977, the linear relation visible in log units between these variables, soon showed an ample curvature towards faint and dwarf objects, suggesting the existence of two different populations of ETGs, the 'ordinary' and the 'bright', following two different relations and therefore possibly originating from two different channels of evolution \cite{Capacciolietal1992}. The 'ordinary' family is bi-parametric ($L\propto I_e R_e^2)$, its members are fainter than $M_B \sim -19$ and their radii are smaller than $R_e \sim 3$ kpc. The 'bright' family is mono-parametric ($I_e$ depends only on $R_e$), it hosts only the brightest cluster galaxies (BCGs) and their members have radii exceeding $R_e=3$ kpc. The bulges of spirals belong to the 'ordinary' family. 

The curved distribution visible in the $I_e -R_e$ plane has been used (among others correlations) several times to argue for distinct formation mechanisms of dwarfs and giants ETGs \cite{Capacciolietal1993,Kormendyetal2009,KormendyBender2012,Tolstoyetal2009,SomervilleDave2015,Kormendy2016}. 
Many authors believe that there is a physical difference between elliptical and spheroidal galaxies. Elliptical and spheroidal galaxies exhibit different parameter correlations. Spheroidals are not low-luminosity ellipticals but rather the result of transformations induced in late-type galaxies by internal and environmental processes. Furthermore, there are possibly two distinct kinds of elliptical galaxies, whose properties differed during the last major mergers, wet or dry, according to whether cold gas dissipation and starbursts occurred or not.

The existence of two physically distinct families of ETGs has been at the center of an ample debate.
Other researches that did not use the effective half light radius parameter, advocated for a continuity among the ETG population \cite{Caldwell1983,BinggeliSandageTarenghi1984,Bothunetal1986,CaldwellBothun1987}.
\citet{Graham2019}  explored a range of alternative radii, showing that the transition at $M_B\sim-19$ mag is likely artificial and does not imply the existence of two different types of ETGs.

The shape of the light profiles of ETGs has been also used to claim a difference between dwarfs and ordinary ETGs: dwarfs have in general exponential light profiles (similar to the disks of LTGs), while ordinary ETGs have $R^{1/n}$ S\'ersic profiles \cite{Sersic1968}, with $n\geq3$. However, exponential light profiles are reproduced by the S\'ersic law when $n=1$. According to \citet{Graham2019}  the curved distribution of ETGs in the KR, is likely associated to the continuous change of the S\'ersic index $n$ with the absolute magnitude (the $M_B - n$ relation \cite{Caonetal1993,Donofrioetal1994}). Along this view \citet{GrahamGuzman2003} argued that the only magnitude of importance in the $I_e -R_e$ plane is at $M_B=-20.5$ mag, where they see a division between ETGs with S\'ersic profiles and core-S\'ersic profiles. This magnitude corresponds to a mass of $\sim2 \times 10^{11} M_\odot$.

\begin{figure}[h!]
	\begin{center}
		\includegraphics[width=12cm]{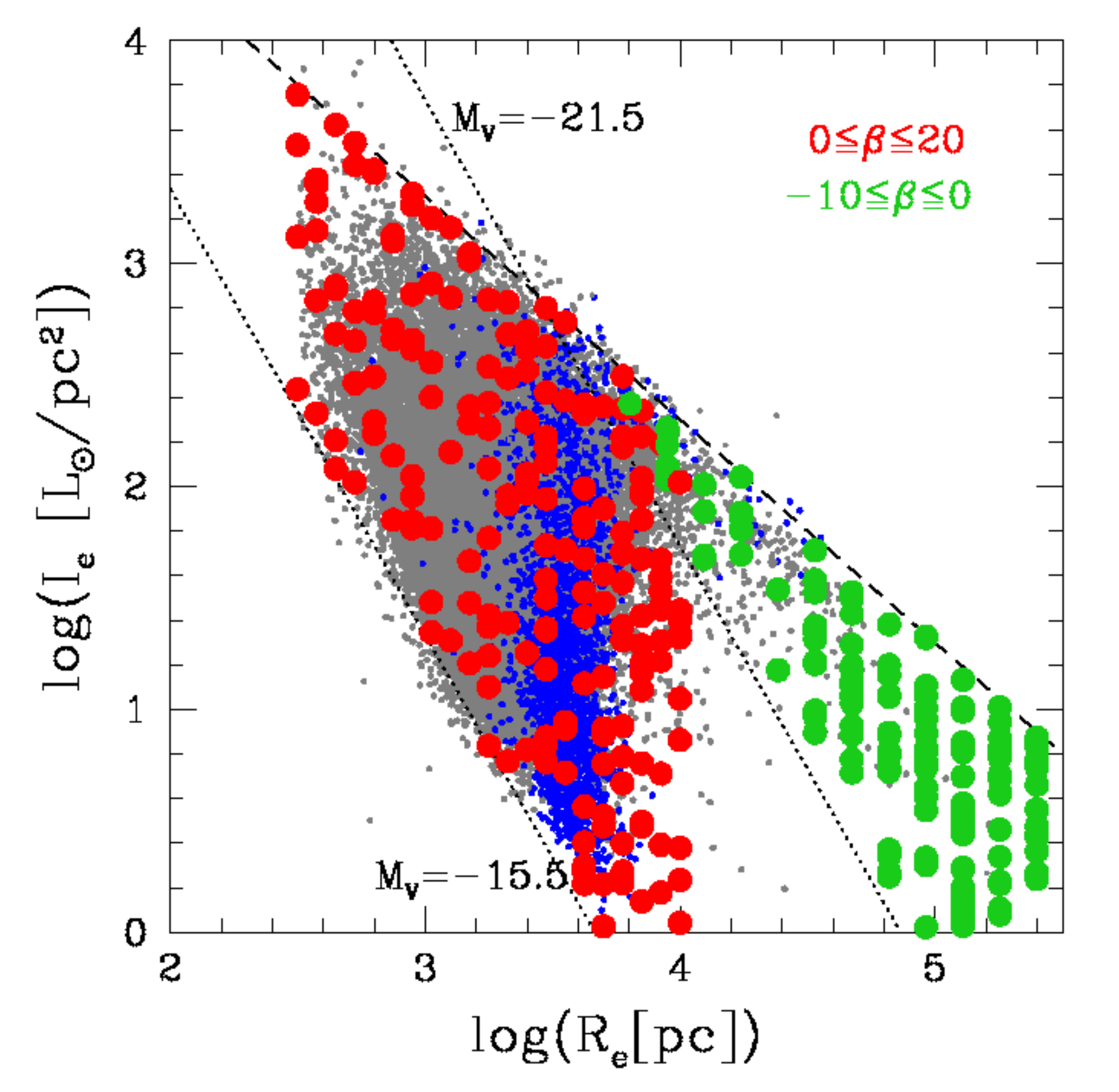}
	\end{center}
	\caption{The KR plane. The gray dots mark the observational data of the WINGS survey. The red  and green dots mark the values of $I_e$ and $R_e$ obtained starting  from eq. \ref{eq1} \citep[see][]{Donofrioetal2021} using different values of $\beta$. The dashed line represents the ZoE. The dotted lines the locus of constant luminosity, respectively at $M_V=-21.5$ and $M_V=-15.5$.}\label{fig:2}
\end{figure}

There are indeed two linear scaling relations involving the structural parameters of ETGs: the  $M_B - \mu_0$ (i.e. total luminosity vs central surface brightness)  and the $M_B - n$ ( total luminosity vs S\'ersic index). These relations do not show evident signs of curvature. The first one is a re-statement of the concentration classes introduced by \citet{Morgan1958}, later quantified by the concentration index $C$ \cite{Fraser1972,BinggeliSandageTarenghi1984,Kent1985,Ichikawaetal1986}. The second is a consequence of the first, being the S\'ersic parameter a measure of the radial concentration of galaxy light. Further examples of the $M_B–n$ diagram can be found in the literature \cite{Caonetal1993,YoungCurrie1994,Grahametal1996,Jerjenetal2000, Ferrareseetal2006b,Kormendyetal2009}. The lack of curvature in these diagrams does not support the view of different formation mechanisms at work for the formation of ETGs.
This debate is controversial: the original “nature” \cite{Eggenetal1962} (monolithic collapse) versus “nurture” (formation through mergers) \cite{ToomreToomre1972,SearleZinn1978,Schweizer1986} idea is still open.

Another interesting feature of the $I_e - R_e$ diagram, well visible in Fig. \ref{fig:2} is the presence of a zone of exclusion (ZoE). Note that there are no galaxies in the upper part of the diagram. The distribution of galaxies seems limited in the maximum surface brightness at each $R_e$. The slope of this line of avoidance is approximately $-1$ in these units, i.e. very close to the slope of the fitted KR for the brightest galaxies ($\sim -1.5$). First noted by \citet{Benderetal1992} in the $k$-space version of the FP, the ZoE was written as $k_1+k_2\leq7.8$.

Recently \citet{Donofrioetal2021} demonstrated that the FJ relation (with $L_0$ and $\beta$ nearly constant) is incompatible with the distribution observed in the KR plane.  On the contrary the use of the modified FJ relation expressed by eq. \ref{eq1} is perfectly compatible with the data (see Fig. \ref{fig:2}).
This means that the parameters $L'_0$ and $\beta$ must be variable factors depending on the mass assembly history of galaxies. Note how the complex distribution of galaxies in the $I_e - R_e$ plane is well reproduced by assuming different values of $L'_0$ and $\beta$. The negative values in particular are able to explain the tail formed by bright and massive objects in a quenched state of evolution.

Under this perspective the appearance of the $I_e - R_e$ plane is also connected to the mass assembly and stellar evolution history of galaxies. The tail of bright galaxies appears only at recent cosmic epochs, when some big objects start to quench their star formation and their luminosity begins to slowly  decrease.

\section{The Mass-Radius relation: a path toward virial equilibrium}
\label{MR}

\begin{figure}[h!]
	\begin{center}
		\includegraphics[width=12cm]{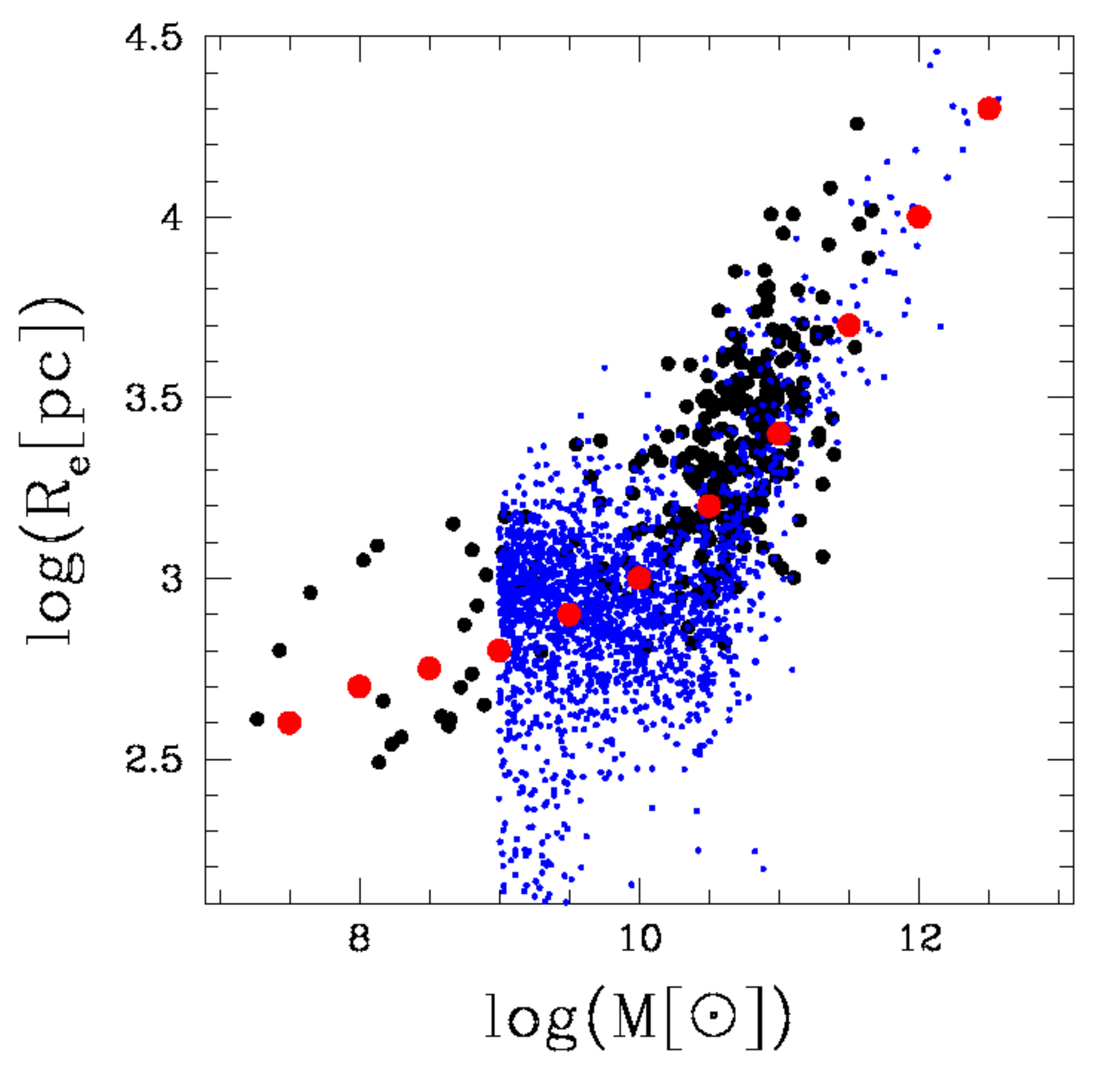}
	\end{center}
	\caption{The MR plane. The black dots mark the observational data of the WINGS survey. The blue dots are the Illustris data of the TNG release shifted by a constant value in $\log(R_e)$ of $-0.45$ (simulations still provide systematically larger radii). The red dots mark the values of $R_e$ obtained from eq. \ref{eqMR} (see text).}\label{fig:3}
\end{figure}

A considerable number of works have been dedicated in the past years to the MR relationship, i.e.  the plot of the stellar mass of the galaxies $M_s$ versus the effective radius $R_e$ in log units
\cite[see e.g.][]{Bernardietal2011,Graham2011,Shankaretal2013,Graham2013, Bernardietal2014,AgertzKravtsov2016, Kuchneretal2017,Huangetal2017, Somervilleetal2018,Geneletal2018,SanchezAlmeida2020,Terrazasetal2020}.
The  increasing interest for the MR relation is due to the difficulty of explaining the observed distribution with the virial theorem and the various models of galaxy assembly predicted by the monolithic and hierarchical scenarios, in particular the curved shape, progressively steeper for the high masses, and the zone of exclusion (ZoE), that is, a region empty of any object on the side of the high masses (see Fig. \ref{fig:1} lower right panel and Fig. \ref{fig:3}). This nontrivial distribution is well apparent even when globular clusters (GCs) and  clusters of galaxies (CGs) are added to the diagram \cite{Chiosietal2020}.

Many papers have already emphasized that the distribution of galaxies in this plane depends on several factors, such as age \cite{Valentinuzzietal2010},  mass-to-ligh ratio \cite{Cappellari2015},  color, S\'ersic index and velocity dispersion \cite[see e.g.][]{Donofrioetal2020,SanchezAlmeida2020}. The distribution of the sizes has been approximated  with a log-normal function \cite{Shenetal2003}, noting that it is clearly different for late- and early-type galaxies. The MR relation is roughly a single power law for the bright ETGs ($M_s > 10^{10} M_{\odot}$), while for the LTGs and DGs the relation is significantly curved, with brighter galaxies showing a faster increase of $R_e$ with $M_s$. For low -mass LTGs the trend is $R_e \propto M^{0.14}$, while for the high-mass galaxies we have $R_e \propto M^{0.39}$. The dispersion around the mean relation is high for low-mass galaxies ($\sim0.5$) and smaller for big objects ($\sim0.3$). For the ETGs the mean relation is $R_e \propto M^{0.56}$, with a slope going progressively toward 1 for galaxies more massive than $\sim10^{10} M_\odot$. Spirals do not seem to have objects along this linear tail \cite{Donofrioetal2020}.

According to \citet{Shenetal2003} the observed MR relation for LTGs can be attributed to the specific angular momentum (AM) of the stars, if it is similar to that of the halo and if the fraction of baryons that form stars is similar to that predicted by the standard feedback models. For ETGs, the observed MR relation is not consistent with the hypothesis that they are the remnants of major mergers, while it seems consistent with that of multiple mergers. {One possibility is that the spheroids below a characteristic mass $M_s \sim 10^{10} M_\odot$ grow from disk instability and mergers, while galaxies above it from dry mergers. Gas dissipation, if present, contribute efficiently to shrink the size of the galaxies \cite{Shankaretal2013}}.

The pronounced curvature of the MR relation suggests again a dichotomy between 'bright' and 'ordinary' ETGs as in the case of the $M_B–\langle\mu\rangle_e$ diagram and the KR plane. A possible explanation invokes the role of supernova-driven winds blowing out the gas from the DGs \cite{MathewsBaker1971,Saito1979,DekelSilk1986}. This feedback effect is one of the most efficient way of puffing-up galaxies sizes. 
However, these studies do not take into account the {gravitational binding energy} of the DM halo \cite{MacLowFerrara1999}, {so that other mechanisms should be sought to explain the discontinuity present in these relations. The discontinuity is not seen in fact in the luminosity-metallicity relation \cite{DekelSilk1986,Mateo1998,Tremontietal2004,Veilleuxetal2005} and is only marginally visible in the $L-\sigma$ relation.} 

More recently \citet{Donofrioetal2020} found a unique explanation for the curved shape of the MR and KR relation in combination with the almost linear trend of the $L-\sigma$ relation. They used 
the modified FJ relation $L=L_0'\sigma^{\beta}$ introduced above that is able to reproduce the curved MR relation and $I_e - R_e$ distribution once coupled with the virial equation. In this case one gets the relation:

\begin{equation}
	R_e = \left(\frac{1}{\frac{k_v}{G}\left(\frac{2\pi\langle I_e \rangle}{L'_0}\right)^{2/\beta}}\right)^{1/\left(4/\beta+1\right)}
	M^{1/\left(4/\beta+1\right)}.
	\label{eqMR}
\end{equation}

and should accept the idea that the parameters $L'_0$ and $\beta$ are variable factors for each galaxy depending on the mass assembly history, with $\beta$ that can assume both positive and negative values (see Fig. \ref{fig:3}). The advantage of this approach is that, in addition to the almost perfect reproduction of the observed SRs, it naturally predicts the existence of the ZoE as the locus of virialized and passively evolving quenched objects. Look at the red dots obtained by eq. \ref{eqMR}. The slope of the MR progressively changes from DGs to giants, converging toward the value of 1 for the bright and massive quenched objects in full virial equilibrium.

In this framework the key role of shaping the SRs is played by the merging and stripping events at play {during galaxy encounters}. These events may change either the luminosity and the radius of a galaxy (increasing or decreasing them). However, while luminosity rarely increases (decreases) by a factor of two ($\sim0.3$ in log units), the radius may change considerably (up to a factor 10). This explains why the $L-\sigma$ relation does not change its linear shape and scatter (that is approximately $\sim0.4$). On the contrary in the SRs where the effective radius $R_e$ is an explicit parameter, a strong curvature distinguishing DGs and giants, is clearly present. \citet{SanchezAlmeida2020} well showed that the MR relation changes its shape and scatter when different radii (probably much closer to the virial radius) are used instead of $R_e$.

When galaxies encounters result in significant stripping of stars and gas,
the total luminosity of the galaxies and the velocity dispersion decrease. The same effect is induced by the quenching of SF and passive stellar evolution, producing values of $\beta$ that can be negative. Notably this scenario is confirmed by numerical simulations \cite{Donofrioetal2021}.
{These also predict that the MR relation evolves with the cosmic epochs, since galaxies are much more dense and smaller in size at earlier epochs}. 
The galaxy size-luminosity relation and the MR relation were then used to argue that the compact ($R_e<2$ kpc) massive ($M_s >10^{11} M_{\odot}$) spheroidal-shaped galaxies at high-redshifts ($z \sim 2 \pm 1$) - known as “red nuggets” \cite{Damjanovetal2009} -  evolved into the large massive ellipticals in the local ($z=1$) Universe \cite{Daddietal2005,Krieketal2006, Trujilloetal2006, vanDokkumetal2008}. 
These massive galaxies (with stellar mass $M_s >3\times10^{10} M_{\odot}$),  evolving passively at redshifts $z \geq 1$, have average sizes smaller by a factor of $\sim3$ with respect to local ETGs with similar stellar mass. Such small sizes are expected if dissipative collapses occur. 

The small objects seen at high redshift are $2\div6$ times more compact than  local galaxies of similar stellar mass \cite{vanDokkumetal2010, Saraccoetal2011}, but observations have now established that many ETGs at high redshifts are not compact and that similar fractions of large and compact galaxies could co-exist \cite{Mancinietal2009, Valentinuzzietal2010}, with a variety of bulge- to-disk ratios \cite{vanderWeletal2011}. 

{From} the analysis of the spectra of 62 ETGs at high redshifts \citet{Saraccoetal2011} found that compact galaxies have most of their stars formed before $z=5$, while larger objects at fixed stellar mass are generally younger. 
\citet{Grahametal2015} identified 24 “compact massive spheroids” as the bulge component of local lenticulars. These bulges have a similar distribution of size, mass, and S\'ersic indices as the high-z compact massive galaxies, and comparable number densities (per unit volume) \cite{delaRosaetal2016}. This similarity strongly suggests that the current evolutionary scenario does not explain the complete picture. 

Another possibility is that the evolution of the red nuggets is driven by the growth of disks \cite{Caldwel1983b,Morgantietal2006,Sancisietal2008, Stewartetal2009,Pichonetal2011, Moffettetal2012,Starketal2013,Grahametal2015,Kleineretal2017}. Gas accretion plays a key role for massive galaxies \cite{Feldmannetal2016}, while less massive objects  accrete a small quantity of gas with time  \cite{Cowieetal1996,Grahametal2017}.

The rapid stripping or ejection of baryonic matter (BM) might inflate galaxies to larger dimensions. The idea came from \citet{BiermannShapiro1979}, who linked the formation of S0s to that of disk galaxies. Recently \citet{Ragone-FigueroaGranato2011} explained the existence of red-nuggets with this mechanisms. The loss of BM could be triggered by quasars (QSO) and/or starburst-driven galactic winds or can be quiet for stars at the end of their evolution. 
In this scheme compact galaxies could transform into less massive and larger systems. Numerical models only approximately follow this scheme: the models show intense episodes of SF and significant galactic winds but, on average, the trend is toward larger masses and almost constant radii. 

\citet{Guoetal2009} and \citet{vanDokkumetal2010} investigated the possibility that the MR relation, at least for the most massive galaxies, is linked to a systematic variation of the S\'ersic index $n$, parameterising the surface density profiles with the redshift. According to \citet{vanDokkumetal2010} the variation of the effective radius $R_e$ (50\% of the light) is:

\begin{equation}
\frac{d\log(R_e)}{d\log(M)}\approx 3.56\log(n+3.09)-1.22
\label{eqRMn}
\end{equation} 

which is accurate to 0.01 dex for $1 \leq n \leq 6$. This means that the radius might increase linearly with the mass if the projected density follows an exponential law, going as $M^{1.8}$ for the de Vaucouleurs profile with $n=4$. A strong evolution in $R_e$ is expected in all inside-out growth scenarios, unless the density profiles are close to exponential. 

{
\section{The MR relation in cosmological context}\label{MR_cosmo}

A new explanation of the existence and curvature of the MR relation (thereinafter MRR) has been given recently by \citet{Chiosietal2020} in the cosmological context of galaxy formation and evolution. They started from the empirical hint that a unique MRR seems to connect objects from Globular Clusters (GCs) to dwarf galaxies (DGs), early type galaxies (ETGs) and Spiral Galaxies (LTGs), and finally Clusters of Galaxies (CoGs), the stellar masses $M_s$  and radii $R_e$ of {which span} about twelve and four orders of magnitude, respectively. 

The data used by \citet{Chiosietal2020} are those of \citet{Bursteinetal1997} for GCs, galaxies in general, and CoGs, of \citet{Bernardietal2010} for ETGs, and of  WINGS for ETGs and CoGs. {The situation is visible} in Fig. \ref{mass_radius}, where the pale-blue filled circles show the observational data with no distinction among the different sources. {The sea-green filled circles are the Illustris models. Note that the minimum mass of the Illustris galaxies at $z=0$ is $10^9 M_\odot$, so the comparison with the observational data should be restricted to this mass limit.} The figure shows the region of the MR-plane populated by real objects of different mass, size and morphological type. Let us quickly summarize  the  main features of the MR-plane:

(i) The family of GCs is well detached from that of normal/giant ETGs (with mass larger than about $ 10^{10}\,\, M_{\odot}$). The region in between is populated by DGs and at the top of the distribution there are the CoGs  with the largest radii and masses. The ETGs are the most numerous and the LTGs occupy more or less the same region, but are not visible in the bright tail. The relative number of objects per group is not indicative of the real number frequencies because severe selection effects are present. 
The best fit of the three samples of data yields linear relations with much similar slopes and zero points (they differ by 0.1 and 1.2, respectively). Therefore, one can consider them as fully equivalent and adopt the one derived from the sample of \citet{Bernardietal2011} as the reference case for his richness 
\begin{equation}
	\log R_{e} = (0.537 \pm 0.001)\,\log M_{s} - (5.26 \pm 0.01).  
	\label{RsMs_Bernardi}   
\end{equation}

(ii) Extrapolating the relation for massive ETGs, eqn.(\ref{RsMs_Bernardi}), downward to GCs and upward to CoGs, one notes that it provides a lower limit to GCs,  passes through  $\omega$Cen  and M32,  marks the lowest limit for the distribution of DGs, and finally reaches the region of CoGs.

(iii) There are no objects in the semi-plane for radii $R_e$ smaller than the values fixed by relation (\ref{RsMs_Bernardi}), independently of mass, but for the ``compact galaxies''  \cite[see][]{Chiosietal2012}.

\medskip
\subsection{The MRR of theoretical models }\label{theo_mod}

The situation is more complicated for galaxy models. The monolithic hydrodynamic models by \citet[][shortly indicated CC-A and CC-B ]{ChiosiCarraro2002} and the early-hierarchical models by \citet[][shortly indicated M-M]{Merlin2012} provide the following MRRs:  
\begin{eqnarray}
	\log R_e& =&   0.331 \log M_{s}  - 3.644       \qquad {\rm \,\, CC-A}    \label{mr_3a}  \\
	\log R_e& =&   0.273 \log M_{s}  - 1.994       \qquad {\rm \,\, CC-B}    \label{mr_3b} \\
	\log R_e& =&   0.241 \log M_{s}  - 1.750       \qquad {\rm \,\,  M-M }    \label{mr_4}
\end{eqnarray}
\noindent
We recall that the three groups of models (identical in the input physics) are calculated with different formation redshift $z_f$ (hence  initial density):  CC-A have $z_f \simeq 5$, CC-B $z_f \simeq 1$, and M-M $z_f \simeq 1-2$.  In the MR-plane they lay on lines with similar slope but different zero point. This suggests  that the slope  is linked to the physical structure of the models while the zero-point is reminiscent  of the initial density. 
Surprisingly, the  slopes of the above relations are not identical to that of ETGs  (eq. \ref{RsMs_Bernardi}), but close to that of DGs. Furthermore, along the sequence of each group, the duration of the star formation activity is long and in a burst-like mode of low intensity in low mass galaxies and short and intense (often a single burst of activity) in the high mass ones. Remarkably, only the most massive galaxies formed in redshift interval $5 \geq  z_f \geq 2$, in which star formation has ceased long ago, may  fall into the region of ETGs. 

\begin{figure*}
	\centerline{
		\includegraphics[width=12cm,height=12cm]{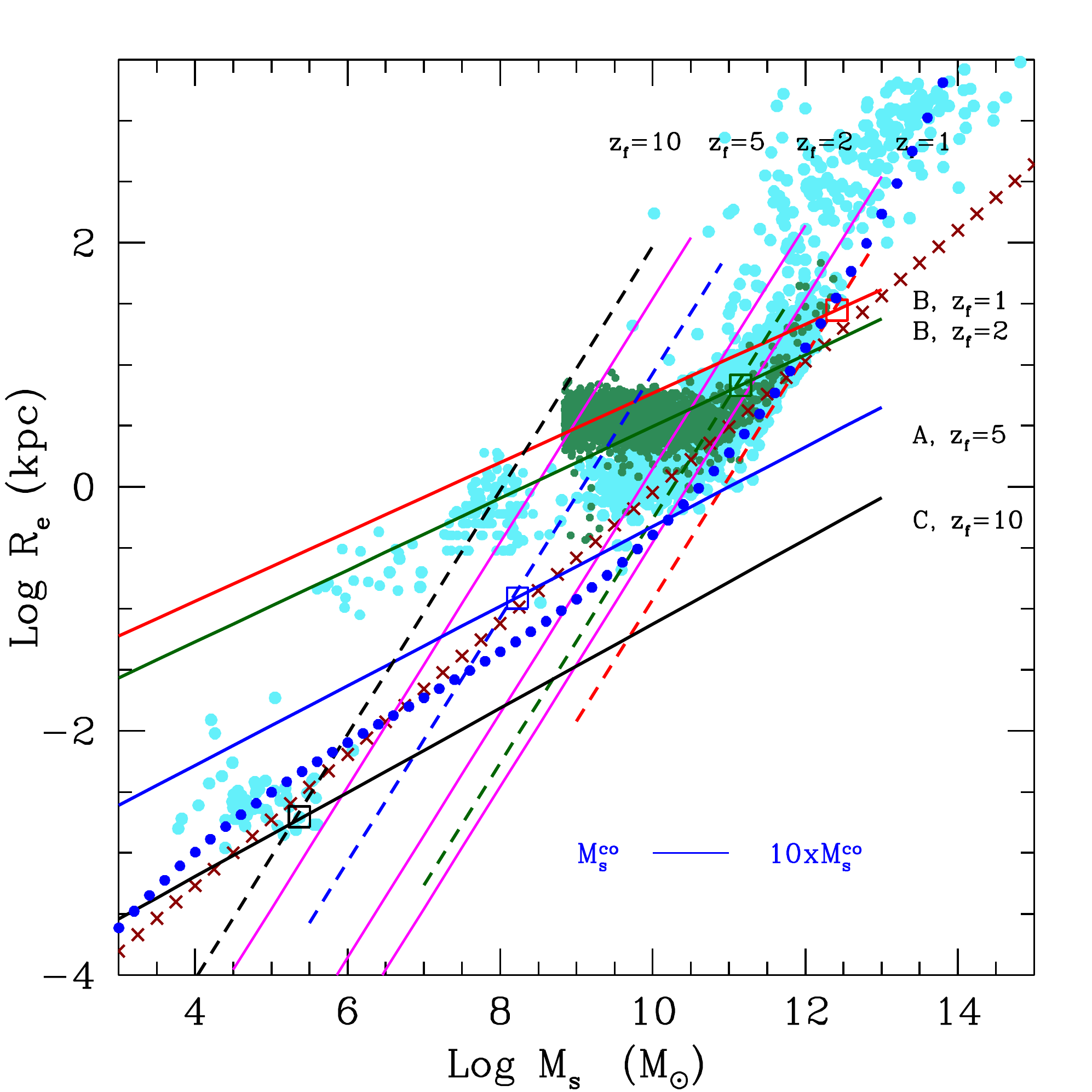}  }
	\caption{ The Mass-Radius plane. Comparison between data and theory. {Radii $R_{e}$ and stellar masses $M_s$ are in kpc and $ M_{\odot}$, respectively. The pale-blue filled circles are the observational data, the sea-green filled circles  the models of Illustris.} The stellar masses of the observational data that refer to objects from GCs to CoGs span the range $10^4$ to $10^{14} M_\odot$ while the theoretical data that are designed to represent  galaxies span the mass range  $10^8$ to $10^{12} M_\odot$. The theoretical data overlap the observational ones for ETGs and partly also for DGs.
		The linear best fit of  normal ETGs ($M_T \geq 10^{10}\, M_{\odot}$) given by eq.(\ref{RsMs_Bernardi}) is the dark-red thick  crossed line that we prolonged down to the region of GCs  and upwards to that of CoGs.
		The four solid lines  labeled A ($z_f =5$, blue), B ($z_f=1$, red and $z_f=2$, dark green) and C ($z_f=10$ black) are the analytical  relationships of eq. (\ref{re_mstar}). They show the loci of galaxy models with different mass but constant initial density for different values of redshift of galaxy formation $z_f$. These lines are the best fit of the  models by \citet{ChiosiCarraro2002}, \citet{Merlin2012} and \citet{Chiosietal2012}.
		The magenta solid lines visualize the locus of  virialized objects on the MR-plane for different values of the stellar velocity dispersion (50, 250, 500 km/s from left to right).
		The dashed black lines for different values of $z_f$ are the MRRs expected for galaxies with total mass equal to $50\times M^{CO}(z)$, the cut-off mass of the Press-Schechter at varying $z_f$ according to relation (\ref{m_knee}).   
		The large empty squares mark the intersections between the lines of constant initial  density and the MRRs for $50\times M^{CO}$ galaxies for equal values of the redshift. All the intersections  lie very close to the relation of eqn.(\ref{RsMs_Bernardi}) shown by the dark-red crossed line. This is the linear interpretation of the observed MRR.
		Finally, the curved blue dotted  line  shows the expected MR relation for the baryonic component of DM halos whose mass distribution follows the cosmological HGF by \citet{Luckic2007}.  The curve has been extended to include the GCs and the CoGs.  
		Note the changing slope of the MRR passing from CoGs to ETGS and GCs. Remarkably the curved line first runs very close to the large empty squares, second to linear fit of the data (crossed line), and third  accounts for the observed MRR passing from GCs to CoGs (about ten orders of magnitude difference in the stellar mass). 
		Finally, the horizontal blue line gives the interval for $M_s$ corresponding to initial masses $M^{CO}(z) < M_T < 10\times M^{CO}(z)$ (the percentage amounts to $\simeq 15\%$). It highlights that at each redshift the high-mass  edge of the MRR has a natural width.
	}
	\label{mass_radius}
\end{figure*}

The Illustris hierarchical models provide similar relationships, once they are split in two groups:
\begin{eqnarray}
	\log R_e &=&   0.297 \log M_s  - 2.513   \qquad {\rm for \,\, \log M_s \leq 10.5,}     \label{mr_5b} \\
	\log R_e &=&   0.519 \log M_s  - 4.492   \qquad {\rm for \,\, \log M_s \geq 10.5. }    \label{mr_5a}
\end{eqnarray}
The first  relation holds for the vast majority of models and reminds that of normal DGs, while the second one holds for a small group of objects and is close to the case of ETGs. In the hierarchical scheme the models of the first group  (in eq. \ref{mr_5b}) are the seeds of those in the second group located along the MRR of eq (\ref{mr_5a}).

Finally, there is the  MRR  proposed by \citet{Fanetal2010}. This is derived in the following way.  Independently of the monolithic or hierarchical scheme, the seeds of galaxies are perturbations of matter made of DM and BM.  These collapse when the density contrast with respect to the surrounding medium reaches a suitable value. Assuming spherical symmetry and indicating with $M_T$ and $R_T$ the total mass and associated radius, and making the approximation  $M_T=M_D+M_B \simeq M_D$ and $R_T \simeq R_D$,   the  mass-radius relation  for each individual galaxy is
\begin{equation}
	R_{D}^3 =   \left( \frac{3}{4\pi} \right) \frac{M_{D}}{ \lambda  \rho_u(z)} \rightarrow   
	R_D \propto \frac{ M_D^{1/3} } {1+z_f}
\label{new_mr_bm} 
\end{equation}
\noindent where $\rho_u(z_f) \propto (1+z_f)^3$
is the density of the Universe at the collapse redshift $z_f$, and $\lambda$  the density contrast of the DM halo. This expression has a general validity, whereas $\lambda$  depends on the cosmological model of the Universe, including the $\Lambda$CDM case. All details and demonstration of it can be found in \citet[][ their Eq. 6]{Bryan1998}. 
The collapse increases the mean density of DM and BM so that, when a critical value of the BM density is reached, stars can form at the center of the system under suitable star formation rates. 
In the context of the $\Lambda$CDM cosmology, \citet{Fanetal2010} have adapted the general relation  (\ref{new_mr_bm}) to provide an equation  connecting the halo mass $M_{D}$, the  stellar mass $M_s$, the half light (mass) radius $R_{e}$,  the shape of the BM $S_S(n_S)$ related to the S\'ersic profile  index  $n_S$,  the velocity dispersion $f_\sigma$ of the BM component with respect to that of DM, and finally the ratio  $m=M_{D}/M_s$. The expression  is
\begin{equation}
	R_{e}=0.9 \left(\frac{S_S(n_S)}{0.34} \right) \left(\frac{25}{m}\right) \left( \frac{1.5}{f_\sigma} \right)^2 
	\left( \frac{M_{D}}{10^{12}  M_\odot} \right)^{1/3} \frac{4}{(1+z_{f})}.
	\label{mr3}
\end{equation}
\noindent where $f_\sigma$ yields the three dimensional stellar velocity dispersion as a function of the DM velocity dispersion  $\sigma_s=f_\sigma \sigma_{D}$ (here we adopt $f_{\sigma}=1$). The typical value for  $S_S(n_S)$ is 0.34.   For more details see \citet{Fanetal2010} and references therein.

The most important parameter of eq.(\ref{mr3}) is the ratio $m= M_{D}/M_s$. Using the Illustris data \citet{Chiosietal2020} investigated how this ratio varies in the mass interval  ${8.5} <  \log M_{D} <{13.5}$ (masses are in $M_\odot$) and from $z=0$ to $z=4$ (see Sect. \ref{stellar_to_halo}). They find that the following relation is good for all practical purposes
\begin{equation}
\log m = \log \frac{M_D}{M_s}  =  0.062\, \log M_D + 0.429.  
\label{final_m}
\end{equation}
The slope of the \citet{Fanetal2010} relation, which visualizes the position on the MR-plane of  systems born at the same redshift once their stars are formed, is 0.333. This is very similar to that of theoretical models, i.e.  eqns. (\ref{mr_3a}), (\ref{mr_3b}), (\ref{mr_4}), and (\ref{mr_5b}). 

The most intriguing question to answer is  {``Why is the observational MRR for ETGs so different from the theoretical one?''}

\medskip
\subsection{ The MRR from the DM Halo Growth Function $n(M_{D},z)$}\label{MRR_DM_HGF}

The observed distribution of astrophysical objects in the MR-plane, going from GCs to galaxies of different mass and morphological type and eventually to CoGs, suggests that a unique relation could exist for all of them and that such relation likely owes its origin to the cosmological growth of DM halos.
The distribution of the DM halos and their number density as a function of redshift has been the subject of several studies which culminated with the large scale numerical simulations of the Universe. We cite here one for all, the Millennium Simulation \citep{Springel_etal_2005}. In parallel the  studies of the \textit{halo growth function, HGF},  as the integral of the \textit{halo mass function, HMF}, appeared in  literature \citep[see for instance][]{Luckic2007,Anguloetal2012, Behroozietal2013}.  
The HGF gives the number density of halos of different mass  per (Mpc/$h)^3$ emerging at each epoch by all creation/destruction events and consequently  yields the  halos that nowadays populate  the MR-plane and generate the observed galaxies. \citet{Chiosietal2020}  adopted the HGF  of \citet{Luckic2007} who, using  the $\Lambda$CDM cosmological model and the HMF of \citet[][]{Warren2006}, derived the number density of halos $n(M_{D}, z)$ over ample intervals of halo masses and redshifts.  Since the $n(M_{D}, z)$ of  \cite{Luckic2007} refers to a volume of 1 (Mpc/$h)^3$, before  being compared with  the observational data, it must be scaled by a suitable factor in order to match the volume sampled by observations. Anyway, the following characteristics of the HGF are worth being noted: (i) for each halo mass (or mass interval) the number density is small at high redshift, increases toward the present, and reach a maximum at a certain redshift. The peak  is either followed  by a descent (for low mass halos) or  a plateau (for high mass halos). In other words, first the creation of halos {outnumbers} the destruction, whereas the opposite occurs in general for low mass halos after a certain redshift. (ii) At any epoch high mass halos are much less numerous than the low mass ones. This implies the existence of a cut-off mass at the high mass side. (iii) The HGF  also implies  that halos of different mass have a given probability of existence at any redshift  \cite[see for more details][]{Chiosietal2012,Chiosietal2020}.

Assuming a certain number density of halos $N_s$ derived from the observational data,  \citet{Chiosietal2020} set up the equation $n(M_{D}, z)= N_s$ whose solution yields the mass of the halos $M_D(z)$ as a function of the redshift  and vice-versa the redshift for each halo mass. In practice for any value $N_s$ one gets a function $M_D(z)$. To each value of $M_D$ along this function, with the aid of eqns. \ref{mr3} and \ref{final_m}, one can associate a value of $M_s$ and $R_e$.  The MRR of luminous galaxies is the result.

Notably for the $N_s$ corresponding to $10^{-2}$ halos per (Mpc$/h)^3$ (roughly the volume surveyed by the SDSS, \cite[see][for details]{Chiosietal2020}), the curve $R_e(M_s)$ falls at the edge of the observed distribution of ETGs in the MR-plane. Higher $N_s$ would shift the curve to larger halos,  the opposite  for lower $N_s$.  
One can therefore draw in the MR-plane the locus of the most  massive $M_D$ and associated $M_s$ imposed by the  halo HGF. The equation $n(M_{D}, z)= N_s$ with $N_s =10^{-2}$  or equivalently $10^6$ halos per $10^8$ Mpc$^3$ rewritten to  derive the halo mass $M_{D}$ as a function of $z$ is
\begin{equation}
	\log M_{D} = 0.0031546\, z^3 - 0.006455\, z^2  - 0.183\, z + 13.287\, .
	\label{eq_ns}
\end{equation}

\noindent 
Starting from this, \citet{Chiosietal2020} associate  $M_s$ and $R_e$ to each $M_D$ for any value of the redshift. The best fit of the resulting MR relation, limited to the mass interval of normal ETGs,
$9.5 \leq \log M_s \leq 12.5$ ($M_s$ in solar units), is

\begin{equation}
	\log R_{e} =  0.048562 (\log M_s)^3 -1.4329 (\log M_s)^2 + 14.544 (\log M_s) -50.898.
	\label{shepherd_rel}
\end{equation}

\noindent  Note that (i) the locus on the MR-plane predicted by $N_s= 10^{-2}$ halos per (Mpc$/h)^3$  nearly coincides with the observational MRR; (ii) the slope gradually changes from 0.5 to 1 going from low masses to high masses in agreement with the observational data \citep[see][and references therein]{vanDokkum2010}; (iii) finally,  eq. (\ref{shepherd_rel}) is ultimately linked to the top end of the halo masses (and their associated baryonic objects) that might exist at each redshift. \citet{Chiosietal2020} named this locus the {\it Cosmic Galaxy Shepherd}. 

The extrapolation of the Cosmic Galaxy Shepherd downward to GCs and upward to CoGs yields the relation 
\begin{equation}
	\log R_e =0.007584 [\log (m\cdot M_s)]^3 - 0.1874\,[\log (m\cdot M_s)]^2  + 1.908 [\log (m \cdot M_s)] - 9.027
	\label{good_relation}
\end{equation}
where $R_e$ and $M_s$ are in the usual units and $m$ is the ratio $m = M_D/M_s$, for which a mean value of $m=25$ is adopted   
\footnote{In \citet{Chiosietal2020}  the same expression is written as $\log R_e =0.007584 (\log M_s)^3 - 0.1874\,(\log M_s)^2  + 1.908 (\log M_s) - 9.027$, in which by mistake the term $(\log M_s)$ does not contain the factor $m$.}. 
As already said this equation represents the cut-off mass of the HDF at different  redshift, however translated into the $R_e$ vs $M_s$. This gives a profound physical meaning to the line splitting the MR-plane in two regions, i.e. the region where galaxies are found, and that of avoidance, the so-called Zone of Exclusion (ZoE) found by \citet{Bursteinetal1997}.

Along the Cosmic Galaxy Shepherd, cut-off masses and redshift go in inverse order: low masses (and hence small radii) at high redshift and vice-versa. More precisely,  halos and their luminous progeny that are born (collapse) at a certain redshift and are now located along the theoretical MRR of eqn. (\ref{mr3}) associated to that redshift. Along each MRR only masses (both parent $M_{D}$ and daughter $M_s$) smaller than the cut-off mass are in place, each of these with  a different occurrence probability. Clearly the low mass halos are always more common than the high mass ones.  We will argue that in the MR-plane, only the most massive  GCs, DGs, and ETGs  are expected to fall along   the Cosmic Galaxy Shepherd. All other objects of lower mass, the DGs in particular, are expected to lie above this limit. This suggests that there are other physical processes concurring to shape the observed MRR. In other words, the question is ``what really determines the position of each galaxy on the MR-plane?''

To answer the above question \citet{Chiosietal2012, Chiosietal2020} argue what follows. The  gravitational collapse of a proto-cloud generating a luminous galaxy is surely accompanied by star formation, energy feed-back, gas cooling and heating, loss of  mass and energy by winds, acquisition of mass and energy by mergers, etc. Therefore the result of all these  processes taking place together may  largely differ from one case to another and also differ from  the ideal case of a dissipation-less collapse. For this latter  \citep{GottRees1975,Faber1984,Bursteinetal1997}  derived the relation
\begin{equation}
	R_{D}   \propto M_{D}^{0.53}. \
	\label{delta_MR4}
\end{equation}
Inside this halo a galaxy with stellar mass $M_s$ and a half-mass radius $R_e$ is built up over the years. 
\citet{Chiosietal2020} take the dissipation-less collapse as the reference case.
Using the data of the Illustris models, they derive the following MRRs 
%
%
\begin{eqnarray}
	\log R_e &=& 0.541 \log M_s  -4.702 + k_m   \quad  {\rm for} \log M_s > 10.5   \label{delta_MR6a} \\
	\log R_e &=& 0.102 \log M_s  -0.017 + k_d   \quad  {\rm for} \log M_s < 10.5   \label{delta_MR6b}
\end{eqnarray}
where the  constants $k_m$ and $k_d$ can be determined by fixing the  initial conditions of the collapsing proto-halo. The slope of eq. (\ref{delta_MR6a}) does not significantly differ from that of the dissipation-less collapse, eqn. (\ref{delta_MR4}) and  that of the empirical MRR of ETGs, eqn.(\ref{RsMs_Bernardi}).  Along  each  MRR of the theoretical manifold, the agreement between data and theoretical models seems to be  possible only for the most massive galaxies. For smaller masses, the slope of the theoretical MRR, eqn. (\ref{delta_MR6b}), is much flatter than the observational one (about a factor of two). 

From the above considerations one could suggest that the Cosmic Galaxy Shepherd and eqn. (\ref{RsMs_Bernardi}) represent the locus in the MR-plane of galaxies formed by quasi dissipation-less collapses.  In contrast,  special conditions ought to hold for all other objects  that  deviate from this condition. The explanation is  different for the monolithic and hierarchical scenarios:

a) In the monolithic view, in addition to star formation, galactic winds are the key ingredient to consider, in particular for low mass galaxies, because DGs  show the largest deviation from the observed MRR, eq. (\ref{RsMs_Bernardi}) or eq. (\ref{delta_MR4}). 
%
%
%
%
The analysis of the problem made by \citet{Chiosietal2020} shows that:
(i) the stronger the galactic wind the larger is the final $R_{e}$. Galaxies depart from the locus represented by eq. (\ref{RsMs_Bernardi}) and/or eq.(\ref{delta_MR4}) at decreasing mass and increasing galactic wind, the low mass ones having the strongest effect; (ii)  the efficiency of winds tends to decrease at increasing initial density. This means that the inflating effect of galactic winds in low mass galaxies of high initial density is low and the final radius of these galaxies will be close to the value predicted by eqs. (\ref{RsMs_Bernardi}) and/or (\ref{delta_MR4}). In conclusion the flatter slope of the theoretical MRR is likely produced by galactic winds.

b) In the hierarchical scenario the situation is more entangled because both mergers and galactic winds  concur to inflate a galaxy. To clarify the issue \citet{ChiosiCarraro2002} discussed the merger between two disk galaxies calculated by \citet{Buonomo2000}. In this case  an elliptical galaxy is generated with twice total mass of the component galaxies, but with  stellar mass and effective radius smaller and higher, respectively, by $\Delta M_s/M_s\simeq -0.9$ and $\Delta R_{e}/R_{e}\simeq 0.5$, with respect to the case of an elliptical of the same mass generated during a monolithic collapse. The reason for that is identified in the enhancement of galactic winds caused by the interaction. More gas is lost, less stars are formed, and the resulting body is in a state of weak gravitational energy.

\begin{figure}
	\centerline{
		\includegraphics[width=12.0cm,height=12.0cm]{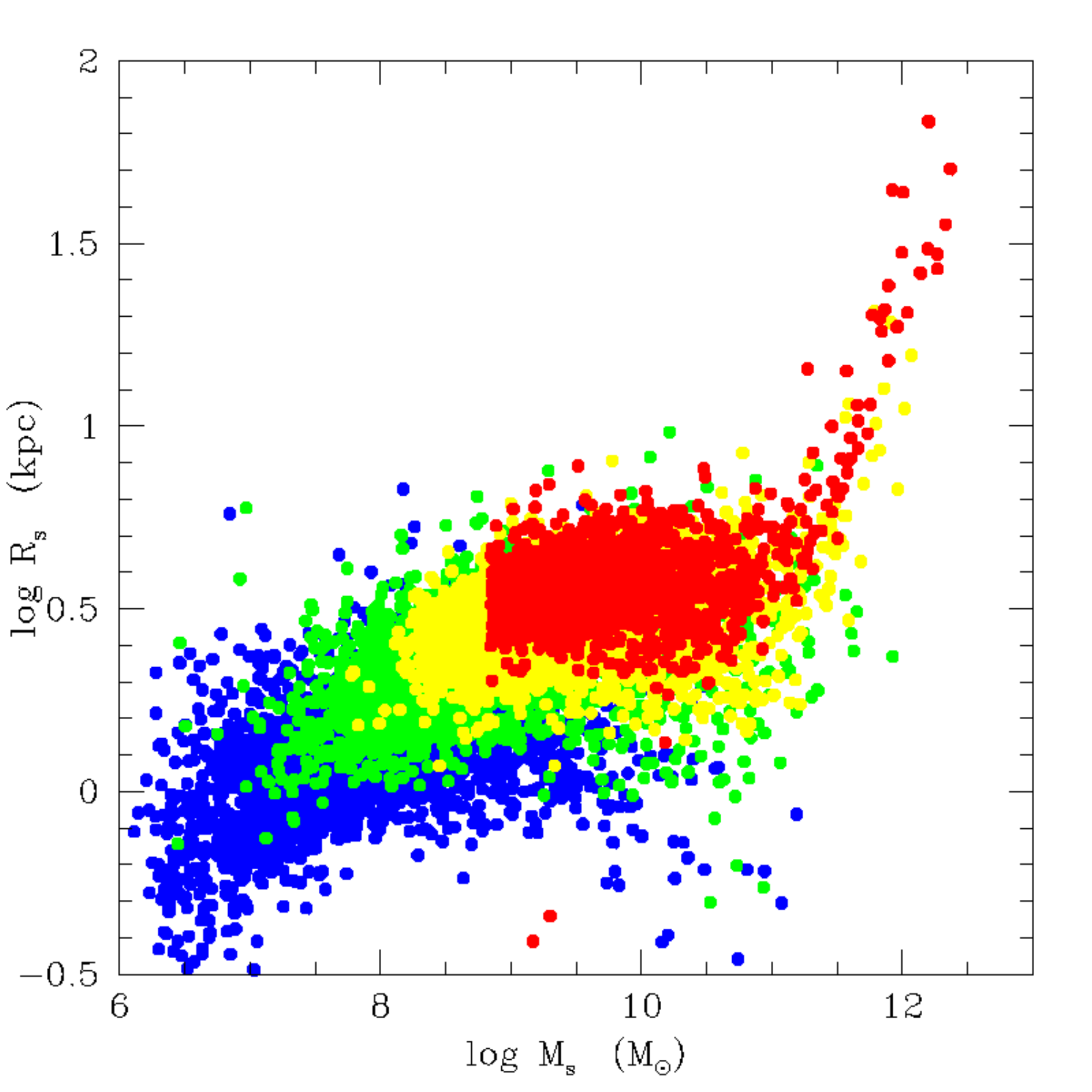} }
	\caption{The stellar  half-mass radius $R_e$ plotted vs the total stellar mass $M_s$ of  galaxy models from the Illustris database at different values of the redshift, i.e.  $z=4$ (blue), $z=2$ (green), $z=1$ (yellow) and $z=0$ (red). }
	\label{mass_radius_4z}
\end{figure}
 
When does the MRR develop in the course of time and evolutionary history of galaxies? In Fig. \ref{mass_radius_4z} we show the $R_e$ vs $M_s$ distribution of the Illustris models at four cosmic epochs. At high redshifts, the distribution is clumpy and irregular. However, starting from $z\sim1.5$ and more clearly at $z=0$, a tail-like feature develops on the side of large masses, say for masses $\gtrsim 2\cdot10^{11}\, M_\odot$.
The best fit at redshift $z=0$, using the relationship $\log R_e = \epsilon \log M_s + \eta$ (masses and radii are in $M_\odot$ and kpc), yields the following values: 
for  $\log M_s >  11.3$  $\epsilon = 0.651$ and $\eta= -6.557$, while for $\log M_s <  11.3$  $\epsilon = -0.005$ and $\eta= 0.592$.
What are the causes of  the cloud-like and tail-like distributions? Why a cloud dominates the low mass range? Why the tail is well visible only for the high masses at low redshifts?  Which is the physical meaning of this distribution?  To cast light on this \citet{Chiosietal2020} examined the history of $R_e$ and $M_s$  for several individual galaxies.
The main conclusion of their analysis is that mergers among objects of low and comparable mass can generate galaxies with larger masses and radii, but exceptions are possible in which either  the mass or the radius or both decrease. In general the galaxies do not leave the cloud region. All this does  not  contradict the previous case of \citet{Buonomo2000} because the monolithic  counterparts to compare with are not available. 
The cloud region is instead roughly coincident with the distribution of DGs of different types \citep[see the discussion by][]{ChiosiCarraro2002}.  At the same time mergers among galaxies with different masses and/or comparable masses can generate objects that shift outside the cloud producing the MR-sequence (actually they define it), the locus of which agrees with the observed distribution for ETGs \citep[see e.g.][and references therein]{Chiosietal2020}.  
The stellar content of massive ETGs suggests that star formation has ceased long ago so that strong energy feed-backs are absent and the systems are close to the virial equilibrium. This implies that important mergers do not longer occur.
At variance DGs are still undergoing frequent mergers, active star formation episodes, and  strong galactic winds. They cannot be therefore in this ideal condition of equilibrium and so they depart from the observed MRR. Nevertheless, there are some DGs that fall along the MRR of massive ETGs and therefore are likely in a similar dynamical and star forming condition, e.g. $\omega$Cen and M32 \citep[see][for more details]{Chiosietal2020}. 

On consideration of these premises, \citet{Chiosietal2020} argued that the observed distribution of ETGs,  inactive DGs and GCs, represents the locus of objects that have reached the ideal situation of mechanical equilibrium and pure passive evolution. They cannot go beyond this limit. Their MRR is therefore in the boundary between the permitted and forbidden regions of the MR-plane. 

\medskip
\subsection{Genesis of the true MRR}\label{genesis}
 
Putting the many tessarae of the mosaic together, the conclusion is that the observational MRR is the intersection of the theoretical manifold of the MRRs (each curve being labelled by the collapse redshift from the past to the present) with the Cosmic Galaxy Shepherd, along which objects in mechanical equilibrium and passive evolutionary state are located.   To prove this statement \citet{Chiosietal2020} resorted to the  method proposed long ago  
by \citet[see][]{ChiosiCarraro2002}, however updating it with recent theoretical and observational data. 
In the MR-plane of  Fig.\ref{mass_radius} they draw two loci and a mass interval as a function of the initial density (redshift):

(a) The first locus  is the MRR  traced by models of different mass but same 
initial density and formation redshift.  Using all models to  disposal \citep{ChiosiCarraro2002,Merlin2006,Merlin2007,Merlin2010,Merlin2012,Chiosietal2012,Chiosietal2020}, this locus is described by the relation
\begin{equation}
	\log R_{e} = [-1.172 - 0.412\, (1+z_f)] +  [0.244 + 0.0145 \, (1+z_f)] \, \log M_{s}.
	\label{re_mstar}
\end{equation}
\noindent
This expression is robust thanks to the regular behavior of the models and the density-mass-radius relationship of eqn. (\ref{delta_MR4}). Relation (\ref{re_mstar}) is compatible with the MRRs predicted by \citet{Fanetal2010} and the models of Illustris by \citet[][]{Vogelsbergeretal2014}. The cases shown in Fig. \ref{mass_radius} are:  $z_f \simeq 1$,  $z_f \simeq 2$, $z \simeq 5$ and  $z_f \simeq 10$. 

(b) The second locus is  the Cosmic Galaxy Shepherd. Among the various HGFs in literature  \cite{Luckic2007, Anguloetal2012, Behroozietal2013}, we adopt the  HGF of \citet{Luckic2007} and make use of the analytical expression for the Cosmic Galaxy Shepherd extending across the whole MR-plane given by eqn. (\ref{good_relation}). However, in order to better illustrate this issue, we present here an  analytical approach based on the  classical halo mass distribution of \citet{Press1974} that is supposed to trace also the mass distribution of  luminous  galaxies (assuming one galaxy per halo).  At each redshift, the HGF of \cite{Press1974} provides the relative number of galaxies per mass bin.  The cut-off mass $M_D^{CO}$ of the \cite{Press1974} function yields the  maximum limit for the galaxy masses at each redshift. In the  \citet{Press1974} formalism, the cut-off mass varies with redshift according to: 
\begin{equation}
	M_{D}^{CO} = M_{N} \times (1+z)^{ - \frac{6}{ n + 3} }
	\label{m_knee}
\end{equation}
\noindent
The exponent $n$  represents the slope of the power spectrum perturbations and $M_{N}$ is a suitable mass scale normalization. At any redshift, most galaxies have total masses smaller than $M_{D}^{CO}$, even if higher values cannot be excluded. It can be easily shown that the fractional  mass in (or the fractional number of) galaxies with mass greater than $M_{D}^{CO}$ is a function of $n$. For $n$=-1.8, the percentage of galaxies in the interval $M_{D}^{CO} < M_D < 10\times M_{D}^{CO}$ is about 15\% while in the range $10\times M_{D}^{CO} < M_D < 100\times M_{D}^{CO}$ is about 1\%.  Therefore,  at any redshift galaxy masses up to say $50\times M_{D}^{CO}$ have a significant occurrence probability. Their radius is derived with the aid of the $M_s$ vs $M_D$ and $R_e$ vs $M_D$ relationships. For   $M_D = \gamma \,\, M_{D}^{CO}$, $\gamma=50$,  and $n=-1.8$, one gets:
\begin{equation}
	R_{e}= 16.9\times 10^{-12} \times \gamma^{-0.79}\times (1+z)^{3.96} \times M_{s},
\end{equation}
\noindent
where $R_{e}$ and $M_{s}$ are in kpc and $M_{\odot}$. These are shown in Fig. \ref{mass_radius} with the dotted lines labelled by the redshifts $z\simeq 1$, $\simeq 2$, $\simeq 5$ and $\simeq 10$. On the MR-plane, they give the rightmost extension of the lines of constant density and hence they identify the maximum galaxy mass. At decreasing redshift this boundary moves progressively toward higher masses. Similar results can be obtained by means of the HGFs of \cite{Luckic2007, Anguloetal2012, Behroozietal2013}, the first of which is the Cosmic Galaxy Shepherd.

\noindent
(c) Finally, the third locus gives the expected interval for $M_{s}$  for objects with total mass $M_T$ between $M_{D}^{CO}$ and $10\times M_{D}^{CO}$ as a function of redshift. Here the relation $M_s(M_T)$ has been plugged into eq. (\ref{m_knee}) for $M_{T}^{CO}$. The permitted intervals are visible in Fig. \ref{mass_radius} by  the horizontal lines labelled $M_s^{CO}$. The interval for $M_s$ going from $10^{10}\,
M_{\odot}$ to $10^{12}\, M_{\odot}$  is fully compatible with the redshift interval for the formation of the majority of stars in a galaxy, i.e. from 2 to 1. This is also the mass range over which at any epoch the probability for the occurrence of massive galaxies falls to a negligible value. In different words, the right-hand border of the MRR has a natural width.

In this context, the relationship for ETGs, see eqn. (\ref{RsMs_Bernardi}), extended to the whole mass range from GCs to CoGs should correspond to the intersection between the lines of constant initial density  and the lines where $\gamma \, M_T=M_{T}^{CO}(z)$ for equal values of the redshift (at least for all values of redshift $>1$). This is what we see in Fig. \ref{mass_radius}, i.e. the straight line marked by the large empty squares.   This line nearly coincides with the Cosmic Galaxy Shepherd  derived from the HGF of \citet{Luckic2007} that is  marked by the crossed dark-red line in Fig. \ref{mass_radius}, i.e. eqs. (\ref{shepherd_rel}) and/or (\ref{good_relation}). Finally, this line is also coincident with the locus  traced by objects that underwent a dissipation-less collapse (or very close to it) and are nowadays in mechanical equilibrium and passive evolutionary state.
This is mainly traced by GCs,  few DGs (the large majority of DGs lie above it), ETGs, and a number of CoGs.  This confirms the result by \citet{Donofrioetal2020}: only passive galaxies (strongly decreasing today in their luminosity) trace the MRR with a slope varying {from 0.5 to 1}, the highest value being reached by galaxies that suffered the  strongest luminosity decrease with the redshift, i.e. those that long ago ceased their stellar activity, i.e. the most massive ones.  Spirals occupy approximately the same location of ETGs in the MR-plane, thus suggesting that their ongoing star formation is not affecting the overall situation of mechanical equilibrium.  Furthermore, it is worth noting that the slope of MRR derived from the HGF is about 1 in the range of massive galaxies (say above $10^{12}\, M_\odot$), i.e. formally identical to the MRR derived from the  virial theorem.  This coincidence might suggest a dependence of the observed MRR slope from the virial condition. The true driver is instead the HGF, more precisely its fall off toward high values of the halos' masses at any redshift. To conclude, all the objects along the MRR are in virial conditions and passive evolutionary state (all mechanical process and star formation activity are at rest).   

}

\section{The Fundamental Plane}
\label{FP}

In the local universe ETGs are seen to lie along a plane, the so-called “fundamental plane” (FP, \cite{DjorgovskiDavis1987,Dressleretal1987}), connecting the surface brightness within the effective radius $\langle I_e \rangle$, the effective radius $R_e$, and the velocity dispersion of stars (central or within the effective radius $\sigma_e$). The intrinsic scatter around the FP is small ($\sim0.05$ dex) \cite{Bernardietal2003, Saulderetal2013}) and the relation appears to extend across all ETGs, DGs, GCs and CGs \cite{MisgeldHilker2011,Donofrioetal2013b}. 

The FP is tilted with respect to the virial prediction. 
{The origin  of the tilt has been debated for several years}. The first attempts to explain it invoked a progressive change of the mass-to-light ($M/L$) ratio of the stellar population with galaxy luminosity, but even systematic changes of the DM fraction and the structural and dynamical non-homology of galaxies can be responsible of the observed tilt \cite[see e.g.][]{Ciotti1991,Benderetal1992, RenziniCiotti1993, Jorgensenetal1996, Cappellarietal2006, Donofrioetal2013}. 

Recently \citet{Donofrioetal2017} proposed another explanation for the tilt of the FP. In their work they demonstrated that the FP can originate from the combination of the virial theorem with the modified FJ relation given by eq. \ref{eq1}. In this case the small scatter of the plane can be obtained if it exist a fine-tuning between the zero-points of the two relations. In other words it must exist a connection between the shape and structure of galaxies and their stellar population content \cite[see also][]{Donofrioetal2011}.

The FP evolves with redshift \cite[see e.g.][]{Treuetal2005, Holdenetal2010, Sagliaetal2010, FernandezLorenzoetal2011, vandeSandeetal2014}. 
{\citet{Beifiorietal2017} using a sample of 19 massive red-sequence galaxies
at $1.39 < z < 1.61$ observed by the K-band Multi-object Spectrograph (KMOS) Cluster Survey, find that the ZP of the FP in the B-band evolves with redshift, from 0.44 (for Coma) to $-0.10\pm0.09$, $-0.19\pm0.05$, and $-0.29\pm0.12$ for clusters at $z=1.39$, $z=1.46$, and $z=1.61$ respectively. Similar results are obtained by \citet{Prichardetal2017}. The properties observed for the high redshift FP suggest an increase of the dynamical-to-stellar mass ratio by $\sim0.2$ dex from $z=2$ to the present. Consequently these data seem to indicate that the fraction of DM contained within $R_e$, compared to that seen in likely descendants objects at low-redshift, was increased by a factor $>4$ since $z\sim2$ \cite{Mendeletal2020}.
The same work suggests the use of the dynamical-to-stellar mass ratio as a
probe of the stellar IMF, finding that high-redshift data can constrain the 
IMF law.}

While the debate is still open on whether the FP coefficients are constant up to $z\sim1$ \cite[see][]{DiSeregoetal2005, Holdenetal2010, Sagliaetal2010, JorgensenChiboucas2013}, there is more consensus about the variation of these coefficients with the magnitude interval of the sampled population \cite[see e.g.][]{Donofrioetal2008} and on the variation of the zero-point with redshift  as a result of an evolving $M/L$ \cite{Faberetal1987} caused by the younger stellar population at high-z \cite{vanDokkumFranx1996, Benderetal1998, Kelsonetal2000, Gebhardtetal2003, Wuytsetal2004, DiSeregoetal2005, Holdenetal2005, Jorgensenetal2006, vanDokkumvanderMarel2007, Holdenetal2010, Toftetal2012, Bezansonetal2013} and by the structural evolution of galaxies with redshift \cite{Sagliaetal2010, sagliaetal16}. Other authors claim that there is not only a dependence of the zero point on redshift, but even the slopes of the structural relations are steeper for high redshift galaxies than for objects of the local Universe \cite{Treuetal2005, Jorgensenetal2006, Fritzetal2009}. 

As discussed in the previous section, several papers have shown that a fraction of intermediate and high-redshift galaxies have smaller sizes \cite{Trujilloetal2007, Newmanetal2012, Houghtonetal2012, vanderWeletal2014, Beifiorietal2014, Chanetal2016} and higher stellar velocity dispersions \cite{Cappellarietal2009, CenarroTrujillo2009, vanDokkumetal2009, vandeSandeetal2013, Bellietal2014} compared to their local counterparts of the same mass  \cite{Brammeretal2011, Pateletal2013, Muzzinetal2013}. Part of this difference might be attributed to environmental effects and can be observed in the FP.
The environment may have a role in accelerating the size evolution in clusters with respect to the field at $z>1.4$ \cite{Lanietal2013, Strazzulloetal2013, Delayeetal2014, Saraccoetal2014, Newmanetal2014}, while in the local Universe there seems to be no significant differences between the mean galaxy sizes in different environments \cite{Cappellarietal2013, Huertas-Companyetal2013}. {The reason for that is not clear; is it because there is not enough time for evolution?}
In clusters central  and satellite galaxies seem to lie on average above and below the FP, possibly for a higher  and lower than average mass-to-light ratio \cite{Joachimietal2015}. 

{Several studies (e.g. \cite{Franxetal2008, CimattiNipotiCassata2012}) have also suggested that the size evolution with redshift is stronger for massive galaxies ($> 10^{11} M_\odot$). This behaviour is consistent with the idea that high-density environments play a major role in size evolution. Galaxies in denser environments probably evolve earlier as indicated by the observed colour–density relation (e.g. \cite{Chuteretal2011}). It is not clear yet if the environment itself  influences the size evolution, since merging events alone do not seem to explain the observed size evolution of ETGs (e.g. \cite{Damjanovetal2009, Nipotietal2012}) or other growth mechanisms are at work, such as the adiabatic expansion due to mass-loss, that could indirectly lead to a correlation of size with environment (if it occurs at earlier epochs within the most massive dark matter halos). There is also the possibility of trends driven by faster quenching in high-density environments (e.g. \cite{Cassataetal2013}).
Whatever the reason of the size evolution, the underlying correlation is likely connected to the halo mass that is strongly related to the number of satellites (e.g. \cite{SkibbaSheth2009, Muldrewetal2012}). A full investigation of this problem requires a careful decoupling of large-scale clustering and small-scale halo occupation (e.g. \cite{Hartleyetal2013}).}

The presumed universality of the FP makes it an appropriate tool for cosmology, e.g. for the Tolman test \cite{Kjaergaardetal1993, Pahreetal1996, Molesetal1998}, or to assess the evolution of $M/L$ with $z$ \cite{Benderetal1992, GuzmanLuceyBower1993, vanDokkumFranx1996, Kelsonetal1997, Benderetal1998, Ziegleretal1999, Jorgensenetal1999, Kelsonetal2000}. 
The usefulness of the FP was recently demonstrated  in the context of weak lensing magnification \cite{HuffGraves2014}, and to map out the peculiar velocity field of galaxies \cite{Springobetal2014}. These are examples of the exploitation of the FP as cosmological probe. In such applications generally one measures the observed galaxy size and predicts it using the FP. The comparisons between predictions and observations are used to get the size changes due to lensing magnification, or the line-of-sight peculiar velocities that modify the redshift and the angular diameter distance used to obtain the physical sizes.

Again we should note that hierarchical numerical simulations, like Illustris, correctly predict a tilt of the FP and an evolution of its coefficients with redshift \cite{Luetal2020}.

\section{The Color-Magnitude relation}
\label{CM}

The color–magnitude relation (CMR) is an important tool used to understand the physical properties of stellar systems.  Its first original application started with the studies of star clusters \cite{Hertzsprung1909, Russell1914}, followed by the analysis of our Galaxy and the Local Group \cite{Baade1944, Sandage1957, Blaauw1959} and by the analysis of the integrated light of galaxies in clusters, in particular in Virgo and Coma \cite{ChesterRoberts1964, Chiosi1967, VisvanathanSandage1977, SandageVisvanathan1978}. The modern CCD instrumentation have provided much richer CMRs \cite[see e.g.][]{Boweretal1992, Kodamaetal1998, Terlevichetal2001, Belletal2004}  allowing the study of the past history of galaxy clusters themselves \cite[see e.g.][]{Cariddietal2018,Sciarrattaetal2019} up to distances of cosmological interest. 

Since colors are independent of distance and are very similar for all cluster members, the CMRs have been considered good cosmological probes \cite{Tullyetal1982, Boweretal1992}, in particular when we look at the fraction of blue and red galaxies and their morphological ratios, the so-called galaxy color bimodality \cite{Baldryetal2004}. Both seem to be different in clusters and in the field \cite{ButcherOemler1978, Dressler1980}. 

In the CMR three main loci are of interest: the first is the red sequence (first noted by \cite{deVaucouleurs1961}), a linear band throughout a broad interval of luminosities mainly occupied by evolved ETGs. The others two  are the blue cloud, in which gas-rich galaxies still form stars at high rates, and the green valley in between, where a complicated interplay between gas conversion and passive evolution is at work \cite{Mencietal2005}. 

Thanks to the large-scale surveys, magnitudes, colors, morphological types and redshifts for thousands of galaxies are now available.  One example is the Galaxy Zoo, derived from the SDSS \cite{Blantonetal2003, Lintottetal2008, Wongetal2012}. These data have amply confirmed the existence and the evolution of the red sequence of galaxy clusters \cite{Stottetal2009, Headetal2014}.
More recently, the faint end of the red sequence has been also  investigated \cite{BoselliGavazzi2014, Headetal2014, Roedigeretal2017}. 

The theoretical analysis of the CMR is difficult because of the age–metallicity degeneracy: stars become red when age and metallicity increase \cite[see e.g.][]{Tinsley1980, SilkMamon2012}.
Understanding the origin of the red sequence, its slope and width has been the subject of several studies \cite[see e.g.][]{Baum1959, Faber1977, Dressler1984, Boweretal1992, Bursteinetal1995, Bursteinetal1997, Gallazzietal2006, Mencietal2008, Valentinuzzietal2011}. The general properties of the CMR have been  investigated \cite{Gladdersetal1998, Tranetal2007, Meietal2009}, within the classical scenario of galaxy formation and evolution with supernova-driven winds \cite{Larson1974, ArimotoYoshii1987, Tantaloetal1996, Tantaloetal1998, KodamaArimoto1997, Chiosietal1998}, within semi-analytical models in the hierarchical scheme \cite{WhiteFrenk1991, Kauffmann1996, KauffmannCharlot1998}, and within N-body-Tree Smooth Particle Hydro-dynamics simulations \cite[see e.g.][]{ChiosiCarraro2002}. 

The most accepted view is that the red sequence is more affected by metallicity than by age, even if the CMR has an age dispersion that increases at decreasing galaxy masses. Reproducing the slope  requires a correct treatment of the chemical evolution  \cite{Kauffmann1996, Nelsonetal2018}.
A crucial element is the knowledge of when and how the red sequence is  formed. The downsizing phenomenon, discovered by spectroscopic analyses of nearby ETGs \cite{Nelanetal2005, Thomasetal2005, Choietal2014}, implies that the red sequence was built over an extended period of time ($\sim5$ Gyr), beginning with the most massive systems  \cite{Tanakaetal2005}. Efforts to directly detect the formation of the red sequence have observed the color bimodality up to $z\sim2$ \cite{Belletal2004, Willmeretal2006, Cassataetal2008}. The data of the legacy surveys  GOODS, COSMOS, NEWFIRM, and UltraVISTA have also shown that massive quiescent galaxies ($M_s\geq3\times10^{10} M_\odot$) begin to appear as early as $z=4$ \cite{Fontanaetal2009, Muzzinetal2013, Marchesinietal2014} and stop assembling by $z=1–2$ \cite{Ilbertetal2010, Brammeretal2011}. \citet{Roedigeretal2017} found that the red sequence flattens in all colors at the faint-magnitude end (starting between $-14\le M_g\leq-13$, around $M_s\sim4\times10^7 M_\odot$), with a slope decreasing to $\sim60$\% or less of its value at brighter magnitudes. This could indicate that the stellar populations of faint dwarfs share similar characteristics (e.g., constant mean age) over $\sim3$ mag in luminosity, suggesting that these galaxies were quenched coevally, likely via pre-processing in smaller hosts. 

In recent times, large-scale numerical simulations of hierarchical galaxy formation in $\Lambda$CDM cosmogony, i.e. including DM and BM, appeared on the scene. In these simulations, much efforts have been made to include star formation, chemical enrichment, radiative cooling/heating, as well as feedback processes of different nature. With these simulations, the variation of the cosmic SF rate density (SFRD), with redshift \cite{MadauDickinson2014, Katsianisetal2017} has been addressed and largely explained \cite[see e.g.][]{Katsianisetal2017, Pillepichetal2018}. Some of these take into account the photometric evolution of the stellar content of galaxies, permitting the analysis of the CMR, in particular for galaxies belonging to clusters. 
As shown by \citet{Sciarrattaetal2019} these simulations nicely reproduce the red sequence, the green valley, and the blue cloud, the three main regions of the CMR.

The major drawback of these massive numerical simulations is  their complexity, high cost in terms of time and effort, and lack of flexibility and prompt response to varying key input physics. 

{Since broadband optical colors are not good discriminants of stellar populations because of the age-metallicity degeneracy, attempts have been made to  break the degeneracy by using stellar absorption lines indexes \cite{Worthey1994, Thomasetal2003}. Recent results suggest that metallicity, $\alpha$-enhancement, and age vary along the mass or velocity dispersion sequence \cite{Caldwelletal2003, Nelanetal2005, Thomasetal2005}, and also vary as a function of environment \cite{Thomasetal2005, Smithetal2006}.
The general impression however, is that the age-metallicity degeneracy cannot be broken.}

Finally we want to remark a notable fact: as shown by \citet{Cariddietal2018}, galaxy clusters share with galaxies in clusters a red sequence that has a similar slope. The mean color of clusters correlates with their total absolute magnitude, in the sense that small and faint clusters are in general bluer than big and luminous clusters. This aspect of the CMR has never been addressed by dedicated studies up to now. It is interesting to note that, independently on the scale of the stellar systems, the behavior of the stellar population seems connected with the structural and dynamical properties of the system, a proof that gravity works in the same way at all scales. In general we can say that the global understanding of the CMR for clusters of galaxies is still in its infancy. 

Great progresses are expected in this field with the new generation of ground and space telescopes, like ELT, JWST, etc., that will reach the faintest galaxies at high redshifts.

{
\section{Star formation in galaxies }\label{sf_mode}
In a galaxy's evolutionary history,  SF  is the starring actor.  Thanks to it, gas  is continuously turned into stars by a number of  not yet fully understood  processes,  so that  within the potential well of DM and BM a shining object is built that is populated by many generations of stars of different mass, age, and chemical composition. In the following we limit ourselves to mention  only the most popular laws for the star formation rate that are customarily used in models of  galaxy formation, leaving aside the much wider subject of the physical processes by which gas can be turned into stars. From an observational point of view, looking at the stellar populations in GCs, DGs,  LTGs, and ETGs, the dominant history of SF changes  a lot passing from one type to another: it is sharply peaked in one or a few initial episodes followed by quiescence in GCs, DSphs, and DEs, a series of bursts and quiescent periods in dwarf Irr, ever continuing in LTGs however showing a spatial and temporal grand design, and an initial dominant episode of high intensity and relatively long duration following by minor activity or quiescence in ETGs.  Can theoretical models reproduce and physically explain this variety of behaviors that apparently is related to the mass and morphological type? To answer the question one has to assume a general law of star formation and look for the physical situations in the history of star formation can change with the morphological type of the host galaxy.

\medskip
\subsection{Star formation in ETGs: mass and/or initial density?}\label{sf_mode_etg} 
In the case of ETGs the best tool highlighting the main driver of the SF and the SFH are the NB-TSPH hydrodynamic simulations, in which the rate of star formation is usually expressed by the \citet{Schmidt_1959} law 

\begin{equation}
    \frac{d\rho_s}{ dt } = - \frac{d\rho_g}{ dt }  =  c^*\frac{\rho_g^k}{t_{g}}
    \label{schmidt}
\end{equation}
where $\rho_s$ is the current mass density of stars, $\rho_g$ is the current mass density of gas, $t_{g}$ a characteristic time scale (typically the free-fall), $k$ a suitable exponent (typically $k \simeq 1$), and $c^*$ 
is the so-called dimensionless efficiency of star formation (typically $c^* =0.01 \div 0.1$).

{Based on simple arguments}, there are at least three prerequisites for gas (likely in form of molecular clouds) to be eligible to star formation: the gas has to be in convergent motion, i.e. the velocity divergence must be negative; the gas must be gravitationally unstable, i.e. it must satisfy the Jeans condition $t_{sound} \geq t_{ff}$
(where $t_{sound}$ is the {time scale related to the} local sound velocity); the gas must be cooling, i.e. it has to verify the relation $t_{cool} << t_{ff}$. Normally, SPH codes treat star formation simply implementing the Schmidt law in the computational language and transforming part of the gaseous particles which satisfy the three conditions above in new, collisionless particles of different mass (“stars”). The characteristic time scale is chosen to be the maximum between $t_{cool}$ and $t_{ff}$ time-scales, however in most  situations $t_g = t_{ff}$ is also a  good choice.
Knowing  $\rho_g$ and $\rho_s$ and integrating upon the current volume of the system one gets the current values of $M_g$ and $M_s$. Nowadays there are numerous galaxy models whose stellar content has been calculated with the above prescription. However they differ in a number of important assumptions, chief among others the cosmological model of the Universe and the scenario in which galaxy formation and evolution is framed. In the following, for the sake of illustration we will summarize here the results of three paradigmatic cases, i.e. the pure monolithic scheme of \citet{ChiosiCarraro2002}, the early hierarchical scheme of \citet{Merlin2010,Merlin2012}, and the full hierarchical  scheme, e.g. the Illustris  case of \citet[e.g.][and references]{Vogelsbergeretal2014}.
It is worth recalling here that care must paid on the link between the Schmidt and Kennicutt-Schmidt SF laws and their implications for numerical simulations \citep{Schaye2008}.

\textsf{The pure monolithic scheme}.  \citet{ChiosiCarraro2002} highlighted the role of over-density of the initial perturbation when exceeding the threshold value. Two groups of models were analyzed  according to the initial over-density: (i) models with mean initial density
$\langle \rho \rangle \simeq 200 \times \rho_{u}(z)$ and collapse redshift $z_f=5$ (shortly named A); (ii)  models  
with $\langle \rho \rangle \simeq 5 \times \rho_{u}(z)$ and $z_{f}\simeq 1$ (named B).  $\rho_u(z)$ is the  density of the Universe at redshift $z$ \footnote{The cosmological parameters were  $H_0=65\, km\,s^{-1}\,Mpc^{-1}$, Baryonic to Dark Matter ratio 1 to 9, i.e. for $M_T= M_{B}+M_{D}$, $M_{B} = 0.1 M_{T}$, $M_{D} =0.9 M_{T}$.}. 
Perturbations with spherical symmetry, assigned mass, mean density exceeding the critical value (and hence suitable radius) are let collapse and form stars. A MonteCarlo procedure is adopted to fix the initial coordinates and velocities of the DM and BM particles.  
The key result of this study is  that the star formation history (rate vs. time) is found to depend on the depth of the gravitational potential well of a galaxy. The following picture can be drawn. In the case of deep gravitational potentials (such as in massive and/or dense galaxies) once star formation has started  energy is injected into the gas by supernova explosions, stellar wind, etc..., but this is not enough to push the gas out of the potential well. The balance between cooling and heating is reached and the gas consumption by star formation goes to completion. Star formation cannot stop until the remaining gas is so little that any further energy injection will eventually heat it up to such high energies (temperatures) that the gravitational potential is overwhelmed. No more gas is left over and star formation is quenched. The star formation history resembles a strong unique burst of activity, a sort of monolithic star forming event, taking place over a certain amount of time, of the order of 1 to 2 Gyr. In contrast, in a galaxy of low mass and/or density and hence shallow gravitational potential, even a small star forming activity will heat up the gas above the potential well. Some of it is soon lost in galactic wind, the remaining one becomes so hot that it will take long time to cool down and to form new stars. The cycle goes on many times in a sort of repeated bursting mode of star formation taking place during long periods of times if not for ever. Out of all this we can derive what follows:
 
(i) The duration, strength, and shape of the SFR as a function of time  strongly depend on the galaxy mass and the initial density:
(a) Galaxies of high initial density and total mass undergo a prominent initial episode of SF  followed by quiescence. (b) The same happens to high mass galaxies of low initial density, whereas the low mass galaxies experience a series of burst-like episodes up to the present. The details of their SFH are very sensitive to the value of the initial density.
The typical dependence of the SFR on time for models B is shown in the left panel of Fig.\ref{histsfr_1}, while the right panel shows the SFR of a low mass galaxy ($M_T = 10^9 \, M_\odot$) for moderate variations of the initial density. Models A are not shown because their SFR is represented by a single initial spike. 

(ii) The gas mass turned into stars (per unit total mass of the galaxy) is nearly  constant. This means that the same engine is at work.

(iii) At increasing total mass of the galaxy the ratio between the left-over gas and the initial total BM decreases. 

(iv) As a result of star formation, large amounts of gas are pushed out from  the central regions  to large distances. When this gas cool, part of it falls back towards the central object.

(v) In general all galaxies eject part of their gas content into the inter-galactic medium, and the percentage of the ejected material  increases at decreasing galaxy masses.

\begin{figure*}
   \begin{center}
		\includegraphics[width=8.0truecm,height=9.0truecm]{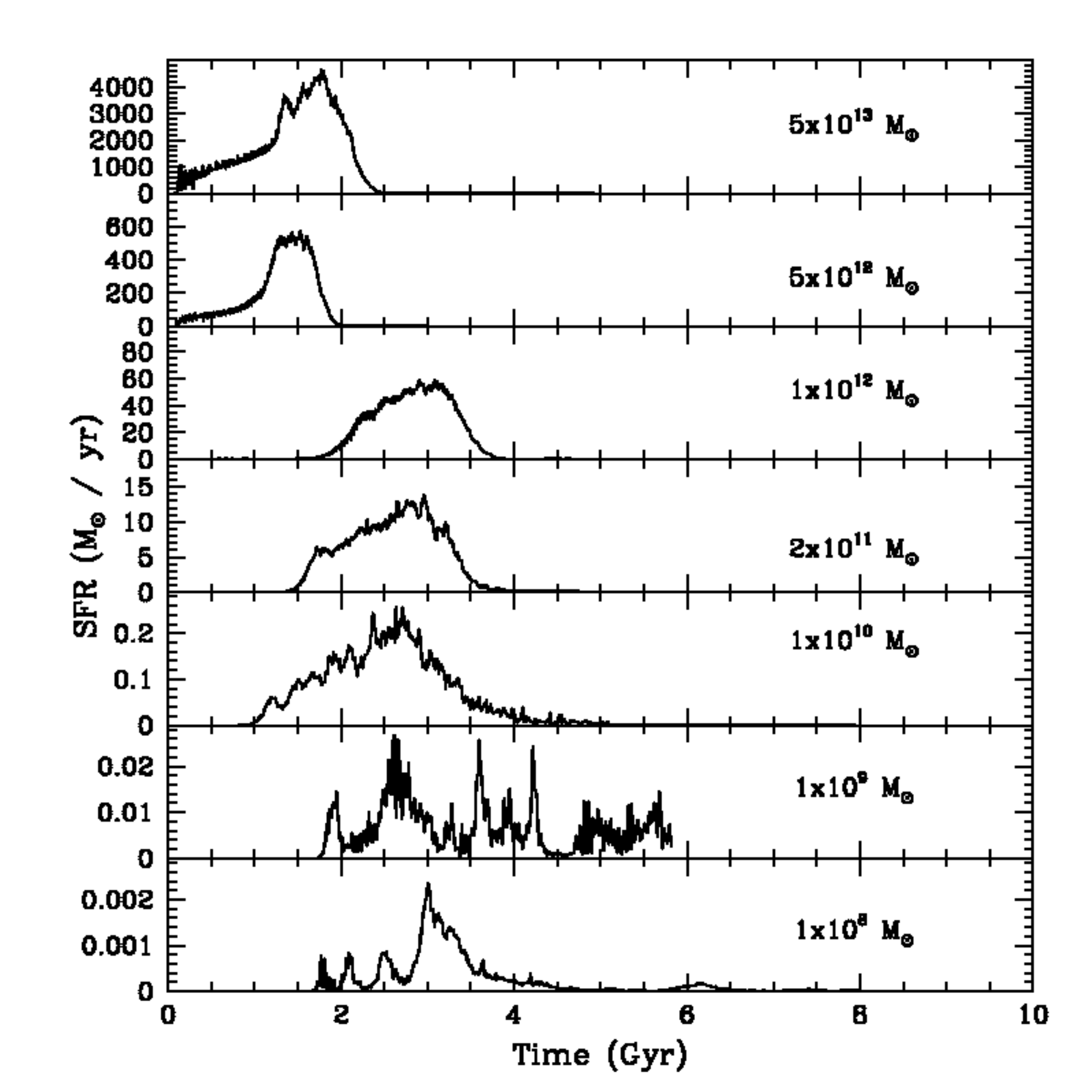}
		\includegraphics[width=8.0truecm,height=9.0truecm]{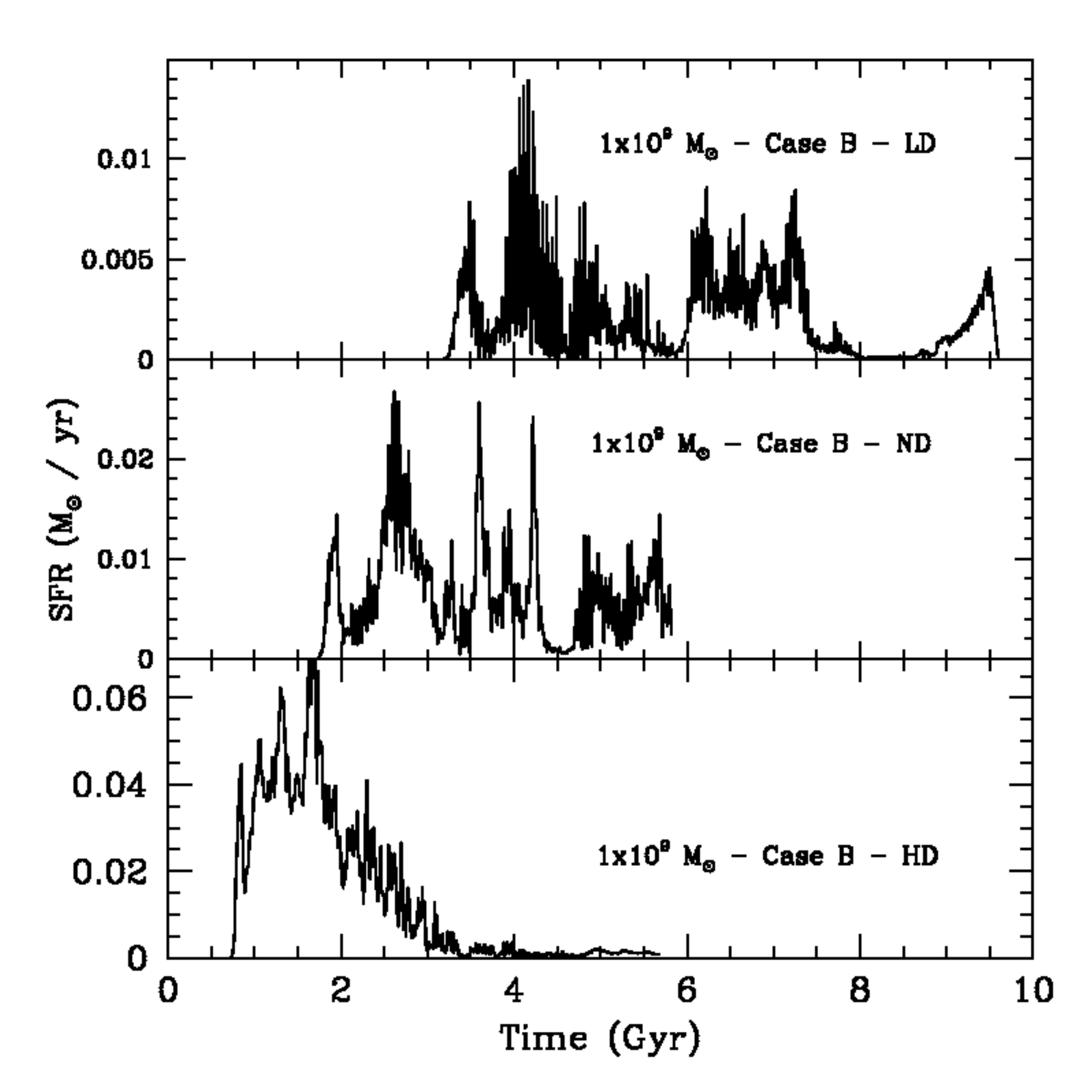}
   \end{center}
 \caption{ \textsf{Left panel}: The SFR as a function of time for the model galaxies of type B of \cite{ChiosiCarraro2002}.
		Their initial conditions are rather simple and grouped according to the initial over-density: models of type A had mean initial density $\langle \rho \rangle \simeq 200 \times \rho_{u}(z)$, whereas  models of type B had  $\langle \rho \rangle \simeq 5 \times \rho_{u}(z)$ where $\rho_u(z)$ is the  density of the Universe at redshift $z$. The Hubble constant was $H_0=65\, km\,s^{-1}\,Mpc^{-1}$ and the redshift of the starting collapse $z_{f}=5$. 
		Only models B are shown here because they are particularly useful to highlight  the effect of the mass at given initial over-density. 
		From the bottom to the top, the SFRs refer to galaxies with $M_T=M_{DM}+M_{BM}$ from 
		$1\times 10^{8}\,\, M_{\odot}$   to  
		 $5 \times 10^{13}\,\, M_{\odot}$.  The initial baryonic and dark mass are $M_{BM}=0.1\,M_h$ and 
		$M_{DM}=0.9\, M_h$, respectively.  
		\textsf{Right panel}: the  SFR of  low mass type B galaxies of the same mass, but different initial over-density. 	
		The mass is $M_T$=$10^9\,\, M_{\odot}$. The initial over-density varies from low (LD) to intermediate (ID) to high values (HD). 
		The figures are reproduced from \citet{ChiosiCarraro2002}. }
	    \label{histsfr_1}
\end{figure*}

\textsf{The early hierarchical scheme}.    \citet{Merlin2010,Merlin2012} using initial conditions derived from large scale cosmological simulations and abandoning the strict monolithic scheme, much improved the  NB-TSPH galaxy models of  \citet{ChiosiCarraro2002}. They also adopted the $\Lambda$CDM cosmology instead of the classical CDM \footnote{The  cosmological background  was the standard $\Lambda$CDM, with $H_0$=70.1 km/s/Mpc, flat geometry, $\Omega_{\Lambda}$=0.721, $\sigma_8$=0.817, and baryonic fraction $\simeq 0.1656$.}.
Without entering into detail,  they cut from large scale simulations calculated with the free code COSMICS by \citet{Bertschinger1995,Bertschinger1998} a spherical portion containing a proto-halos of DM and BM in cosmological proportions with the desired over-density, mass and size (the reference proto-halo with the highest mass to consider). The same is made for halos with lower mass and smaller dimensions at fixed mean density. The procedure to obtain halos with the same mass but different initial mean density is more complicated and will not reported here \cite[see][for the details]{Merlin2010,Merlin2012}.  These proto-halos contain a number of distinct lumps of matter that will merger together later on. The cosmological simulation provides  the initial positions and velocities of all the particles in the proto-halos.  The expansion of the Universe is taken into account.   The proto-halos are followed through their expansion (caused by the Hubble flow), down to their collapse and aggregation into a single objects. The redshift of the collapse varies from model to model and, inside the same model, from the center to the periphery. In general the collapse occurs in the redshift interval $4 > z > 2 $, it starts in the central regions and gradually moves outwards. The collapse is complete at redshift $z\simeq  2$. 
All models develop a stellar component. The more massive halos  experience a single, intense burst of star formation (with rates $\geq 10^3 M_{\odot}$/yr) at the early epochs. The intermediate mass halos ($M_{T}\simeq 10^{11} M_{\odot}$) have star formation histories  that strongly depend on the initial over-density, i.e. with a single or a long lasting period of activity and strong fluctuations in the rate. The small mass halos ($M_{T}\simeq 10^{9} M_{\odot}$) always have fragmented star formation histories: this is the so-called ``galactic breathing'' phenomenon. 
These models are classified as \textit{early hierarchical} because they experience repeated episodes of mass accretion at very early epochs and then evolve in isolation ever since. 
They confirmed the correlation between the initial properties of proto-halos and the star formation history found by \citet{ChiosiCarraro2002}. The models have morphologies, structures and photometric properties similar to real galaxies \citep[see][\, for all other details]{Merlin2012}. 

\textsf{The fully hierarchical scheme}. This is the most difficult case to discuss because of  mergers among galaxies of different mass, size, and age. If gas is present recurrent episodes of stronger star formation activity may  occur. It is conceivable that seed galaxies prior to any encounter behave like the general scheme envisaged before and governed by the initial density and mass. Mergers among objects of similar mass would likely  enhance the rate of star formation in a sort of burst of short duration and the fold the two histories together. Mergers among objects of much different mass, would simply generate a temporary perturbation on the star formation history of the most massive one, while less massive object simply loses its identity. In the hierarchical scenario, tracing the star formation history of a single galaxies is a hard task. Anyway,  the observational evidence provided by the stellar content of  galaxies of different mass strongly support  the mass-density scheme  we have described. 

The general trends of the SFR described in this section agree with the picture envisaged  long ago by  \citet{Sandage1986}, examining the SFR in galaxies of different types  
\citep[see also][]{Tinsley1980,Chiosi2014,Matteucci2016}. This scenario has been confirmed by studies of SF histories based on absorption line indices \cite{Thomasetal2005}, and by the recent  study of 
\citet{Cassara_etal2016}. A good agreement also exists with other independent numerical NB-TSPH models of galaxy formation and evolution by \citet{KawataGibson2003a,KawataGibson2003b,Kobayashi2005}.
}

\medskip
{
\subsection{The rate of star formation in disk galaxies } \label{SK}

According to \citet{Matteucci2016} the most common parameterization of the SFR in LTGs is the \citet{Kennicutt_1998} generalization  of  the original \citet{Schmidt_1959} law, where the SFR is proportional
to the gas density $\rho$. \citet{Kennicutt_1998} suggested  that the SFR can be written as: 

\begin{equation}
	SFR(t) = \nu \Sigma_g^{\kappa} 
\end{equation}
where $\Sigma$ is the gas surface mass density, $\nu$  the efficiency of star formation (the SFR per unit mass of gas) and $\kappa = 1.4\pm 0.15$, as deduced by the data of the star forming galaxies \citep[see also][]{Kennicutt_Robert_1998}. Other parameters, such as gas temperature, viscosity and magnetic field are not considered. 

Actually, the ``Kennicutt law'' was in use long before its discovery. In the mid seventies \citet{Larson1975,Larson1976} developed the first modern hydrodynamic models of formation and structure of elliptical and spiral galaxies, showing that a rate of star formation strongly declining during the latest stages of collapse was necessary to form a massive disk in spiral galaxies. However, once the gas has settled onto the equatorial plane and built up the disk, the rate of star formation should increase  to a peak value and then decline again. The duration of this phase and the height  of the peak were found to depend on the position on the disk. Larson envisaged several physical mechanisms that might strongly suppress star formation during the latest stage of collapse, e.g. velocity dispersion of the gas, tidal forces exerted on the remaining gas by the already formed spheroidal component, and dependence on the cloud-cloud collision frequency. The same processes were also invoked to control the second phase of star formation. Starting from this \citet{Talbot_Arnett_1975} correlated the process of star formation with the surface mass density of the gas in an already  flattened disk, whose thickness is regulated by the balance between the gravitational attraction and the increase of the scale height by energy injection by short-lived stars (e.g. type II supernova explosions by massive stars). They  proposed a star formation rate proportional to the surface mass density of gas. \citet{Chiosi_1980} folded the \citet{Larson1976} results into  the \citet{Talbot_Arnett_1975} mechanism and incorporated all this into a new model for the chemical evolution of galactic disks in presence of infall. In this model the disk is described by a series of concentric rings (no mass exchange among them), whose surface mass distribution at the present time  $t_g$ is given by an exponential law of type  $\Sigma(r)=\Sigma_d exp(-r/R_d)$, where $\Sigma_d$ and $R_d$ are two scale parameters.  The formation of the disk is supposed to occur by rapid infall of the gas left over by the formation of the halo and the central spheroidal component. The temporal and spatial dependence of the infall rate is given by 
\begin{equation}
 \frac {d\Sigma(r,t)}{dt} = A(r) exp(-t/\tau)
 \label{sigma}
\end{equation}
where $A(r)$ is a suitable function to be determined. This is derived by integrating eq.(\ref{sigma}) with respect  to time and by equating it to the present day mass distribution. We obtain
\begin{equation}
 A(r) = \Sigma_d exp(-r/R_d) \tau^{-1}[1- exp(-t_g/\tau)]^{-1}
 \label{aoferre}
\end{equation}
for $r_B \leq r \leq r_D$, where $R_B \simeq 2$ kpc and $r_D \simeq 20$ kpc are the  typical radius of a bulge and of a disk of spiral galaxies respectively.  The scale parameters $\Sigma_d$  and $R_d$ are determined by knowing the rate and the surface mass at certain position of the disk (e.g. the solar vicinity in our case).
Thanks to  the short time scale of the energy input from massive stars (a few million years), compared to the mass accretion time scale by infall (from hundred to thousand million years) the disk was supposed not to differ from an equilibrium state so that the \citet{Talbot_Arnett_1975} formalism could be applied. \citet{Chiosi_1980} and \citet{Chiosi_Matteucci_1980} proposed and used the SFR: 

\begin{equation}
	\frac{d \Sigma_s (r,t)}{dt}  = - \frac{d \Sigma_g (r,t)}{dt} =
	\tilde {\nu} \left[\frac{\Sigma(r,t) \Sigma_g(r,t) } {\Sigma (\tilde {r}, t)}  \right]^{\kappa -1} \Sigma_g (r, t) 
	\label{sigma_sfr}
\end{equation}
where $\Sigma_g(r,t)$ and $\Sigma_s(r,t)$, are the surface mass densities of gas, stars at the position $r$ or  and time $t$, respectively. The quantities $\Sigma(\tilde{r}, t) $ and $\tilde  \nu$  are the total surface mass density  at a particular  distance from the galaxy center, and an efficiency parameter. They play the role of a particular radial scale  controlling star formation. In the Larson's view they might be associated to the radial distance at which the central spheroidal component and the innermost regions of the disk exert  their tidal effect on the residual external gas. The spatial and temporal dependence of the relation \ref{sigma_sfr} in the infall model for the disk of the Milky Way and disk galaxies in general is such that at any time the SFR is strongly inhibited at distances $r> \tilde{r}$, while at any given $r$ the SF starts small, increases to a peak value, and then declines again. This behaviour of the SFR is typical of all infall models, where because of interplay between gas accretion  and consumption, the SFR starts low, reaches a peak after a time approximately equal to $\tau$ and then  declines.
Independently of the position, the net temporal dependence of the SFR  is the time delayed exponentially declining law:

\begin{equation}
SFR \propto \frac{t}{\tau}\exp\left(-\frac{t}{\tau}\right).
\label{timedelayed}
\end{equation}

The Schmidt law  is  the link between gas accretion by infall and gas consumption  by star formation. Thanks to the infall model by varying $\tau$ (time scale of the galaxy formation process) one can recover all types of star formation indicated by observational data going from GCs to LTGs and ETGs.
The infall scheme and companion SFR  have been widely used in many studies on the subject  of galactic chemical evolution \citep[e.g.][for a recent review and references]{Matteucci2016}. The infall galaxy model is very flexible and {can be adapted} to a wide range of astrophysical problems. Suffice to recall that it has been used by \citet{Bressan_etal1994} to model the spectro-photometric evolution of ETGs reduced to point mass objects, 
extended by \citet{Tantaloetal1998} to the case of spherical systems  made of BM and DM mimicking ETGs, adapted by  \citet{Portinari2000} to include  radial flows of gas in disk galaxies, and recently used by \citet{Chiosi_etal_2017} to study the cosmic star formation rate and by \citet{Sciarrattaetal2019} to investigate the color-magnitude diagram of galaxies in general.
}

\medskip
\subsection{The Mass-SFR relation}
\label{SFRM}

The connection between the structure and dynamics of galaxies and their stellar population, that we have encountered addressing the FP problem, is also part of the SF problem of galaxies. 
Observations have in fact revealed that the SFR and the stellar mass ($M_s$) of active star-forming galaxies are tightly correlated ($SFR \propto M_s^{0.6}$). This trend is known as the galaxy "main sequence" (MS)
\cite{Brinchmannetal2004, Salimetal2007, Noeskeetal2007, Elbazetal2011, Salimetal2012, Whitakeretal2012, Rodighieroetal2014, Speagleetal2014, Schreiberetal2015}. 
Adopting different samples the MS may be different, either in slope and scatter, {primarily for selection effects on  the adopted SF indicator used \cite{Popessoetal2019}.} One may select galaxies according to their mass and/or color, picking preferentially the blue cloud objects, or using the BzK color selection \cite{Daddietal2004, Daddietal2007, Pannellaetal2009} or the UVJ  selection \cite{Williamsetal2009, Whitakeretal2012}, or adopting a minimum threshold for the specific SFR ($sSFR=SFR/M_s$) \cite{Karimetal2011}.

The presence of a main sequence, with a scatter of  0.3 (in log units for active star-forming objects), indicates that these galaxies {have a SFR that spans a factor of two}. This can be explained by the self-regulating nature of the SF process, that is by the interplay between gas accretion, SF and feedback \cite{Schayeetal2010, Daveetal2011, Haasetal2013, Lillyetal2013, Tacchellaetal2016, Rodriguez-Pueblaetal2016}.

\begin{figure}[h!]
	\begin{center}
		\includegraphics[width=15cm]{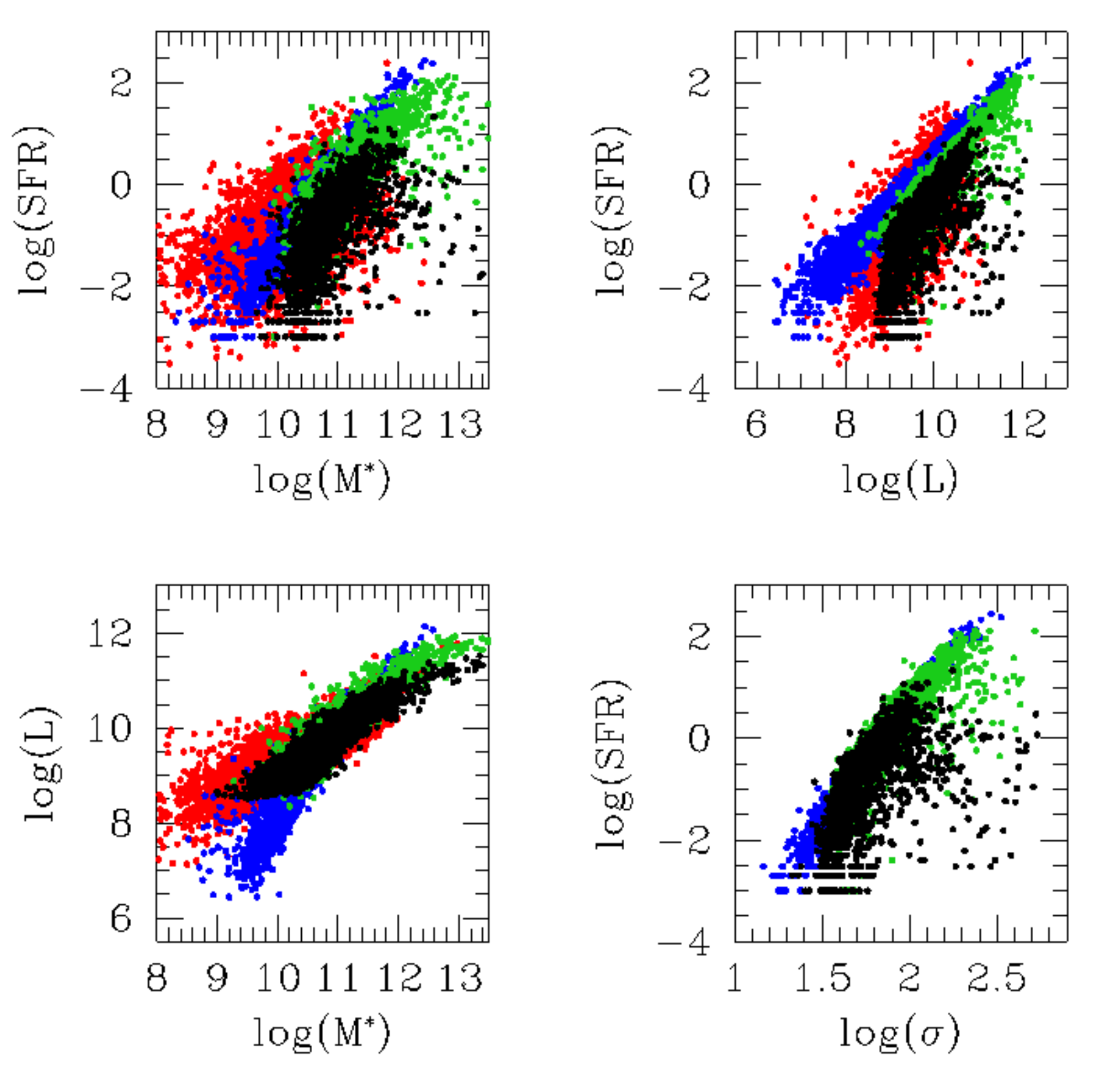}
	\end{center}
	\caption{The correlations between $M_s$, $L_V$, $SFR$ and $\sigma$. {Mass and luminosity are in solar units, SFR in $M_\odot/yr$ and $\sigma$ in \kms.} The red dots are the observational data of the WINGS survey for all morphological types \cite{Fritzetal2007, Fritzetal2011}. The blue dots are the prediction of the Illustris simulation for galaxies at $z=4$. The green dots the prediction at $z=1$ and the black dots the prediction at $z=0$. Note the lack of objects with mass below $10^9 M_\odot$ at $z=0$.}\label{fig:4}
\end{figure}

The scatter however is much larger ($\sim 0.6$) if all types of galaxies are considered. We can see it in Fig. \ref{fig:4}. The red dots in the various panels represent the data of the WINGS database \cite{Fritzetal2007, Fritzetal2011}. The SFR in the last 20 Myrs, measured from the spectral energy distribution in more than 3000 objects of all morphological types, is plotted versus the stellar mass.  The artificial data coming from the Illustris simulations are also represented for different redshift epochs: $z=4$ (blue dots), $z=1$ (green dots) and $z=0$ (black dots). 
{
Before drawing any conclusion, it is worth recalling that the model galaxies of the Illustris simulation were chosen to have stellar masses above  $10^9 M_\odot$  at $z=0$. Therefore the comparison between theory (black dots) and data (red dots) in Fig \ref{fig:4} is possible only for $M_s \geq 10^9\, M_\odot$. We note that in the common region the simulations predict the correct slope and quite a similar scatter at $z=0$. 
They also predict that the slope mildly changes with redshift and that the scatter increases going to the present epoch. This limit on $M_s$ does not exist for the samples at higher redshifts.}

The origin of the scatter and its amount might be different for dwarf and giant galaxies \cite{MattheeSchaye2019} and can be attributed both to short-time and long-time processes, such as the competing effects of inflows and outflows, the variation of the halo mass, the variation of the SFE and the feedback effects from the active nuclei.
\citet{Daddietal2007, Elbazetal2007, Noeskeetal2007} claim that the correlation is present even at redshift $\sim2$, with a nearly constant slope and a dispersion similar to that observed for galaxies in the local universe \cite{Brinchmannetal2004}.

{The information that one can draw from the MS is still under debate} \cite{Kelson2014, Abramsonetal2015}. There are many open questions: does the main sequence imply that the SFH of galaxies of the same stellar mass is similar? Is the MS a median “attractor-solution” \cite{Pengetal2010, Behroozietal2013}?  Is it an average sequence for a population at a certain age of the Universe \cite{Gladdersetal2013, Abramsonetal2016}? Do galaxies of the same mass have different SFHs on longer time-scales? What effects are most significant at different mass and time-scales?

The slope and scatter of the MS might encode such crucial information. What makes the growth of galaxies different? Which are the important time-scales of SF? Which are the systematic and stochastic effects behind the scatter? Which is the role played by the environment and by the DM in the assembly accretion history? 

The MS measured at higher redshifts shows a positive correlation evolving only a bit in slope and scatter \cite[see e.g][]{Elbazetal2007, Noeskeetal2007, Whitakeretal2012}. This might support the idea  that the link between structure and stellar population in galaxies is already in place at $z\sim 2.5$ \cite{Wuytsetal2011}.

The SFR in a galaxy depends on a variety of factors, such as the rate at which the galaxy accretes mass from the IGM, the rate of shocking and cooling of gas onto the galaxy, the details of how the inter-stellar medium (ISM) converts gas into stars, the amount of galactic fountain and outflow, etc. This complex non-linear physical mechanism is difficult to understand, in particular if one wants to discover what processes dominate, and if and how these change over time.
The PCA reveals that neutral gas fraction $f_{gas}$, stellar mass $M_s$ and SFR form a nearly flat 2D surface \cite{delPLagosetal2016}. The location of the plane varies with redshift, and galaxies can move along it when $f_{gas}$ and SFR drop with redshift. Their position along the plane is correlated with gas metallicity. This is a sort of “fundamental plane of SF” whose curvature is determined by the dependence of the SFR on gas density and metallicity. 

\section{The Mass-Metallicity relation}
\label{MZR}

It has been  known for a long time that the mean metallicity of galaxies correlates with the mass (and luminosity) \cite{Faber1973, Lequeuxetal1979, Skillmanetal1989, BrodieHuchra1991}.  
{By metallicity astronomers mean the abundance of heavy elements in the gas phase of the ISM.}
Such relation is observed either in gas-rich and gas-poor galaxies and suggests a similar physical mechanism behind the origin of the phenomenon \cite{Zaritskyetal1994}. Recently, the data of the SDSS have permitted the analysis of the mass–metallicity (MZR) relation over a wide interval of masses and metallicities \cite{Tremontietal2004, Maiolinoetal2008}. 

All studies confirm the trend of decreasing metallicity towards lower stellar masses, but the true form of the MZR is not yet well established. This depends on the strong systematic uncertainties affecting the measurement of the metallicity. There are a variety of methods to determine the metallicity \cite{Kewleyetal2008}. Some are based on the photoionization models for HII regions by reproducing some emission-line ratios, like $([O II]\lambda3727 + [O III]\lambda\lambda4959, 5007)/H\beta$ \cite{KobulnickyKewley2004} and $[NII]\lambda6583/[OII]\lambda3727$ \cite{KewleyDopita2002}. Some others on the fits of the electronic temperature ($T_e$), using strong-line ratios for HII regions and galaxies, like $([OIII]\lambda5007/ H\beta)/([N II]\lambda6583/H\alpha)$ and $[N II]\lambda6583/H\alpha$ \cite{PettiniPagel2004}. However, there are some problems when using these strong-line metallicity calibrations. For example, the MZRs with different calibrations have different shapes and normalization  \cite{Kewleyetal2008}. Furthermore, for high-z star-forming galaxies, these calibrations may not be valid, since their physical conditions in terms of gas density, ionization, N/O abundance, etc. might significantly be different from those in the local universe \cite{Lyetal2016}.

\citet{Kewleyetal2008} have shown that the method used to measure the oxygen abundance ($\log(O/H)$), typically assumed to trace the ISM metallicity, affects the shape and normalization of the MZR.  Differences up to 0.7 dex in the abundances at fixed stellar mass, using different emission-line methods are measured. This difference is not constant with the stellar mass and can give significant differences in the shape of the MZR. Possible origins for these discrepancies are discussed in \cite{Stasinskaetal2002, Kewleyetal2008, Lopez-Sanchezetal2012, Blancetal2015, Bresolinetal2016}. 

{The observations are commonly explained by gas outflows that are much stronger in dwarf galaxies than in giant elliptical galaxies}. The massive galaxies are able to retain the gas much longer than low-mass objects. This permits an increase of metallicity, because the new generations of stars are formed in a metal enriched environment. {At the same time}, low-mass objects loose their gas through galactic winds. Alternative explanations invoke a variable SF efficiency (SFE). This is larger in more massive systems, that formed most of their stars in a short time at high redshift, quickly enriching the ISM to solar or super-solar metallicities.

The MZR clearly depends on how gas accretion, SF and outflows proceed with time and therefore it contains important information about these processes. Several examples of the MZR have been published adopting samples of massive star forming galaxies at different redshifts ($0 < z < 3.5$) \cite[see e.g.][]{Tremontietal2004, Kewleyetal2008, Erbetal2006, Maiolinoetal2008, Zahidetal2011, Henryetal2013, Maieretal2014, Steideletal2014, Sandersetal2015}, while only few studies, mostly at $z \sim 0$, have extended the MZR to low mass DGs \cite{Lequeuxetal1979, Leeetal2006, Vaduvescuetal2007, Zahidetal2012, AndrewsMartini2013}. The luminosity-metallicity (LZR) relation has also been studied by several authors \cite{Lequeuxetal1979, RicherMcCall1995, MelbourneSalzer2002, Salzeretal2005, Sweetetal2014}.

Recently the chemical evolution models of \citet{DeLuciaetal2004}, in a hierarchical context, have also explained the observed MZR and the Tully–Fisher relation . This was possible by including feedback processes into the cosmological simulations. The drawback is that the feedback includes some free parameters, such as the efficiency or the yield, {that can be chosen to match the observations}.

Another difficulty of the outflow scenario is that different amounts of {DM can play a key role} in stopping the outflow of gas \cite{DekelSilk1986}. The works of \citet{Leeetal2006} and \citet{Dalcanton2007} have for example shown that the simple outflow of the gas  does not reproduce correctly the yields observed in the ISM of DGs.  The large variations in the effective yields and the dispersion in the relation are difficult to understand using only superwinds or outflows, in particular for the low metallicities observed at low masses and luminosities. It seems that neither the simple infall nor the outflow models are able to reproduce the low effective yields of low-mass galaxies.

In nearby galaxies, in the $10^6 \div 10^{9.5} M_{\odot}$ range,  the MZR follows a shallow power-law ($Z\propto M_s^{\alpha}$) with slope $\alpha = 0.14\pm$0.08. Approaching $M_s \sim 10^{9.5} M_{\odot}$ the MZR steepens significantly, showing a slope of $\alpha = 0.37\pm0.08$ in the $10^{9.5}\div10^{10.5} M_{\odot}$ range. Finally a flattening towards a constant metallicity is observed at higher stellar masses because the metallicity of the most massive galaxies saturate.

The evolution with redshift of the MZR \cite{Maiolinoetal2008, Yuanetal2013, Zahidetal2014} is a tool to trace the history of chemical enrichment in the different cosmic epochs. At high redshifts  the MZR has a steeper slope.  
The MZR at $z\sim3.5$ seems to evolve much stronger than at lower redshifts \cite{Maiolinoetal2008}. This is an epoch of strong  SF activity and metal enrichment also for massive systems. The metallicity evolution of low-mass systems seems stronger with respect to that of high-mass systems, an effect that reminds the "downsizing" of galaxies in a chemical framework. {Recent results concerning the evolution with redshift of the MZR up to $z \simeq 2.7$ are those by \citet{Wuyts_etal_2016}. Using the Integral Field spectroscopy they obtained data in good agreement with the old long-slit spectra, except for the slope of the relation at $z\sim2.3$ in the low-mass regime, where they measured a steeper slope than in previous literature results. }

The ISM of galaxies can be enriched by different effects: the accretion of gas from the inter-galactic medium (IGM), the injection and mixing of metals coming from the SF, the removal of these metals when they are locked into long lived stars and stellar remnants, the ejection of these metals when galactic outflows are at work, the mixing of high and low metallicity gas in the circum-galactic medium (CGM) and the removal/re-accretion of this gas out of the halo or back in the galaxies \cite{Lynden-Bell1975, Larson1976, LaceyFall1985, Edmunds1990, Dalcanton2007, Oppenheimeretal2010, Lillyetal2013, Maetal2016}. All these processes play an important role in shaping the evolution of galaxies. 

Not surprisingly the links between the gas mass ($M_g$), the  SFR, the stellar mass ($M_s$), and  the metallicity Z, is evident in a number of observed correlations. The most notable examples, in addition to the MZR, are: (i) the $M_s - SFR$ correlation  (dubbed the “Main Sequence”, MS \cite{Brinchmannetal2004, Noeskeetal2007, Daddietal2007, Elbazetal2011, RenziniPeng2015}; (ii) the correlation $M_g - SFR$ correlation between  (the so-called “Schmidt-Kennicutt”, SK, relation; \cite{Schmidt1959, Kennicutt1998, Bigieletal2008, Leroyetal2009} that has already been discussed in previous sections.

{
Before leaving the subject of the mass-metallicity relation in galaxies, we would like to briefly touch upon the companion, long debated subject of the age-metallicity relation for the stellar population. Age, metallicity, stellar mass are indeed the key parameters to play with to reconstruct the past history of formation and evolution of galaxies of any type. Unfortunately, the optical colors of old populations are affected by the age-metallicity degeneracy \citep{Worthey1994a, Worhey1994b,Worthey1999}: it implies that the spectro-photometric properties of an unresolved stellar population can not be distinguished from those of another population three times older and with half the metal content (the so-called 3/2 degeneracy, i.e. in the space color(s)-age the axis are not each orthogonal). Many efforts have been made over the past twenty years to break the degeneracy. \citet{Worthey1994a} analysing some optical features of the spectrum built up the so-called Lick system of indices and found that if on one side the indices decrease the age degeneracy, on the other side the age degeneracy is still there. The Lick system has been improved \citep{Trager2000a,Trager2000b, Vazdekis2010}, other features have been added e.g. the CaII IR triplet of \citep{Cenarroetal2001a,Cenarroetal2001b}, other  high resolution features have been introduced \citep{RodriguezMerino2020}.   Spectral windows, in particular the mid-UV, seem to be more promising \citep{Yi2003, Dorman2003, Kaviraj2007}.
The overall results indicate that the UV indeed helps to better constrain the age of unresolved systems (as would be expected since the MS turn-off are much more sensitive to age than the red giant branch), but the determination of chemical composition is still better determined by the more sensitive optical features. 
\citet{Lietal2007} to bypass the difficulty suggested the PCA method based on a large number of indices.
The problem became even more complicated by recognizing that another parameter played an important role. i.e. the so-called {$\alpha$-enhancement} measured by the ratio $[\alpha/Fe]$, where $\alpha$ is the abundance of elements like C, O Mg, Ti etc. 
\citep[see][for a thorough discussion]{Tantalo_etal_1998,Tantalo2004,Tantalo_etal_2004a,Tantalo_etal_2004b, Tantalo_etal_2007}.  The enhancement factor adds another degree of freedom to the age-metallicity degeneracy that now becomes the age-metallicity-enhancement degeneracy. The new degeneracy has size comparable to the old one. The whole issue is still open, \citep[see][for a recent review]{Chiosi2014}. Despite the large uncertainties, the broad band color and line indices technique has been largely used to infer the age, metallicity and degree of $alpha$-enhancement in galaxies of different morphological type. In relation to ETGs, the most massive objects of the galaxy population and the expectation from the classical hierarchical view of galaxy formation, \citet{Jimenez_etal2007} analysed the spectra of a larger number of ETGs from the SDSS  to infer the ages, metallicities and star formation histories and  found clear evidence of ``downsizing'', i.e. galaxies with large velocity dispersion and hence mass have older stellar populations. Most of the ETGs seem to complete  their stellar content at redshift $z > 2.5$, to increase their metallicity on a rather short time scale, and to possess subsolar $[\alpha/Fe]$ ratios.  This finding cannot be easily reconciled with the hierarchical scenario while it agrees with the early hierarchical models of  \citet{Merlin2012}. The issue is still open.
}

{
\section{Relationships between DM-halo and BM-guest galaxy}\label{stellar_to_halo}

\medskip
\subsection{The stellar-to-halo mass ratio}\label{ratio_ms_md}

The previous sections have clearly demonstrated that the observed properties of  galaxies are regulated by a complex series of physical effects tightly intertwined.
Last but not least is the ratio between the stellar mass in a galaxy and its dark matter component $M_s/M_D$ (and its inverse $M_D/M_s$). The ratio $M_s/M_D$ is a quantity that ultimately affects the half-luminosity radius  $R_e$ of the stellar component of a galaxy, see Sect. \ref{theo_mod}. The analysis  of the Illustris data and the theoretical galaxy models of   \cite{ChiosiCarraro2002,Merlin2006,Merlin2007,Merlin2010,Merlin2012,Chiosietal2012}   led \citet{Chiosietal2020}  to suggest that the ratio $M_s/M_D$  depends on the total  mass of the galaxy $M_T \simeq M_D$  and the redshift $z_f$ at which  the bulk of SF occurs. This is shown by Fig. \ref{md_ms_zeta} for the Illustris data.    For low values of the redshift (say below 0.6), the ratio smoothly  decreases with mass $M_{D}$ (low mass galaxies are slightly more efficient in building their stellar content);  the opposite occurs for higher redshifts, where $M_s/M_D$ increases with $M_{D}$. 
\citet{Chiosietal2020} give the following analytical expression for the ratio  $M_s/M_D$ as function of $M_D$ and $z$
\begin{equation}
	\log \frac{M_s}{M_D}= [0.218\,z - 0.101] \, \log M_D + [0.169 \, z - 2.227 ]  
		\label{ratio_ms_md_eq}
\end{equation}
where the halo mass goes from $10^4 M_\odot$ to $10^{14} M_\odot$ and the redshift from 0 to 4. The ratios $M_s/M_D$ predicted by eq. (\ref{ratio_ms_md_eq}) are indicated by the small black dots   of Fig. \ref{md_ms_zeta}. 
	
Other relationships for the inverse ratio $m= M_D/M_s$ can be found in the literature \citep[see for instance][]{Fanetal2010, Shankar_etal_2006, Girelli_etal_2020}. 
For $M_D \geq 10^{11}\, M_\odot$ \citet{Fanetal2010} propose the relation:
	
	\begin{equation}
		m = \frac{M_D} {M_s}= 25 \left( \frac{M_D}{10^{12}} \right)^{0.1} \left( \frac{1+z}{4} \right)^{-0.25}
		\label{md_ms_fan}
	\end{equation}
from which one derives the ratio $M_s/M_D$ shown in Fig. \ref{md_ms_zeta} by the red circles. In practice there is no dependence on redshift.
	
Notably, the curve of  \citet{Fanetal2010} agrees with the one  derived by \citet{Chiosietal2020} using the Illustris models for values of the redshift smaller than  about 1.6 (the slope is nearly identical). 
\citet[][and references there in]{Shankar_etal_2006} presented a detailed analysis of the dependence of $M_s$ on $M_D$. First, they claim that for  $M_D < 10^{11}\, M_\odot$  the relation should be
	\begin{equation}
		m= \frac{ M_D}{M_s}  = C\, M_D^{-2/3}
		\label{md_ms_shan}
	\end{equation}
with $C$ a suitable proportionality constant to be determined. Assuming equality between the values of $m$ derived with the two above relationships (at the transition mass $M_D \geq 10^{11}\, M_\odot$), the  proportionality constant is $\log C = 9.044$. The ratios $M_D/M_s$ resulting by eq.(\ref{md_ms_shan}) are shown in Fig.\ref{md_ms_zeta} with the dark golden circles. Note that the relation of \cite{Shankar_etal_2006} agrees with that of the Illustris models for redshifts in the range from 2 to 4. 

It is also worth noting that  the linear extrapolation of  the \citet{Fanetal2010} relationship (red circles) in Fig. \ref{md_ms_zeta} to lower masses and the linear extrapolation of the \citet{Shankar_etal_2006} curve (dark golden circles) to higher values of the mass  encompass the predictions derived from the Illustris models for all the values of the redshift. 

\citet{Shankar_etal_2006} derived a second analytical expression for the relation between $M_s$ and  $M_D$:  
	\begin{equation}
		M_s = 2.3 \times 10^{10} M_\odot \frac{    (M_D/3\times10^{11} M_\odot)^{3.1}  } 
		{1 + (M_D/3\times10^{11} M_\odot)^{2.2}  }
		\label{md_ms_shan2}
	\end{equation}
	
\noindent for $M_D \geq 10^{11} M_\odot$. In this relation there is not an explicit dependence on the redshift. The ratios $M_s/M_D$ predicted by eq.(\ref{md_ms_shan2}) are visible in Fig. \ref{md_ms_zeta} with the black filled squares.  Eq. (\ref{md_ms_shan2}) predict ratios $m(M_D,z)$ that are in agreement with those from eq.(\ref{ratio_ms_md_eq}) derived from the Illustris data, eq.(\ref{md_ms_fan}) from  \citet{Fanetal2010}, and eq.(\ref{md_ms_shan}) only in the region around $\log(M_D) \simeq 12$ and $z \simeq 0$. 

	\begin{figure}
		\centerline{
			\includegraphics[width=10.0cm,height=10.0cm]{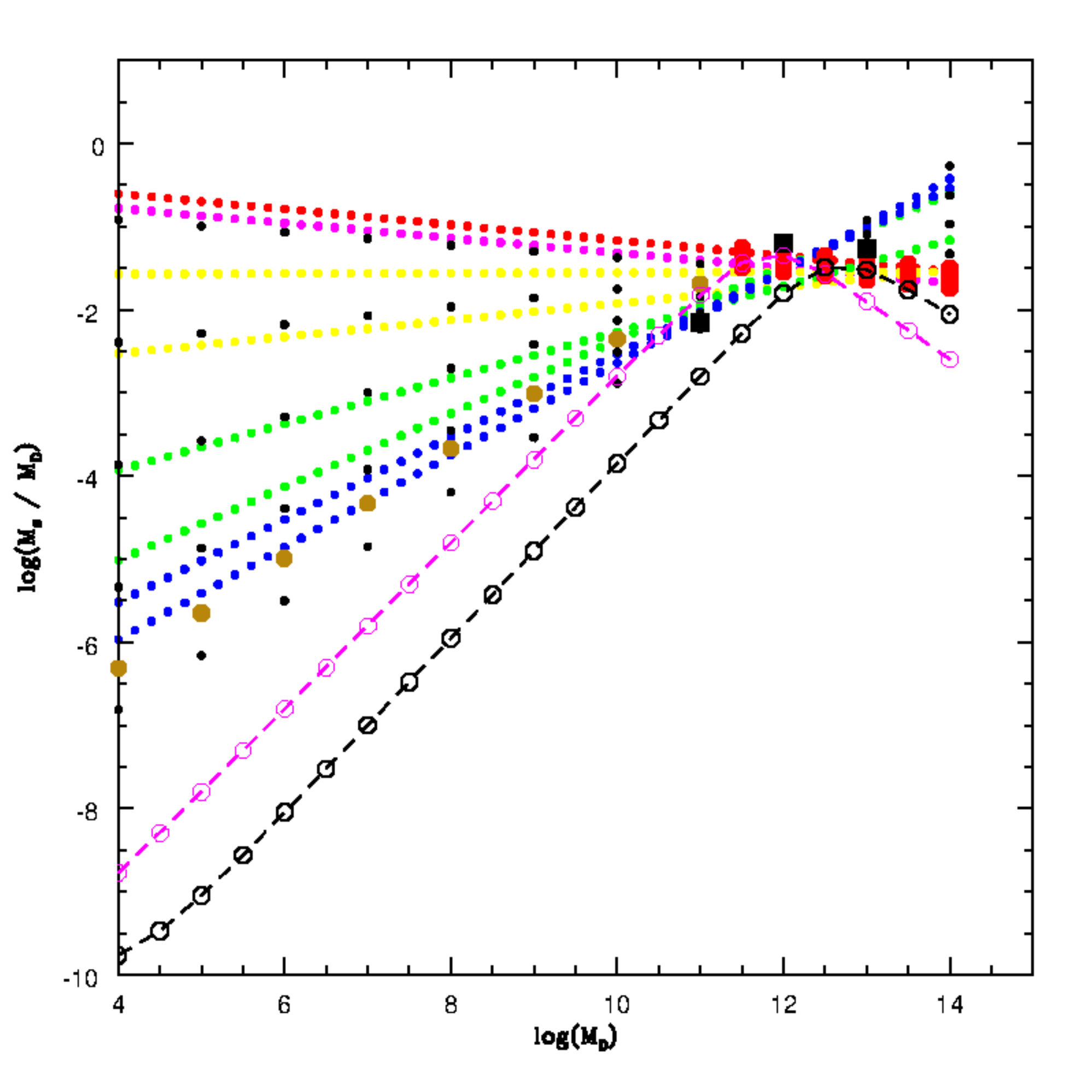}  }
		\caption{ The relations between $M_s/M_D$ and $M_D$ at different redshifts for different theoretical models (all masses are in solar units). The colored dotted lines correspond to eight values of the redshift $z=0$ and $z=0.2$ (top, red), $z=0.6$, $z=1.0$ (intermediate, yellow), $z=1.6$ and $z=2.2$ (intermediate, green), $z=3$ and $z=4$ (bottom, blue). The black dots are the values resulting by eq. (\ref{ratio_ms_md_eq}) at varying $\log M_D$ (from 4 to 14 in steps of 1) and redshift $z$ (from 0 to 4 in steps of 1), respectively. The large red and golden circles are the combination of eq. (\ref{md_ms_shan}) and eq. (\ref{md_ms_fan}).   The open magenta ($z=0$) and dark-olive ($z=3.95$) open circles are the relations $M_s/M_D$ vs $M_D$ at different redshifts according to \citet{Girelli_etal_2020}. Note that all relations agree at $log M_D \simeq 12$, while they badly disagree at lower values of $M_D$. Reproduced from \citet{Chiosietal2020}. }
		\label{md_ms_zeta}
	\end{figure}
	
In a very recent study \citet{Girelli_etal_2020} have thoroughly investigated the  stellar-to-halo mass ratio of galaxies ($M_s/M_{D}$)  in the mass interval $10^{11} < M_{D} < 10^{15} $ and  redshifts from $z=0$  to $z=4$. They  use a statistical approach to link the observed galaxy stellar mass function on the COSMOS field  to the halo mass function from the $\Lambda$CDM-Dustgrain simulation  and derive an empirical model to describe the variation of the 
stellar-to-halo mass ratio as a function of the redshift. Finally they provide  analytical expressions for the function $M_s(M_D,z)$. The relations $M_s/M_D$ vs $M_D$ as function of the redshift obtained with the formalism of  \citet{Girelli_etal_2020} are also shown in Fig. \ref{md_ms_zeta} (the magenta and dark-olive-green open circles joined by dashed lines of the same color). See also for a similar analysis the study of \citet{Engler_etal_2020}.
	
It is soon evident that while all studies agree on  the $M_s/M_D$ ratios for objects with halo mass in the interval $11.5 \leq log M_D \leq 12.5$ nearly independently of the redshift, they badly disagree each other going to lower values of the halo mass. Furthermore, they also disagree with the theoretical results predicted by Illustris. 
The problem is open to future investigations.  

\medskip
\subsection{Redshift evolution of DM-halos and their BM-guests} \label{DMH}
 
When galaxy formation started DM and BM were in cosmological proportions (i.e. $M_{D} = \omega M_{B}$ with $\omega \simeq 6$). Then the SF gradually stored more and more BM into stars. 

Here, exploiting again the  Illustris library of model galaxies \citep{Vogelsbergeretal2014} we show the relationships  between the stellar mass $M_s$ (as a proxy of the BM component) and the dark mass $M_D$, and that between $R_e$ and $R_D$ for four different values of the redshift ($z=4$, $2$, $1$, and $0$). They are visible in the left and right panels of Fig. \ref{ms_md_rs_rd}, respectively. Masses (in $M_\odot$) and radii (kpc) are in log units and the color code indicates the redshift ($z=4$, blue; $z=2$, green; $z=1$, yellow; $z=0$, red). 

It is clear that the efficiency of SF over the Hubble time, i.e. the transformation of gas in stars, is different in galaxies of different masses. Since $M_B < M_D$, $M_s$ is always smaller than $M_D$. However, galaxies of different total mass can build stars at different efficiencies, and the ratio $M_s/M_D$ is therefore expected to vary with $M_D$ and redshift. In the left panel of Fig. \ref{ms_md_rs_rd}, we note that $M_s$ increases with $M_D$, so that low mass galaxies build up less stars than the more massive ones. The slope of the relation however decreases as the redshift goes to zero. In more detail, for redshifts $z\gtrsim 2$ and masses $M_D \simeq 10^{12}\, M_\odot$ the slope decreases at decreasing redshift so that more and more stars are present at given $M_D$. More precisely, for $z\lesssim 2$ and $M_D \leq 10^{12}\, M_\odot$ the above trend holds, but above this limit the opposite occurs, at a given $M_D$ less stellar mass is present than expected. In other words, massive galaxies are less efficient builders of their stellar content. We can approximate this relation between $\log M_s$  and $\log M_D$ with the linear dependence $\log M_s=\alpha \log M_D + \beta$, where $\alpha$ and $\beta$ may vary with the mass range and the redshift.  
From the linear fit we obtain: 
(z=4, $\alpha=1.55 $, $\beta= -8.19$),
(z=2, $\alpha=1.44 $, $\beta= -6.78$),
(z=1, $\alpha=1.16 $, $\beta= -3.37$, for  $M_D < 12.0$),
(z=1, $\alpha=0.76 $, $\beta= -2.30$, for  $M_D > 12.0$),
(z=0, $\alpha=0.93 $, $\beta= -0.43$, for  $M_D < 11.5$), and
(z=0, $\alpha=0.79 $, $\beta= -1.22$, for  $M_D > 11.5$).
The ratio $M_s/M_D$ varies  from 0.2 to 0.05 when the mass $M_D$ increases from 
$10^7$ to $10^{12}$  $M_\odot$ with mean value $\simeq 0.10$. The overall process of star formation is not highly efficient, large amounts of gas remain unused and likely expelled into the external medium through galactic winds partially enriched in metals. Similar results are given by \citet[][and references]{Merlin2012}.    
The efficiency of star formation is customarily measured by the ratio $M_D/M_s$ as a function of $M_D$. This is simply given by:

\begin{equation}
	\frac{M_D}{M_{s}} = 10^{-\beta} M_{D}^{1-\alpha} 
	\label{inverse_md_ms}
\end{equation}
that has already been discussed in sect. \ref{stellar_to_halo}.

\begin{figure*}
	\centerline{
		\includegraphics[width=8.0cm,height=8.0cm]{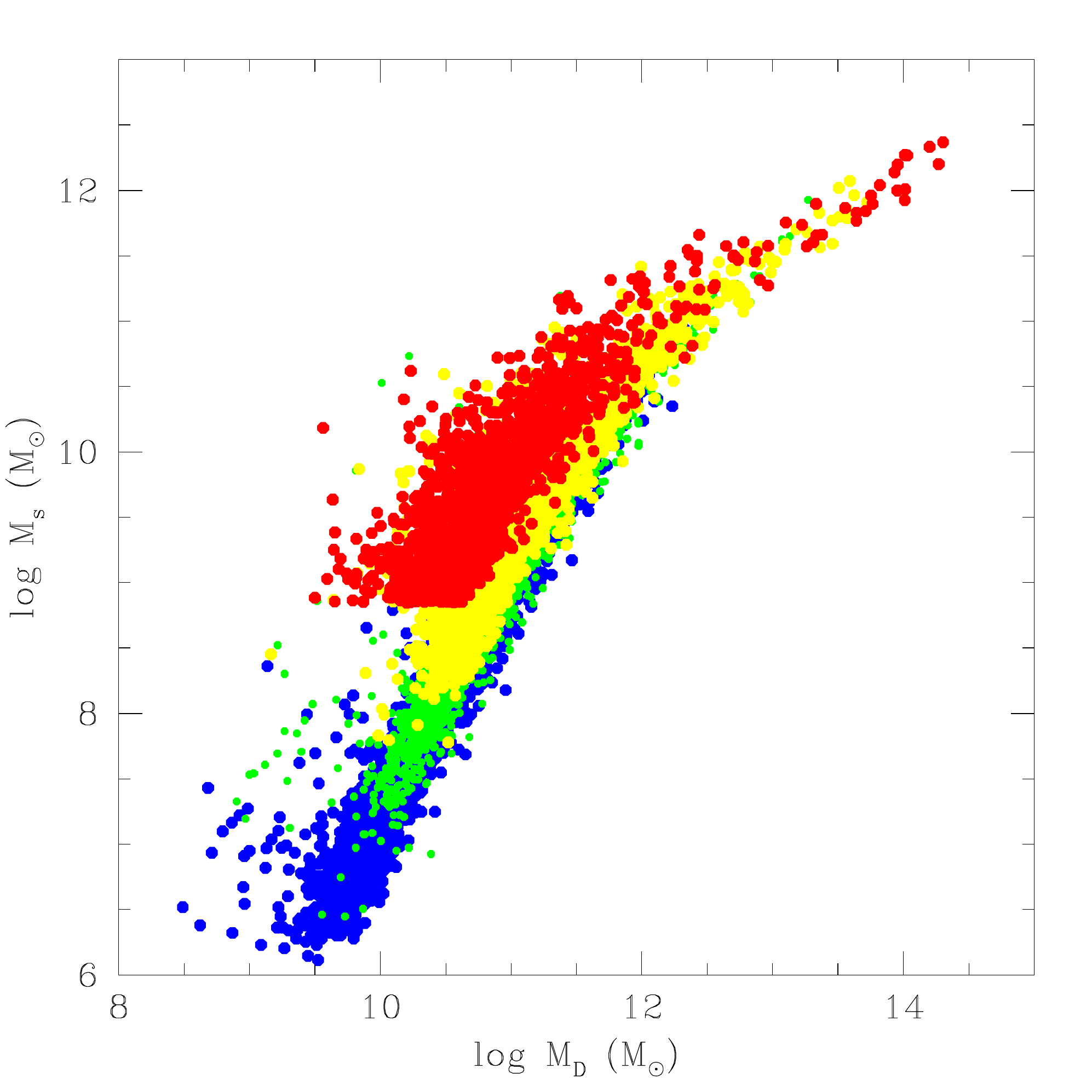}
		\includegraphics[width=8.0cm,height=8.0cm]{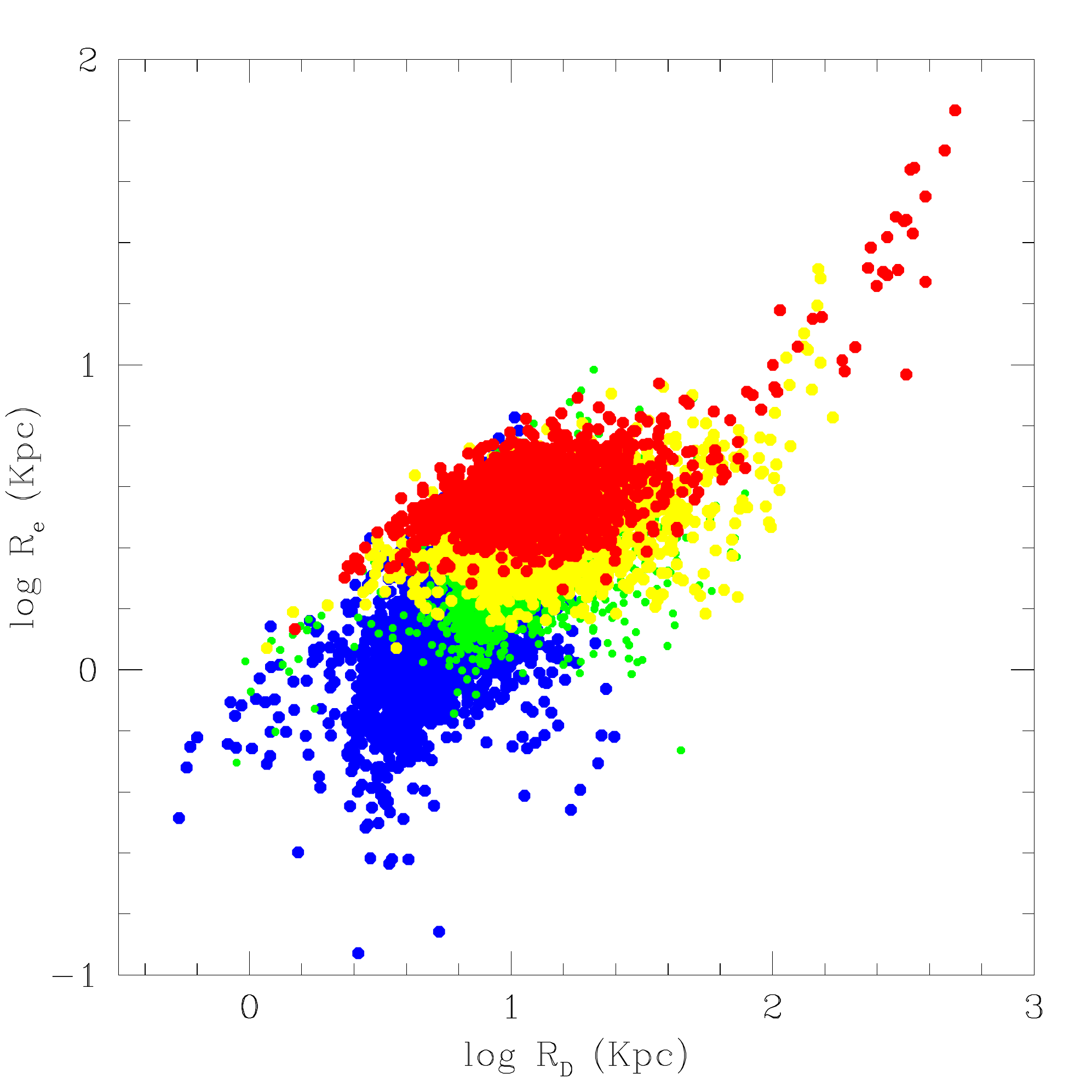}  }
	\caption{ {Left Panel}: The $M_s - M_D $ relations at different redshifts ($z=4$, blue; $z=2$, green; $z=1$, yellow;  $z=0$,  red). Masses are in solar units. The solid  lines are the best fits discussed in the text.  {Right Panel}: the same as in the left panel but for the $R_e - R_D$ relations. Radii are in kpc.}
	\label{ms_md_rs_rd}
\end{figure*}

Similarly we can derive the relations: 
$\log R_e = \gamma \log R_D + \eta$
($R_e = \eta R_{D}^{\gamma}$) that  are shown in the right panel of Fig. \ref{ms_md_rs_rd}. From the linear fit we obtain:  
($z=4$, $\gamma=0.39 $, $\eta= -8.19$),
($z=2$, $\gamma=0.30 $, $\eta= -6.78$),
($z=1$, $\gamma=0.22 $, $\eta= -3.37$, for $M_D < 12.0$),
($z=1$, $\gamma=0.22 $, $\eta= -2.30$, for $M_D > 12.0$),
($z=0$, $\gamma=0.29 $, $\eta= -0.43$, for $M_D < 11.5$), and
($z=0$, $\gamma=0.29 $, $\eta= -1.22$, for $M_D > 11.0$).

The radius $R_D$ is larger than $R_e$ by a factor of 3 to 10 as the galaxy mass increases from $10^9 \, M_\odot$ to $10^{13}\, M_\odot$. 
The slope $\gamma$ of the $R_e$-$R_D$ relation (in log units) first decreases by about a factor of 2, from $z=4$ to $z=1$, and then increases again at $z=0$.  What is important is that while at high redshifts (our $z=4$, $z=2$ and $z=1$ cases) the galaxy distribution on the $R_e$-$R_D$ plane is a random cloud of points, at $z=0$ a regular trend appears and $R_e$ increases with $R_D$ on the side of large values of $R_D$ (largest masses). However, in the region of low radii and masses a cloud of points is still visible. The reason must be attributed to the effect of strong galactic winds and mergers among galaxies of similar mass in the hierarchical process  that strongly perturb the mechanical equilibrium of these systems \citep[see][]{Chiosietal2020}. Finally,  note that the ratios $M_s/M_D \simeq 0.1$ and $R_e/R_D \simeq 0.1 - 0.3$
confirm the predictions of \citet{Bertinetal1992,Sagliaetal1992} based on analytical models for galaxies made of DM and BM.

}

{
\section{The angular momentum - mass correlation}
\label{AMM}

We now turn back our attention again to the correlations observed among galaxies. First we want to explore the correlation between angular momentum $J$ and mass $M$ that is one of the most fundamental SRs of galaxies. {It is at least as important as} the SRs between rotation velocity, velocity dispersion, characteristic size, and mass. {The correlation between angular momentum and mass} largely determines another basic property of galaxies, i.e. the characteristic size (e.g. the half-mass radius $R_h$) of disk-dominated galaxies.

Operationally one defines the stellar specific angular momentum $j^* = J/M_s$ (the angular momentum per unit mass), the stellar mass $M_s$, and the bulge fraction $\beta^* = M_b^*/(M_d^* + M_b^*)$, where $M_d$ and $M_b$ are the mass of the disk and the bulge respectively. In a plot of $\log j^*$ against $\log M_s$ galaxies of different morphological types and bulge fraction $\beta^* $ follow nearly parallel sequences. Over the mass range $8.9 \leq \log(M_s/M_{\odot}) \leq 11.8$  disks and bulges follow SRs of the form $j^* \propto M^{\alpha}$  with $\alpha = 0.67\pm0.07$. The different sequences have a maximum offset in zero-point by a factor of $8\pm2$ \cite{FallRomanowsky2018}.  
 
A similar result was obtained by \citet{ObreschkowGlazebrook2014}, who discovered a strong correlation between the baryon mass $M_b$, $j_b$, and the bulge mass fraction $\beta$, fitted by
$\beta = -(0.34 \pm 0.03) \log (j_b M^{-1}_b /[10^{-7}$ kpc\, km s$^{-1} M^{-1}_\odot]) - (0.04 \pm 0.01)$ over a range of 
$0\leq\beta\leq0.3$ and $10^9 M_\odot < M_b < 10^{11} M_\odot$. This $M-j-\beta$ relation likely originate from the proportionality between $j M^{-1}$ and the surface density of disks. 

The above picture seems to indicate that disks and spheroids are independent structures, formed by distinct physical processes: disks are likely formed by diffuse gas settling down on a flat surface within DM halos, while spheroids formed more violently by merging and collisions of cold gas clumps.
In this scenario, disk-dominated galaxies are not affected by major mergers, while spheroid-dominated galaxies have properties substantially linked to stripping and merging. The interesting thing is that this relation offers a natural explanation of several classical SRs, such as the FP of spiral galaxies, the TF relation, and the MR relation. It can also be the basis for an objective classification scheme alternative to the Hubble sequence.

In CDM models, galaxies get their angular momentum in the initial phases of density perturbation growth, when the collapsing DM clouds are tidally torqued by neighboring overdensities \cite{Hoyle1951, Peebles1969, Doroshkevich1970, White1984}. The classical theory of disk galaxy formation \cite{FallEfstathiou1980, RydenGunn1987, Dalcantonetal1997, Moetal1998} predicts that gas acquires nearly the same specific angular momentum of the host DM halo. This angular momentum sets the disk size, and largely determines the final morphology \cite{Fall1983, FallRomanowsky2013}. The baryons increase their rotational support by falling into the potential wells of the DM halos conserving their angular momentum. To what extent the baryons preserve the angular momentum during this process is one of the key issues in our understanding of disk galaxy formation.

The angular momentum of the DM halos is often expressed with the dimensionless spin parameter $\lambda = j/\sqrt{2}R_{vir}V_{vir}$, where $R_{vir}$ and $V_{vir}$ are the virial radius and virial velocity of the halo, and $j$ the specific angular momentum  inside $R_{vir}$ \cite{Bullocketal2001}. 
The spin parameter of DM within $R_{vir}$ is found to have log-normal distribution with a median $\lambda \sim 0.04$ and rms variance of $\sigma \ln \lambda \sim 0.55$ \cite{Bullocketal2001, Vitvitskaetal2002, Bettetal2010}, while BM seems to have a spin higher than the halo’s average \cite{Pichonetal2011, Kimmetal2011, Tillsonetal2011, Danovichetal2012, Codisetal2012, Stewartetal2013, Ubleretal2014, Danovichetal2015}. In a set of zoom-in simulations \citet{Danovichetal2015} have shown that $\lambda$ of the cold gas grows when crossing the virial radius (see also \cite{Pichonetal2011}).

From the side of numerical simulations we should highlight the long suffered problem of the “angular momentum catastrophe" \cite{NavarroWhite1994, NavarroSteinmetz2000}.
The problem emerged from the comparison with {observations} of disk galaxies. While the observed disks have shown a specific angular momentum $j$ lower by a factor of two, modelled disks appear to have radial scale-lengths smaller by a factor of 10, resembling bulges rather than disks \cite{NavarroSteinmetz2000}. 
In the last years however, simulations seem to have solved the problem by inserting an efficient stellar feedback \cite{Scannapiecoetal2008, Zavalaetal2008, Salesetal2010}. For example, \cite{Brooketal2011}, but see also \cite{Brooketal2012, Ubleretal2014, Christensenetal2014} found that supernova feedback can selectively remove low angular momentum gas via outflows, leading to disk formation more in line with observations.

{In a recent paper \citet{PengRenzini2020} argued that the stellar angular momentum of galaxies increased by a large factor over the last $\sim10$ Gyr (i.e. $z\sim2$), starting from an epoch when the majority of galaxies acquired their ordered rotation. The size of $J$ follows directly from the SRs of spiral galaxies, i.e. from the connection:

\begin{equation}
    J \propto M^* R_e V_{rot}
\end{equation} 

between stellar mass, effective radius and rotational velocity.
This behavior could be driven by the baryonic gas vorticity of the circum-galactic filaments that might drive the galaxy evolution. In this framework, the gas in the filaments regulate the fluctuations in the specific SFR of galaxies, offering an explanation for the existence of the main sequence \citep{Lillyetal2013}.

For what concern the angular momentum of galaxies at 
high redshift, we refer to the paper of \citet{Burkertetal2016}. This work analyze a sample of $\sim360$ massive star-forming galaxies at $z \sim 0.8–2.6$. They found a $J$ distribution broadly consistent with the theoretical prediction for the dark matter halos, either in terms of spin parameter
$\langle\lambda\rangle\sim0.037$ and its dispersion ($\sigma_{\log\lambda}\sim 0.2$). These data support the hypothesis that on average, at high redshifts, the specific angular momentum of spirals is the same of dark matter halos ($j_d = j_{DM}$). Including the molecular gas, these authors measured a total BM to DM mass ratio of $\sim5\%$ for halos of $\sim10^{12} M_\odot$, which corresponds to $\sim31\%$ of the available baryons. This means that high-z disks are strongly baryon dominated.
}

}

\section{The scaling relations of black-holes and galaxies}
\label{BHM}


Today the idea that the history of the massive black-holes (BHs) at the center of galaxies and that of galaxies themselves is strictly entwined is widely accepted, after the discovery that the BH mass correlates with various properties of the host galaxies \cite{FerrareseFord2005, KormendyHo2013, Graham2016},  such as bulge mass \mbu\ \cite{KormendyRichstone1995}, total stellar mass $M_s$ \cite{HaringRix2004}, velocity dispersion $\sigma_\star$ \cite{FerrareseMerritt2000, Gebhardtetal2000}, light concentration \cite{Grahametal2001}, and halo circular velocity \cite{Ferrarese2002}. The ensuing paradigm of BH and host bulge/spheroid co-evolution is today widely accepted and supported by these well-known correlation for quiescent and almost quiescent galaxies. Unfortunately, the physical nature of this connection is still obscure \cite{SilkRees1998, Shapiro2005} despite intense observational efforts. 


For galaxies whose nuclei are currently active, there are basic observational issues that remain open at the time of writing. The necessity to resort to type-1 AGN for studying the \mbh\ -- {$\sigma_\star$\ or \mbh\ and \mbu}\footnote{It is yet unclear which of the two relation is the most fundamental, albeit the relation $\sigma_\star$\ with \mbh\ has been considered as the primary one in several past works.  The two relations will be considered as interchangeable when the \mbh\ - bulge relation is mentioned in a generic context.} relations outside of the local Universe raises two overarching  questions.  The first one is whether  the \mbh -- bulge relations are observationally consistent with the one obtained for quiescent galaxies at very low redshift.  A related issue  is about the selection effects specific to the \mbh-- bulge relation for type-1 AGN with respect to the one of non-active galaxies. The second question is whether there is a significant  evolution of the \mbh -- bulge relation with cosmic epoch. 

Some general considerations are in order, before focusing on the analysis of the scaling relations and on the two main questions above. The most accurate black hole mass determinations are the ones that probe the truly central regions of a galaxy, where the gravity of the black hole is the dominant force. This occurs within a distance from the BH $r_\mathrm{h} = GM_\mathrm{BH}/(\sigma_\star)^2 \approx 43 M_\mathrm{BH,8} \sigma_{\star,100}^2$ pc, where $M_\mathrm{BH}$\ is in units of $10^8$\ solar masses, and the $\sigma_\star$\ of 100 km s$^{-1}$. The BH sphere of  influence has been resolved in several nearby galaxies, presumably hosting  the most massive BHs that were shining at $z \approx 2$, where the most luminous quasars are observed \cite{lynden-bell69}. In the local Universe, these galaxies mostly appear as spent or almost-spent active nuclei \cite[e.g.,][]{lynden-bell69,kingnealon19}. {As long as a galaxy has a central black hole, there is no such a thing as a quiescent galaxy: some nuclear activity occurs, even if  at extreme low level, and detected only in the nearest cases (i.e., Sagittarius A) and under particular circumstances. We consider here weakly active sources whose Eddington ratio is too low to enter into the domain of radiative efficient accretion mode (a typical  example could be M87).} 


\subsection{Massive black holes at the center of quiescent (or weakly active) galaxies}

\label{sd}

{The method employed for modelling  stellar system in dynamical equilibrium is that of orbit superposition \citep{schwarzschild79}. The gravitational potential is defined as the sum of the central black hole (assumed a central point whose mass is to be determined) and of the stellar mass density derived from the stellar mass-to-light ratio.    What is computed is the combination of orbits  compatible with the spatially resolved stellar {kinematics} and photometric profiles. For the kinematically hot galaxies the early way to get the BH mass was based on the fit of the line-of-sight velocity dispersion of spherical galaxies assuming that the stellar distribution function is isotropic \cite{Youngetal1978}. In more modern approaches,    the fit is made over  the entire line-of-sight velocity distribution \cite{Rixetal1997, Gebhardtetal2007} for arbitrary   galaxy models whose gravitational potential  includes the effect of dark matter, and of triaxiality \citep{gebhardtthomas09,vandenboschetal08}.  The most general and accurate possible models, with the highest resolution of spectroscopic observations are reputed to be most accurate \citep{KormendyHo2013}, provided that the BH sphere of influence is adequately resolved. }
A case in point is the estimate of the black hole mass in M87: early estimates yielded a mass  $\sim5 \times 10^9 M_\odot$ from spherical, isotropic models  \citet{Youngetal1978}. More recent analyses based on stellar dynamics yielded \mbh\ in the range $\approx (6. - 6.5) \cdot 10^9$ \msol\ \citep{gebhardtthomas09,gebhardtetal11}.  {The stellar dynamics mass value has been spectacularly confirmed by the Event Horizon Telescopes observations that yielded \mbh $\approx 6.5 \pm 0.2| \mathrm{stat} \pm 0.7| \mathrm{sys} \cdot 10^9$ \msol\ from the inference of the angular size of the black hole gravitational radius  \citep{evht19}.  
}


{One of the most promising developments in the last years has been the increasing number of dynamical mass estimates obtained with ALMA \citep[e.g.,][for mildly active and quiescent  galaxies]{barthetal16,boizelleetal19}. ALMA has the capability to resolve cold molecular gas kinematics on angular scales well below 1 arcsec. \citep{woottenthompson09}. This is becoming instrumental to high-precision measurements of black hole masses in the ``intermediate" mass domain, a previously uncharted territory. For instance, sub-parsec resolution ALMA observations revealed a black hole with mass $\sim 5 \cdot 10^5$ \msol\ in the dwarf galaxy NGC404 \citep{davisetal20}. }


Space-based long-slit spectra of optical emission lines   yield a velocity cusp \citep{macchettoetal97}. A striking example is provided by the radial velocity curve of NGC4374 \citep{boweretal98}: the STIS spectra show a Keplerian swing beginning at $\pm 0.5$ arcsec and culminating at $\pm 0.1$ arcsec, with a radial velocity difference of $\delta v_\mathrm r \approx$ 400 \kms, implying an \mbh\ $\approx (1.5_{-0.6}^{+1.1})\cdot 10^9$ \msol{}.  The main concern is  that gas motions could be affected by radiation forces, shocks, turbulence, and magnetic fields,  and not only by gravitation.  {Relatively few galaxies have been found to have regular disk-like profile suggestive of a velocity field dominated by Keplerian motion in a dynamically cold disk \citep{KormendyHo2013}. In addition, the Keplerian assumption is not  consistent with  gas flow toward low-accretion-rate SMBHs and at variance  with observations of the Galactic Center \citep{jeteretal19}. For M87, both an early and a more recent  analysis based on HST data  suggest  a black hole mass of $\approx (3 - 3.5) \cdot 10^9$ \msol{} \citep{Harmsetal1994,macchettoetal97,walshetal13}, and very close to the value obtained by modeling the jet boundary shape \cite{Nokhrinaetal2019}, but always at variance with   the values obtained from stellar dynamics (Section \ref{sd}). }


\medskip

\subsection{Relations \mbh\ vs \mbu\ and $\sigma_\star$\ for quiescent galaxies }

{As mentioned earlier, the correlation between \mbh\ and host galaxy bulge properties -- \mbu\ and $\sigma_\star$\ or even bulge luminosity -- is now an established fact since more than 20 years \citep[see e.g.][]{KormendyRichstone1995,FerrareseMerritt2000,Gebhardtetal2000}. Widely used forms of the relation between  \mbh\ and $\sigma_\star$\ based on sources for which there is a dynamical \mbh\ determination are the ones of \citet{mcconnelletal11} and of \citet{KormendyHo2013} for early-type   galaxies. \citet{KormendyHo2013} derived a power law:   
\begin{equation}
\log (M_\mathrm{BH}) \approx 8.491 \pm 0.049 + (4.384 \pm 0.287) \log \sigma_{\star,200},    
\end{equation} 
where the mass is in solar units and $\sigma_\star$\ is units of 200 \kms. For both early type and spiral galaxies  \citet{mcconnellma13} yield a significantly steeper slope $\gtrsim 5$, with a lower intercept for spiral galaxies, implying that \mbh\  in ETGs is about a factor $2$ higher than in LTGs at a given $\sigma_\star$ \citep{mcconnellma13}.   Equivalent relations (i.e., with similar scatter, around 0.30 dex) have been defined with the bulge mass, and infrared luminosity, usually suggesting a power-law relation between \mbh\ and \mbu\  with an exponent $\approx 1$\ or larger 
\citep{mcconnellma13,KormendyHo2013,bennertetal21}.  

There is still much ongoing research considering the linearity of the relation, its slope, and the origin of its dispersion.   Theories that connect  galaxy evolution and black hole  growth predict the existence of a second parameter which may account for the dispersion in the \mbh\ - $\sigma_\star$\ correlation. Black hole - spheroid coevolution models would require that the BH mass scales with the gravitational binding energy of the spheroid host, $\sim $ \mbu/$r$\ \citep{hopkinsetal07}. The correlation can be easily turned into bivariate relations \mbh $\propto $\mbh$^{0.6} \sigma_\star^{1.2}$\ and $\propto r_\mathrm{bulge}^{0.6} \sigma_\star^{2.4}$\ that imply  correlations between \mbh\ and $\sigma_\star$\ and \mbu\  consistent with the observed ones {\citep{sagliaetal16}}. A correlation with the binding energy of the host galaxy \citep{barwaykhembavi07,allerrichstone07} implies the presence of a second parameter that may compensate for the changes in the galaxy structural parameters occurring at fixed \mbh. {\citet{sagliaetal16}} found significant bivariate correlations consistent with a connection between \mbh\ and binding energy and with bulge kinetic energy, although the scatter remains comparable to the one for the \mbh\ - $\sigma_\star$\ correlations obtained with the best dataset of \mbh\ dynamical mass measurements. }

The $\log(M_{BH})-\log(M_{s,sph})$ relation  reported by \cite{KormendyRichstone1995, Franceschinietal1998, Magorrianetal1998, McLureDunlop2002, MarconiHunt2003, HaringRix2004} is almost linear, but
the inclusion of low-mass spheroids revealed departures from linearity. \citet{Laor1998, Laor2001, Wandel1999} and \citet{Ryanetal2007} obtained a much steeper power law with a slope of $1.53\pm0.14$. The mean $M_{BH}/M_{s,sph}$ ratio is probably not a universal constant, as it drops from $\sim0.5\%$ in bright ($M_V\sim-22$) ellipticals to $\sim0.05\%$ in low-luminosity ($M_V\sim-18$) bulges.
\citet{Saluccietal2000} claimed that the $M_{BH}-M_{s,sph}$ relation is significantly steeper for spiral galaxies than for (massive) elliptical galaxies. \citet{Graham2012} suggested that the relation between luminosity ($L$) and stellar velocity dispersion ($\sigma$) for low-luminous ETGs is inconsistent with the $M_{BH}-L$ and $M_{BH}-\sigma$ relations. They prefer a broken $M_{BH}-M_{s,sph}$ power-law relation, with a near-linear slope at the high-masses and a near-quadratic slope at the low-masses.
In a recent review article \citet{Graham2016} analyzed the consequences of this steeper relation, that can be {rich of} implications for the theories of galaxy–BH coevolution. \citet{Scottetal2013, GrahamScott2013}   offered an interpretation for the curvature of the $M_{BH}-M_{s,sph}$ relation invoking the presence of core-S\'ersic  and S\'ersic spheroids at the high- and low-mass ends of the distribution respectively. 


{
The highest degree of correlation is obtained for ETGs and for bulges that follow a S\'ersic - surface brightness profiles. Galaxies obeying a S\'ersic photometric profile down to the resolution limits of their surface brightness profiles are believed to be the product of wet mergers, i.e., merger of gas rich galaxies that provide material to sustain accretion on the central black hole and trigger a period of sustained nuclear activity. The ensuing feedback effects (both radiative and mechanical) on the host, due to the active nucleus and to the merger-induced star formation, make it possible to couple the growth of the central black hole to the host spheroid mass \citep{zubovasking12,kingpounds15}: the feedback forces  by the quasar expel so much  gas to quench both star formation and stop black hole growth, ultimately accounting for  the relation  between \mbh\ and $\sigma_\star$, and \mbh\ and \mbu\ \citep{dimatteoetal05,robertsonetal06}.} 

{However, the most massive elliptical galaxies often exhibit surface brightness profiles that are  flatter than the  extrapolation of S\'ersic-like profiles. Sources showing a deficit with respect to the S\'ersic profile are contributing to the scatter in the  \mbh\ -- \mbu relation \citep{kormendybender09}. Core profiles  are believed to be due to dissipationless mergers of galaxies that have central black holes. N-body simulations show that merging of two galaxies with a sharp cusp may results in a merger remnant with a shallower core  \citep{milosavljevicmerritt01,kulkarniloeb12,bortolasetal16}. The formation of a core has been ultimately linked to a bound binary black hole system, which produces a depletion of the stellar component in the nucleus  due to  slingshot ejection of stars on nearly-radial orbits.  }

{The size of the core and the starlight and mass deficits in the centers of core galaxies (i.e., the mass ejected by the binary) have been found to scale approximately with the mass of the central black hole \citep{graham04,KormendyHo2013}, in agreement with theory that predicts a mass deficit  \citep{merritt06} to be 0.5  $n$ \mbh, with $n$\ as the number of major merger events. The luminosity deficit  correlation provides an independent way  to estimate \mbh\  in core ellipticals. Core radius is most strongly correlated with the black hole mass and  correlates better with total galaxy luminosity than it does with velocity dispersion \citep{ruslietal13}. In addition, core scouring changes the orbit distribution.  Only radial orbits allow for close passage past the galaxy center and thus only those stars  can reach the vicinity of the central binary black hole. Consequently, the orbital structure   in the core after core scouring is predicted to be strongly biased in favor of tangential orbits, while the ejected stars contribute to enhanced radial motions outside the core \citep{quinlanhernquist97,milosavljevicmerritt01}.   For example, the orbital structure of the S0  NGC524 shows  tangential anisotropy right at the SMBH radius  of influence, corresponding to the core region in the photometric profile \citep{krajnovicetal09}.  Similar results apply to the elliptical galaxy NGC1600 \citep{thomasetal16}, and agree well with predictions from numerical simulations where core profiles are the result of SMBH binaries impoverishing  the central nuclear regions \citep{rantalaetal18}.} 

Recent work emphasizes the presence of substructures in the \mbh -- $\sigma_{\star}$\ relation \citep{sahuetal20}. Pseudo-bulges are associated with spiral galaxies, and studies  of their photometric profiles  reveal that they are disk-like with a different surface brightness profile than classical bulges \citep{kormendykennicutt04,kormendybender12}. Pseudo-bulges are known to be offset in the \mbh\ -- $\sigma_\star$ relation  in the sense of systematically lower \mbh\ \citep{sagliaetal16}. In the case of pseudo-bulges, the growth of the central black hole may be decoupled from the growth of the host spheroid and not associated with galaxy merger, but instead with mechanisms of secular   evolution not related to gravitational interaction with other galaxies; {in observational terms, some studies \citep{KormendyHo2013} find weak \mbh\ correlations for pseudo-bulges, see, however, e.g., \citep{bennertetal21}.} 

The most massive BHs have been detected only in the more luminous galaxies ( $-22\leq M_B\leq -18$) \cite{FerrareseFord2005} and it is not clear yet if fainter and less massive systems host massive BHs and whether they follow the extrapolations of the SRs defined by the brightest objects. 
Searches for BHs in less luminous galaxies of the Local Group have produced ambiguous results, as in the case of M33 \cite{Merrittetal2001, Gebhardtetal2001}, NGC205 \cite{Vallurietal2004}, and M32 \cite{Verolmeetal2002}. Some galaxies exhibit
a compact stellar nucleus (with half-light radius $r_h\sim2-4$ pc) in their center. This is $\sim20$ times brighter than a typical globular cluster \cite{KormendyMcClure1993, ButlerMartinez-Delgado2005}. 
In the Virgo and Fornax Clusters $\sim 25\%$ of dE galaxies contain such nuclei \cite{BinggeliSandageTammann1985, Ferguson1989, BinggeliCameron1991}, but the observations with the Hubble Space Telescope  revealed that these structures are far more common: about 50-80\% of the less luminous galaxies contain a distinct nuclear star cluster \cite{CarolloStiavelliMack1998, Matthewsetal1999, Bokeretal2002, Balcellsetal2003, GrahamGuzman2003, Lotzetal2004,  Grantetal2005, Coteetal2006}.
{Nuclear star clusters are not a replacement for  black holes. On the contrary  low mass galaxies  ($10^9 -  10^{10}$ \msol) show a high incidence of nuclear star clusters coexisting with massive black holes \citep{greeneetal20}. However, nuclear star clusters are rare in high mass galaxies \citep{grahamspitler09}, suggesting that the growth of BH during activity may lead to the demise of the star cluster itself \citep{antoninietal19}. }

{The low mass end of the  relation \mbh\ - $\sigma_\star$\ for quiescent galaxies  is still poorly sampled and of uncertain interpretation \citep{grahamscott15}. Tidal disruption events (TDEs) provide an independent methods for \mbh\ estimation. First, TDEs are luminous flares predominantly detected in quiescent galaxies (very few events have been detected in AGN, as the luminosity of the AGN obliterates the brightness increase associated with the TDE). Flares are produced by the tidal debris that fall back toward the black hole and that form an accretion ring or a disk around an otherwise inactive black hole.   Second, a TDE can take only with relatively low black hole masses,  \mbh $\lesssim 10^8$ \msol\ for a solar-mass star \citep{gezari21} to avoid that the star crosses the black hole event horizon without being tidally disrupted. The central BH mass is recovered via synthetic multi-band optical light curves based  on hydrodynamical simulations of polytropic tidally disrupted stars   \citep{mockleretal19}. The method is not yielding yet a good agreement with other \mbh\ estimates, as it is not clear which parameter should be correlated with \mbh, in a model in which the TDE luminosity is powered by fall-back accretion \citep{gezari21}.   }

\medskip
\subsection{\mbh\ measurements in AGN}


Stellar dynamical determinations for AGN has been possible only for weakly or mildly active Seyfert 1 galaxies. Presently, only $\lesssim  100$ dynamical \mbh\ measurements have been obtained by modeling stellar kinematics of  quiescent and active nuclei \citep[e.g.,][]{KormendyHo2013,sagliaetal16}.
A case in point is the intermediate Seyfert 1 galaxy NGC3327 \citep{daviesetal06} which illustrates the complexity of the nuclear regions of a mildly active AGN, even if an application of the Schwarzschild method of orbit superposition allowed for a meaningful estimate of the stellar dynamic mass $\sim 10^7$ \msol. A Population B source (see \S \ref{ms} and  Fig. \ref{fig:grid}) radiating at modest Eddington ratio, NGC3227 shows a  nuclear stellar distribution within a few parsecs of the central black hole affected by intense bursts of star formation occurring in its recent past. Similar considerations apply to the stellar dynamics results on  MCG–6-30-15 \citep{raimundoetal13}.  In general, stellar populations in the closeness of the active nucleus are not easy to model, also because of the uncertain distribution of obscuring dust, even in the least active nuclei, such as NGC4258. Nonetheless the stellar dynamical and maser masses agree very well for this source  \citep{siopisetal09}. A dynamical mass estimate for Cen A also agrees with the estimate derived from cold molecular H$_2$\ gas \citep{neumayeretal07,Cappellarietal2009}, suggesting that molecular gas  could provide mass estimations as accurate as the ones based on stellar dynamics. 



The most reliable method to probe distances within $r_\mathrm{h}$ from the black hole of galactic nuclei that are currently mildly active involves observation of H$_{2}$O masers \cite{miyoshietal95,herrnsteinetal05,greeneetal10}.  The  H$_{2}$O emission profile shows a radial velocity ``cusp'' at distances where the velocity field is governed by gravity of the black hole i.e., when $r \lesssim r_\mathrm{h}$. This method is not exempt by issues that could bias the results.  A maser disk with  Keplerian rotation could have a non-negligible disk mass comparable to the black hole mass \citep{hureetal11}, and the effects of disk self-gravity might lead to large systematic errors in the derivation of the black hole mass \citep{kuoetal18}. Disturbed morphology and kinematics are an ubiquitous feature of  maser systems especially above Eddington ratio  $\lambda _{\rm Edd}\, \gtrsim$ 0.1 \citep{kuoetal20}.

Outside of the local Universe (i.e., beyond 2.5 Mpc), VLBI observations of {\it mega-maser} systems can probe   within the sphere of influence for BHs down to $10^6$ \msol\ even at 100 Mpc \cite{vandenbosch16}.   {\it Mega-masers} are more frequently associated with active galaxies \cite{constantin12}, and they should be more common at high redshift \cite{vandenbosch16}.  The  exploitation of {\it mega-masers} is however difficult, because the maser signal of high-redshift sources is lost in noise, and major surveys have until now failed to detect a {\it mega-maser} in the wide majority of galaxies.  Mass determinations based on the resolved maser systems are completely independent from any other method, and best suited to cross-check the estimates obtained from stellar and gas dynamics.  Several $\mathrm H_2 \mathrm O$\ masers have been detected from the circumnuclear regions of quasars also at relatively high redshift  \citep{staceyetal20,broomeetal19}. The hope to go beyond modest distances rests with SKA — because of its unprecedented	$\mu$Jy sensitivity — and in  future space based radio interferometry with $\mu$arcsec spatial resolution.  Assuming Earth-space baselines of about 30,000 km, angular resolution of 2 $\mu$-arcsec   would be achievable  at 8 GHz \cite{taylor08}.  This angular scale corresponds to a projected linear size of $\approx 0.02 $  pc at $z=2$, therefore allowing to probe within the BH sphere of influence even at the remote epochs of the “cosmic noon,” a key epoch of galaxies AGN when a population of most luminous quasars was shining bright and producing the maximum feedback on their host galaxies.


{ALMA is today the most powerful tool to yield \mbh\ also for AGN \ \citep[e.g.,][]{onishietal15}. However, the CO J = 2-1 kinematics in a sample of nearby AGN reveals non circular motions in the inner kiloparsec of all  galaxies in the sample, although  molecular gas and stellar kinematics show an overall agreement. The CO observations of nearby radio galaxies  detect molecular disks, but also cautions about the possibility of  asymmetries and disruptions due to interactions with the radio jet \citep{ruffaetal19}.   
}



{Several studies have employed the capabilities of STIS on board HST to study the dynamics of line emitting gas in proximity of the central black  hole \citep{pastorinietal07}, or the sub-arcsec  spatial resolution of imaging spectrometers or IFU units operating with adaptive optics \citep{hicksmalkan08}. The concern is    that radiation forces within the inner 100 - 1000 pc  of the central black hole may be affecting the dynamics of the line emitting gas even more than in the case of cold gas dynamics, especially if the AGN is radiating at high Eddington ratio \citep{zamanovetal02,xukomossa10,marzianietal16,schmidtetal18,bertonetal16}.

Spectro-astrometry is another promising tool: the approach is based on the different   photocenter position of emission lines at different velocity  \citep{gneruccietal11}. Whereas a relatively modest spectral resolution ($\sim 10$ \kms) is sufficient, sub-arcsec spatial resolution is required, obviously the higher the better, achievable only from space or from ground using active optics \citep{gravitycollaboration21}. This is the approach  exploited by GRAVITY, an instrument of the Very Large Telescope Interferometer (VLTI) \citep{gravitycollaboration17}. } After first light in 2017, GRAVITY  detected a spatial offset (with a resolution of 10 micro-arcseconds corresponding to about 0.03 parsecs) between the red and blue centres of the Paschen-$\alpha$\ line of 3C273  \cite{sturmetal18}.  This offset corresponds to a gradient in velocity and implies that the gas is orbiting the central supermassive BH.  With the new capabilities of GRAVITY \cite{abuteretal17} and with the wave-front corrections of an adaptive optics system, it will be possible to repeat this feat in many low-$z$\ type-1 (i.e., broad line) AGN \citep{boscoetal21}. The broad line region velocity field has been spatially resolved and modelled even in NGC3783 \cite{amorimetal21} to provide an \mbh\ estimate, although this achievement will likely remain restricted to low-$z$ Seyfert-1 nuclei for the time being. 
 


If we exclude masers, for which BH masses can be inferred from rotation   \cite{FerrareseFord2005} and spectro-astrometry, for the vast majority of Type-1 AGN the BH masses are derived from the (presumed) virial motions of the broad line region (BLR) gas clouds orbiting in the vicinity of the central compact object. If the motion in the emitting gas is in virial equilibrium, we can write the central black hole mass \mbh\ as:
\begin{equation}
M_{\rm BH} = \frac{r_\mathrm{BLR} \delta v_{\rm K}^{2}}{G}.
\label{eq:vmbh}
\end{equation}
Here $\delta v_{\rm K}$ is the virial velocity module, $r_\mathrm{BLR}$\ the radius of the BLR, $G$\ the gravitational constant. Eq. \ref{eq:vmbh} can be useful if we can relate $\delta v_{\rm K}$\ to the observed velocity dispersion, represented here either by the dispersion $\sigma$ or by the FWHM of a suitable broad emission line: 
\begin{equation}
M_{\rm BH} = f_{\rm S} \frac{r  {\rm_\mathrm{BLR} FWHM}^{2}}{G}
\label{eq:vir}
\end{equation}
via the structure factor (a.k.a. form or virial factor)    whose definition is given by:
\begin{equation}
 \delta v_{\rm K}^{2} = f_{\rm S}{\rm FWHM}^{2}.
\end{equation}

\medskip
{Mildly ionized gas dynamics i.e., gas motions within the broad line regions of type-1 AGN, is the basis of the estimate of the \mbh\ for large samples of quasars up to the highest $z$, following Eq. \ref{eq:vir}. In addition to a measure of the virial broadening provided by the emission line width, a measure of the line emitting gas distance from the central black hole is needed. Under the assumption that the main source of line emission is provided by photoionization \citep{shuder81}, the distance is measured by the time lag of the emission lines with respect to continuum variations \citep{peterson93}: $r_\mathrm{BLR} \approx c \tau$, where $\tau$ is the time delay. Recent observations measure $\tau$\ as a function of wavelength across the line profile in an attempt to resolve the velocity field of the emitting region \citep{brothertonetal20,williamsetal21}. The "reverberation mapping" technique has been described in several reviews that also include a critical discussion of the technique shortcomings \citep{horneetal04,marzianietal06,peterson14}. The $r_\mathrm{BLR}$\ estimates have been carried out mainly for the HI Balmer line \hb\ for $\sim 100$ type-1 AGN, recently supplemented by the monitoring of the SDSS Reverberation Mapping Project that yielded data for 144 quasars \citep{lietal17}. 
The reverberation mapping determinations of $r_\mathrm{BLR}$ offer a sort of  primary step over which a correlation between  $r_\mathrm{BLR}$\ and luminosity is built (Section \ref{rl}), in turn instrumental to the determination of the \mbh\ in large samples of quasars (Sect.  \ref{slmass}). }

\medskip
\label{rl}
\begin{figure}[h!]
	\begin{center}
		\includegraphics[width=10cm]{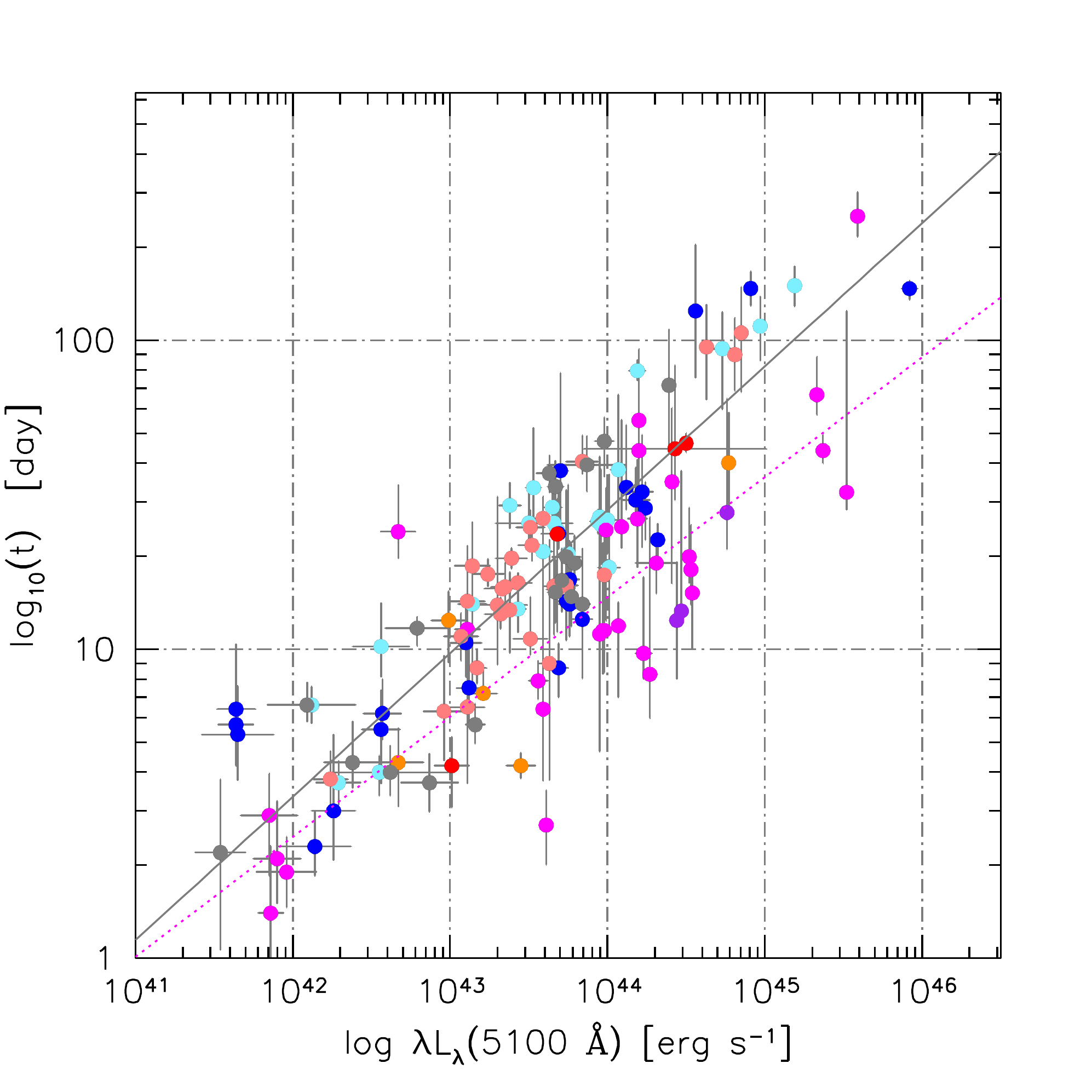}
	\end{center}
	\caption{The radius-luminosity  relation, expressed as the relation between the time lag derived from reverberation mapping and the optical luminosity. Data are from \cite{duwang19}, and include, in addition to the sources of \cite{bentzetal13} also the xA sources monitored in dedicated campaigns \cite{duetal16,duetal18}. Sources are color coded according to the spectral types identified along the quasar main sequence: B1$^{++}$ (red), B1$^{+}$ (orange), B1  (rose), B2 (grey), A1 (aquamarine), A2 (blue), A3 (magenta), A4 (purple), {and roughly correspond to a sequence of increasing Eddington ratio}. The grey line traces an unweighted least square fit for the full sample, the dotted magenta line refers to an unweighted lsq but for sources radiating at extreme Eddington ratios (A3 and A4) only.    }\label{fig:r}
\end{figure}

\subsection{The Radius - Luminosity relation}
\label{rl}

A correlation between radius of the emitting regions and continuum  luminosity is expected on the basis of the spectral similarity of quasars. Even if this is an oversimplification, we observe always the same lines, and their relative intensities change only within a limited range, also in response to continuum variation. The ionization parameter should remain roughly constant, implying that $r \propto L^a$, with an exponent  $a$\ at any rate close to 0.5 \cite{kaspietal00,bentzetal13}. The scaling relation has been derived from spectroscopic monitoring of emission lines (mostly the HI Balmer line \hb) that yield the time delay $\tau$ of the emission line response to continuum variations \cite{peterson93,peterson17}.  A sufficient number of sources is available for a correlation analysis since the early 2000s   \cite{kaspietal00}. The consideration of various aspects (host galaxy subtraction and removal of the line narrow component believed to be emitted in a different region) and the increase of the number of monitored sources has led to a standard $r - L$\ relation with an exponent consistent with 0.5 within the uncertainties \cite{bentzetal06,bentzetal09a,bentzetal10,bentzetal13}. However,  the  $r - L$\ {relation suffers} of  significant scatter because  it  was derived neglecting the diversity of type-1 quasars organized by the quasar Main Sequence (MS, see below), and is biased in favour of  sources radiating at relatively low Eddington ratio. It is not difficult to account for this preferential selection: such sources are relatively low accretors and therefore more prone to variability associated with an unsteady accretion flow. Recent work \cite{duetal16,duwang19} has shown that sources that radiate at high \lledd\ significantly deviate from the correlation of \cite{bentzetal13}: their radius is shorter than the one expected on the basis of their luminosity. Including high Eddington ratio sources in the correlation creates a cluster of data points that increases the scatter in the correlation.  Fig. \ref{fig:r}\ shows that the relation for  sources radiating at the {highest value of the Eddington ratio} is significantly offset from the one of other spectral type along the quasar main sequence discussed in Sect. \ref{ms}.  Linear combination with the dimensionless accretion rate {(i.e., the mass accretion rate normalized by the Eddington accretion rate)} or Eddington ratio leads to a significant reduction of the scatter \cite{duwang19,martinez-aldamaetal20}.

\medskip
\subsection{Scaling laws for AGN black hole mass estimates}
\label{slmass}

The virial theorem can be conveniently rewritten as $\log$\mbh =  $\alpha \log L$+$\beta \log$ FWHM +  $\gamma$, where $\beta = 2$.  The luminosity term comes from the use of the radius — luminosity relation, $r - L^a$. Several different scaling laws based on this expression have been defined for the width of different lines, and  for different continuum and line luminosity as well. The most widely used has been perhaps the one formulated by \citet{vestergaardpeterson06} for \hb\ and continuum luminosity at 5100 \AA.

The main underlying assumptions in the use of the virial theorem are that the broadening is due to Doppler effect because of the line emitting gas, and that the velocity field is such that the emitting gas remains gravitationally bound to the black hole. Early UV and optical inter-line shift analysis provided evidence that not all the line emitting gas is bound to the black hole \cite{gaskell82,tytlerfan92,brothertonetal94,marzianietal96,leighlymoore04}. The emerging scenario is that outflows are ubiquitous in AGN, they occur under a wide range of physical conditions, and are detected in almost every band of the electromagnetic spectrum and on a wide range of spatial scales, from few gravitational radii to tens of kpc (e.g., \cite{capettietal96,colbertetal98,everett07,carnianietal15,crescietal15,bischettietal17,komossaetal18}. 

For z$>$4, \mbh\ estimates historically rely on the \civ\ high-ionization line, and the highest-$z$ sources appear almost always high-accretors \cite{banadosetal18,nardinietal19}. The source of concern is that high-ionization lines such as \civ\ are subject to a considerable broadening and blueshifts associated with outflow motions already at low redshift \cite{coatmanetal16, sulenticetal17,marinelloetal20a,marinelloetal20b}. Overestimates of the virial broadening by a factor as large as 5 -- 10  \cite{netzeretal07,sulenticetal07,mejia-restrepoetal16,mejia-restrepoetal18} for SMBHs at high $z$\ may pose a spurious challenge to concordance cosmology \cite{trakhtenbrotetal15} and lead to erroneous inferences on the properties of the seed BHs believed to be fledgling precursors of massive BHs. The solution is either to carry out \hb\ observations at high redshift (a feat that is becoming easier  as more NIR spectrometers are being installed at the focus of large telescopes) or to use a surrogate line whose profile is also virially broadened. The \aliii\ and \ciii\ lines  could be much robust estimator of the BH mass. These lines, in a blend at 1900 \AA, can be easily observed with optical spectrometers up to redshift $z \approx 4$. Similar considerations apply to the use of \mgii\ \cite{trakhtenbrotnetzer12,shenliu12}  which however can be observed only up to $z \approx 2.5$\ without the use of NIR spectrometers.  Another approach has been to apply corrections to the \civ\ line width \cite{coatmanetal17,marzianietal19}, although such corrections, to be effective, require the knowledge  of the quasar rest frame, that remains poorly known from rest-frame UV observations only.  \citet{shenliu12}  propose scaling laws in which the virial assumption is released  that is, with $\beta \neq 2$. For \civ, this means correcting for effect associated with the emission component due to an outflow, that overbroadens (and shifts) the line. The  scaling law introduced by \citet{parketal13} follows this approach assuming $\gamma = {0.5}$, that is, a FWHM dependence that is much weaker than the one of the virial law. The scaling law suggested by \citet{parketal13} applied to  a high luminosity sample properly corrects for the overbroadening of the \civ\ line profiles of high Eddington ratio  sources of a high-luminosity sample, but overcorrects the width in case of sources radiating at modest Eddington ratios, yielding a large deviation from the \hb-derived \mbh\ values \cite{marzianietal19}.

\medskip
\subsubsection{The virial factor: orientation and radiation effects}
\label{angle}

{The application of Eq. \ref{eq:vir} requires the knowledge of the $f_\mathrm{S}$, a quantity of $\mathcal{O} \sim 1$\ but that can be significantly different from source to source. The presence of a rotating accretion disk and a spin axis for the central black hole guarantees that axial, and not spherical symmetry, is satisfied for AGN \citep{antonucci93,urrypadovani95}. Accordingly, unification schemes distinguish between sources that are observed with the line of sight oriented not very far from the disk axis, and sources that are seen almost edge-on, for which the observation of the BLR is precluded by obscuration (type-2 AGN). Leaving apart obscured sources, there is a considerable range of orientation angles (from 0 to 45 - 60) that are possible for type-1 AGN.}  The effect of orientation can be quantified  by assuming that the line broadening is due to an isotropic component + a flattened component whose velocity field projection along the line of sight is $\propto 1/\sin \theta$ \cite{mclurejarvis02,collinetal06,decarlietal11,mejia-restrepoetal18}. Even with this assumption, it is not known how to connect the viewing angle of the black hole + accretion disk system and the parameters measured on the optical and UV spectra.  Only in a few special cases this feat has been possible. In such cases the viewing angle is constrained by data unrelated to the spectra, such as the radio morphology or the jet beaming \cite{willsbrowne86,decarlietal11,punslyetal20}. {The dependence on orientation can be overcome by spectropolarimetric measurements: if the emission line light is scattered by an equatorial scatterer, then the width of the polarized line flux should be related to the velocity field as measured by an observer in the equatorial plane of the accretion disk, i.e., as if the viewing angle were $\theta = 90$\ from the disk axis, de facto removing the orientation effect. Spectropolarimetric measurements allowed for the estimate of the black hole mass in a few tens of type-1 AGN \citep{savicetal18,afanasievetal19,savicetal20,capettietal21}. The technique  requires large-aperture telescopes even for nearby, bright AGN, whose polarization is notoriously low ($\lesssim $1\%; Sniegowska et al. 2021, in preparation). } 


{A parameterization of the virial product dependent on the Balmer \hb\ line has been suggested \citep{mejia-restrepoetal18} in the form $f_\mathrm{BLR} \propto $ FWHM$^{-1.17}$\ and exploited in several works \citep{martinez-aldamaetal19,bonetal20}. This relation is however especially risky in samples covering a wide range of luminosity, since it is not accounting for the increase in line width expected with increasing mass, if line broadening is predominantly virial (Sect. \ref{virial}). In addition, orientation is not the only variable affecting $f_\mathrm{S}$. Radiation forces act on   gas motions and make  the $f_\mathrm{S}$\ dependent on Eddington ratio \citep[e.g.,][]{netzermarziani10,khajenabi15}. The effect can be as large as a factor $\approx 2$ and, perhaps more importantly, the efficiency of radiation forces is dependent on the gas column density, leading to the preferential expulsion of gas of  lower column density \citep{netzermarziani10}. Recent attempts to derive the $f_\mathrm{S}$\ from dynamical models still do not consider the role of radiation pressure on the gas motion \citep{pancoastetal14,pancoastetal14a,pancoastetal18, williamsetal20}. In addition,  there are basic difficulties in modelling the BLR. One of the main issues is whether there are indeed gas clouds or whether the broad lines are emitted directly by a continuum-illuminated accretion disk \citep{collinsouffrinetal88,dumontcollinsouffrin90d}. If clouds are indeed present, the mechanism of confinement is  unclear, although confinement by an external magnetic field is favored  \citep{rees87,bottorffferland00,chelouchenetzer01,shadmehri15,esseretal19}. The quasar main sequence discussed in Sect. \ref{ms} provides a focus for these questions, but the physical processes of line emitting gas dynamics have not yet been contextualized for different accretion modes (Sect. \ref{ml}).  

The virial factor $f_\mathrm{S}$\ has been estimated by scaling the {\em virial product} $r_\mathrm{BLR}\delta v^2$\ to the \mbh\ -- $\sigma_\star$\ for quiescent galaxies obtaining an average $f_\mathrm{S} \approx 5.5$\ if the velocity dispersion of the broad emission line is used ($\approx 2.3$\ from the FWHM). This approach provides a test of consistency for the reverberation mapping technique \citep{bennertetal21} within a factor 2-3 uncertainty.  {In principle, the  $f_\mathrm{S}$\ uncertainties could be reduced, if a careful separation of different morphological types and of different accretion modes is carried out. For instance, the technique applied to NLSy1s yields $f_\mathrm{S} \approx 1.1$\ (for FWHM; \cite{wooetal15}); \citet{duwang19} show that sources accreting at high rates do not obey the \citet{bentzetal13} relation. }     
}

\smallskip
\subsection{\mbh\ vs \mbu\ and $\sigma_\star$ for AGN and its consistency with quiescent galaxies at low-$z$} 


There is a {general consensus} that most galaxies host a massive BHs that went through phases of activity. This latter had a role in the BH growth and in the regulation of the SF activity of the host galaxy by means of wind/jet driven feedback mechanisms \cite{Shankaretal2009a, Shankaretal2009b, AlexanderHickox2012}. The theoretical models show that an AGN and its host may coevolve \cite{SilkRees1998, Granatoetal2004}, leading to characteristics (such as the $M_{s,sph}/M_s$ ratio and/or the central stellar velocity dispersion $\sigma$) related to black hole mass ($M_{BH}$).

An early answer to the question  ``do  galaxies hosting an AGN share the same $M_\mathrm{BH} - M_\mathrm{Bulge}$\ correlation of normal galaxies?"  was affirmative:  AGN have the same BH-bulge relation as ordinary (inactive) galaxies \cite{wandel02}. Fast forward twenty years,  there is not yet an established view. {A most recent work, based on  state of the art  surface photometry, and  spatially-resolved kinematics  to measure $\sigma_\star$,  find  that  correlations between \mbh\  and host galaxy properties hold for AGN within the limits of an intrinsic scatter 0.2 - 0.4 dex, and are consistent with the ones  of quiescent galaxies \citep{bennertetal21}. 
}

Recent works {also} point {toward a} complex scenario involving selection biases \cite{schulzewisotzki11} and a better appreciation of the active galaxies diversity.  We may represent the distribution of objects in the \mbh\  -- \mbu\ (or $\sigma_\star$)  diagram    by the bivariate distribution function of bulge mass \mbu\   and \mbh\ $\Psi$(\mbh,\mbu).   The $ \Psi$\ distribution can be factorized as {$ \Psi  =  \gamma$(\mbh $|$ \mbu)$\phi$(\mbu)}\ where $\phi$(\mbu) is the spheroid mass function and $\gamma$\   represents the \mbh - \mbu\ correlation i.e., the probability of having the black hole mass \mbh\ for a given \mbu.  A correct evaluation of $\gamma$(\mbh $|$ \mbu)\ relies  on: (1) the knowledge of  $\phi$(\mbu), which is not a trivial task to achieve even in the local Universe, and needs a separate consideration of purely spheroidal (i.e., diskless) galaxies and galaxies with pseudo-bulges or with a bulge/disk system;  (2)   the absence of biases affecting $\gamma$(\mbh $|$ \mbu).  

Both the determinations of \mbh\ and bulge parameters are challenging, when derived from conventional optical and NIR measurements. At present, black hole masses  for  type-1  AGN are more frequently derived   through  the so-called, single epoch “virial broadening” estimation i.e., through the measurement of the  radial velocity broadening term that appears squared  in Eq. \ref{eq:vir} from single epoch spectra. In practice, it is the measurement of  the  FWHM or $\sigma$\footnote{Not to be confused with $\sigma_\star$.}  of broad emission lines \cite[e.g.][]{mclurejarvis02,vestergaardpeterson06}. To obtain \mbh, an estimate of the radius $r_\mathrm{BLR}$ is also needed, and a rather poorly-defined scaling law of $r_\mathrm{BLR}$\ vs luminosity is applied.   

The bulge estimates in AGN samples are  hampered by the luminous  source associated with the active nucleus, which may well outshine the entire galaxy. The \mbu\ has been frequently computed from  the host galaxy luminosity  \citep{pengetal06a,pengetal06b,treuetal07,mcleodbechtold09,decarlietal10,bennertetal10,targettetal11,schrammetal08}.  
However, type-1 AGN remain offset from inactive galaxies in the \mbh\ - $L_\mathrm{bulge}$\ relation: AGN have more luminous bulges  at a given black hole mass \cite{nelsonetal04,bennertetal21}. There are   evidences that the $M_{BH}-L_\mathrm{bulge}$ relations defined by quiescent BH samples differs from that defined by the galaxies in the SDSS  \cite{Bernardietal2007}. Interestingly, the offset is larger for AGN of larger Eddington ratio \citep{barthetal21}. This suggests that the central regions of galaxies hosting an AGN  have in general lower mass-to-light ratios than inactive galaxies, most likely for the presence of a young stellar population in the bulge of active systems \citep{kimho19}.  


The  $\sigma_\star$\  has been measured either   directly i.e., from the width of absorption lines associated with the stellar component of the host galaxies \citep{wooetal06,wooetal08,shenetal08} or by using   the widths of narrow emission lines a proxy of the stellar velocity dispersion  \cite{shieldsetal03,shieldsetal06,salvianderetal07}. The latter approach is  fraught from systematic effects. In the case of quasars and AGN radiating at moderate and high Eddington ratio, the \oiii\ broadening is strongly affected by non-virial motions \cite{zamanovetal02,marzianietal06,mathur11,marzianisulentic12,craccoetal16}. 

More reliable results  for dynamical mass measurements of the host galaxy from spatially-resolved images have been obtained with adaptive optics \citep{inskipetal11}. CO emission profiles have been used to estimate dynamical masses for individual objects since the early 2000s \cite{walteretal03} even at fairly high redshift, and nowadays  ALMA is rapidly adding to the available dynamical mass measurements for the host galaxy \cite[e.g.,][]{tanetal19,molinaetal21}, considering that the velocity field of the molecular gas is often  regular and consistent with rotation.    State-of-the-art  surface photometry of the AGN host galaxies in the NIR  achieve  decomposition  in spheroid, disk and bar component, as most of the host of nearby Seyfert galaxies are of morphological type Sa/SBa. As mentioned, a most recent work didn’t detect   significant differences in the scaling with \mbh\ and $\sigma_\star$ between active and non-active galactic nuclei \cite{caglaretal20, bennertetal21}, and didn’t find difference between pseudo and classical bulges or barred and non-barred galaxies in the \mbh\ -- \mbu\ relation \cite{bennertetal21}, although this result is still controversial \cite{kormendyetal11,hokim19}.  In addition,   $\gamma$  is still computed with the single epoch technique, without consideration of the diversity in accretion structure (and hence virial factor) that is expected in type-1 AGN samples.  For type-1  active nuclei radiating at Eddington ratio above 0.01, the geometry and structure of the emitting region are affected by the accretion mode, which in turns affects the expression of the virial factor that is dependent on  kinematics, geometry, and viewing angle  \cite{collinetal06,mejia-restrepoetal16,mejia-restrepoetal18,parketal12,shankaretal19}. A study separately considering sources in different accretion modes  and the statistical bias introduced by orientation effects is not available has yet. 

Keeping  the attention focused on $\gamma$, the radius of influence $r_\mathrm{h}$  is of the order of parsecs, and insufficient resolution may prevent reliable BH mass estimates or forces to target only the largest BHs \cite{gultekinetal09,gultekinetal11}, leading to a selection effect that yields an increase in the $M_\mathrm{BH}$ -- $\sigma$\ relation for quiescent galaxies  by a factor of a few \cite{shankaretal16,shankaretal19}.  AGN will on average host more massive BHs than in the volume-limited case \cite{laueretal07}, determining a Malmquist bias  towards more massive BHs at a given spheroid mass, shifting $\gamma$\ upwards and causing an offset in the zero-point of the \mbh\ -- \mbu\ relation.  There are competing effects:  the fraction of active galaxies among SMBHs varies considerably with mass (high-mass BHs are likely less active than low-mass BHs \cite{schulzewisotzki11}). The strength of the bias depends on the limit in luminosity, the shape of the distribution function of spheroids, the scatter of the \mbh-\mbu\ relation, and the Eddington ratios. If, as mentioned, the active fraction decreases as the BH mass increases, then for a given spheroid mass it will be more probable to find small-masses BHs in an AGN sample, causing a bias towards lower \mbh/\mbu ratios, and a change in the slope of the relation.  

The low-\mbh\ end of the correlation is especially problematic, as it is for quiescent galaxies. Narrow-line Seyfert 1s nuclei (NLSy1s,  low-$z$\ type-1 AGN several of which are accreting at high rate, \cite{marzianisulentic14}), often  hosted in dwarf high surface brightness galaxies \cite{krongoldetal01} and in barred spirals \cite{crenshawetal03,ohtaetal07}, possess under-massive BHs   \cite{mathuretal01,chaoetal08}.  NLSy1 nuclei often reside in disk-dominated galaxies with pseudo-bulges \cite{orbandexivryetal11,mathuretal12,ermashkomberg12,olguin-iglesiasetal17,jarvelaetal18,doietal20}. These types of bulges are more closely associated with the evolution of disks and may be typical of systems that did not experience a minor or major merger capable to lead to a real bulge development.  Several studies found that disk-dominated  galaxies deviate from the $M_\mathrm{BH}$\  -- $M_\mathrm{bulge}$\  correlation, and, if considered as a distinct class, may not follow a \mbh -- \mbu correlation \citep{kormendyetal11,davisetal18,sahuetal20}.  However, if one applies a correction for the disk component, and considers only the bulge,  the AGN in  the low black hole mass ranges \mbh $\lesssim 10^8$ \msol\  might follow a relation consistent with the local   \mbh -- \mbu\  correlation  \citep{sanghvietal14,bennertetal11}. At any rate, the relation between $M_\mathrm{BH}$\  and  $\sigma_\star$\ or \mbu\ should be taken with special care in particular in the lower $M_\mathrm{BH}$\  range. Relatively few objects are unobscured type-1 AGN.  Chandra observations are detecting a wealth of black holes in star-forming galaxies, in the range between 10$^6$  -- 10$^7$ \msol, even at high $z$ \cite{mezcuaetal16,fornasinietal18,zuoetal20}.  They are low mass by supermassive black hole mass standards, and most likely still growing in an obscure phase. It is not known how they would be located in the \mbh -- \mbu\ plane. These elusive AGN are potential targets for JWST  \cite{satayapaletal21}.

\medskip
\subsection{Over-massive and under-massive black holes}

At  the time of its discovery, the luminous quasar   HE0450-2958 appeared as an oddity:  a quasars without a host galaxy! \cite{magainetal05}. Understandably enough, the source attracted a lot of interest, and perhaps even a revival of the non-cosmological interpretation of quasar redshifts \cite{sulenticarp79,arpetal79,sulenticarp83}.  HE0450-2958  appears  hosted by a galaxy much fainter than that inferred from the correlation between BH mass and bulge luminosity \cite{kimetal07}.  In the case of quiescent galaxies, compact dwarf galaxies whose BH has a mass reaching even 15\%\ of the total galaxy mass  \cite{reinesetal14,sethetal14,vanloonsansom15} are observed. A possible explanation is that their outer parts may have been stripped by repeated encounters with other galaxies and produced an ultra-compact dwarf galaxy.  The EAGLE cosmological and hydrodynamical simulations suggest that these kind of objects are outliers resulting from the combination of stellar tidal stripping and the early formation epoch, that leaded to a rapid BH growth at high redshift, with the first mechanism being the most relevant for 2/3 of these sources \citep{barberetal16}.   However,   the disk/bulge decomposition is a delicate procedure. A careful reanalysis of the most striking cases,  Mrk1216, NGC1277, NGC1271, and NGC1332, suggests that a proper re-evalution of the disk size with an ensuing increase in spheroid mass will bring these sources in better agreement with the \mbh -- \mbu\ relation \cite{savorgnangraham16}.  The case of  HE0450-2958 has not been fully explained to-date. Past works have considered intriguing lines of evidence suggesting high \lledd\ and BAL outflow \cite{merrittetal06,liparietal07}.   However  HE0450-2958, that  appears as  a mini-BAL from a FOS spectrum, shows modest optical \feii\ emission, and a spectrum similar to the one of PG1211+143 \cite{merrittetal06}. According to the main sequence trends (see \S \ref{ms}), the object should not be highly accreting \cite{marzianisulentic14,duetal16a}. It is also unlikely that 
HE0450-2958 is a recoiling black hole ejected by a companion galaxy at about 7 kpc of projected linear distance, on the ground of the strong narrow line emission of \oiiiopt\ \cite{merrittetal06}. HE0450-2958 does not appear as an extraordinary powerful quasar. The upper limits on the host galaxy luminosity are not very constraining, so that this object could be well within the limits set by the scatter in the \mbu -- \mbh\  correlation \cite{kimetal07}.

However, recoiling black holes — provided that they are the active member of the binary,  as suggested by numerical simulations \cite{nguyenbogdanovic16} —  may systematically lower \mbh\ and ultimately increase the scatter of the observed BH–host galaxy bulge relation due to ejected BHs \citep{volonteri07}. Recoiling BHs have lower masses than their stationary counterparts, but the deficit in mass depends on   kick speed and  merger remnant properties \citep{blechaetal11}. The effect is of an overall downward shift in the normalization and an increase of the scatter in the \mbh — \mbu\ relation:  the offset between  the stationary and the  recoiling BH population can reach $\delta \log g \approx $ 0.4 dex, if the rotational velocity of the secondary BH is close to its escape velocity. The amplitude of the downward offset depends on the recoil velocity as well as on the accretion history of the stationary black hole, and can be lower, yielding a $\delta \log \phi \approx $ 0.2 dex. This scenario is not as yet contextualized: a large fraction of type-1 AGN shows evidence that they do not host a sub-parsec binary black hole with a significant mass ratio between the secondary and the primary (say $q \gtrsim 0.1$). Conclusive evidences in favor of such binary systems are very rare at the time of writing.  

\medskip
\subsection{Evolution of the ${M}_{\mathrm{BH}}\mbox{—}$\mbu\ relation} 

Active galactic nuclei and quiescent bulge-dominant galaxies do not show  strong evidence of evolution in the ${M}_{\mathrm{BH}}$ -- \mbu\ relation up to $z \sim 0.6 - 1$ \cite{salvianderetal07,schulzewisotzki14,lietal21}.  At higher redshift, there is an increasing evidence of evolution, in the sense of high-$z$ SMBHs that are overmassive at a given bulge mass than expected from the local scaling relation \citep{mclureetal06,decarlietal18}.  Between redshift 1 and 2, \citet{merlonietal10} suggested a significant increase  of the $M_\mathrm{BH}/M_{\rm Bulge}$ ratio ($\propto (1+z)^{0.68}$).  Studies at even higher redshift  used the velocity dispersion of the gas as a proxy of the stellar velocity dispersion and dynamical mass measurement from inclined disk models \citep{vayneretal21}. They suggest over-massive black holes   \cite{targettetal11} with respect to the local scaling law.  The most recent results confirm that quasars host galaxies are  under massive relative to \mbh, and detect a large difference, even by an order of magnitude, with  systems at redshift in between 1.4 and 2.6 residing off the local scaling relation.  Several quasar host galaxies have been resolved {in their} [C II] emission on a few kpc scale at redshift $\approx 6$. Even in this case, the dynamical mass estimates for the host galaxies give masses  more than an order of magnitude below the values expected from the local scaling relation \citep{decarlietal18}, in agreement with the results for galaxies at $z \approx 7$ derived from cosmological hydrodynamical simulations \cite{marshalletal20}.  

The evolution of the \mbh\ -- \mbu\ relation with the cosmic epochs can be interpreted in several ways: the most straightforwards is that of a rapid growth of SMBHs at high redshift \cite{lupietal21}.  Also a variation of structural properties of AGN hosts remains possible \cite{shankaretal13,zhuetal21}: elliptical galaxies are not really monolithic spheroids, but  have undergone significant late-time dissipation-less assembly.  There are intriguing caveats with the interpretation of a rapid black hole growth.  First,  very massive seed black holes need to be formed at $z\approx$ 20 to account for masses $\sim 10^9$ \msol\ observed at redshift $z\gtrsim 4$\ (\cite{volonteri10,trakhtenbrot21}, and references therein).   Second, BH  masses (unlike the masses of galaxies!) can only increase with cosmic epoch.  If  the merger-driven hierarchical scenario that implies the parallel growth of bulges and BHs is taken literally, the larger $M_\mathrm{BH}/M_{\rm Bulge}$\ ratio at high $z$\ means that {mergers} affect more bulge than BH masses (at cosmic epochs associated with $z \gtrsim 1$),  an implication consistent with the anti-hierarchical growth and downsizing of the nuclear activity at low-$z$ \cite{hirschmannetal12}. {If pseudo-bulges follow the same \mbh\ scaling relations as that of classical bulges \cite[e.g.,][]{bennertetal21}, hierarchical growth might not be the only mechanism that drives the relation between  \mbh\ - \mbu: in spiral galaxies, secular evolution might lead to a parallel growth of bulge and central black hole}. Clearly, this issue should be analysed in connection to ongoing star formation properties of the pseudo-bulge hosts \cite{zhaoetal21}.  Host   and black hole properties are different for different masses, and the relation between galaxy color and black hole mass is different for the red and blue sequence quiescent galaxies, suggesting different channels of black hole growth for the two   sequences \cite{dulloetal20}.

In conclusion,  AGN with ``coreless" elliptical/bulge-dominated hosts may straightforwardly follow a relation similar to the one of normal galaxies. In other words, the \mbh -- \mbu\ relation may strictly holds for massive evolved systems, also if the nucleus is active, in a form that is as yet {indistinguishable} from the one of quiescent galaxies. It remains to be tested whether these sources could be mainly  AGN accreting at relatively low rate and radiating at modest Eddington ratios (Population B, Section \ref{ms}).  Significant deviations may be associated with disk dominance, but a careful assessment of the relative disk and  bulge  contribution might bring the system with the over-massive BHs in agreement with the established relation \cite{caglaretal20,zhaoetal21}.   The local NLSy1s -- all of which are Population A (Section \ref{ms}), with a significant fraction of high accretors -- are instead believed to be with black holes under massive with respect to their host masses. In  this respect they are different from the high-$z$\ quasars with over-massive black holes. However, the observational properties of low-$z$\ AGN accreting at relatively high rate can still be regarded as typical of very high $z$ quasars, when massive bulges were not yet formed, as originally suggested by \citet{mathur00,sulenticetal00a}. The analogy is based on the optical, UV, and X-ray AGN {spectroscopic} properties that are mainly governed by the Eddington ratio. In addition, modest masses of low-$z$ quasars can  grow by a factor $\sim 10$ on time scales shorter than  timescale of the cosmic evolution of quasar accretion rates, and therefore bring under massive BHs in line with the \mbh -- \mbu\ relation  \cite{fraix-burnetetal17}.

\section{The fundamental plane of AGN and the type-1 AGN main sequence}
\label{AGNSRs}

Some general considerations are in order when restricting the attention to the nuclei of galaxies. First, the central engine of nuclear activity is contained within a few parsec from its prime mover, the accreting massive black hole. Several scaling laws that are widely applied  in the study of galaxies are not considered in the study of AGN:   the Kormendy relation loses its meaning in the context of a system that is observed without  spatial extension. Or, they might connect different physical bodies: when we speak about the $r - L$ relation for AGN, $r$\ is the radius of the line emitting region, and $L$\ is the luminosity of the AGN. The two parameters do not refer to  cospatial entities.  Similar consideration apply to the \mbh\ -- $r$, or the \mbh\ -- $L$, or the \mbh\ -- metallicity relations. 

The virial equation (Eq. \ref{eq:vir}) is yielding the same FWHM for the same $r$/\mbh; what matters is the radius in units of gravitational radii, a dimensionless quantity. A smaller mass can give the same line width of a larger mass provided that $r$\ scales with \mbh. This is why we need an estimate of the linear size $r$ to recover a value of \mbh\ in physical units.  This scale invariance is obviously not applicable  to radiative phenomena: the flux reaching a distance $r$ will decrease with the inverse of the square of $r$\ on a {\em dimensional} scale.    The BLR radius $r_\mathrm{BLR}$\ subtends such a small angle that has not been directly resolved if not in the last few year thanks to the GRAVITY instrument \cite{amorimetal20,amorimetal21}. The foundations of any AGN diagnostics therefore rest on the scale invariance of gravitational forces, and on  electromagnetic phenomena instead lacking such scale invariance.  These considerations can be translated in mathematical terms to provide at least a self-similar framework that includes the fundamental plane of black holes, the modelization of jets  \cite{heinzsunyaev03} {and the quasar }main sequence (MS).

\medskip
\subsection{The fundamental plane of black hole activity}

The  fundamental plane of black hole activity can be written as a correlation between black hole mass, X-ray and radio luminosity. The correlation defines a plane in the space of parameters defined by the mass and the radio and X-ray luminosities. In its original formulation, the  fundamental plane was written as \cite{merlonietal03}:
\begin{equation}
log L_\mathrm{R} = (0.60\pm 0.11) \log L_\mathrm{X} + (0.78^{ +0.11}_{-0.09}) \log M_{BH} + 7.33^{+4.05}_{-4.07}. \label{eq:ss}
\end{equation}

The scatter is large, implying that a fourth variable might be involved, for instance black hole spin \citep{unalloeb20}.   The salient point is however that the relation holds over a huge range of black hole masses, from a few times solar (i.e., from the domain of the so-called micro-quasars) to the largest black hole masses detected in the Universe $\sim 10^{10}$ \msol\ \cite[e.g.,][]{schindleretal21,valtonenetal12}. \footnote{As stressed by \citet{sulenticetal04,sulenticetal06,sulenticetal07}, \mbh\ much in excess of $\sim 10^{10}$ \msol\ are unrealistic and probably the results of  the use of a high-ionisation line affected by wind kinematics as a virial broadening estimator (VBE). This makes the primary black hole of OJ287 \cite{sillanpaaetal88} as the most massive {active} black hole known to-date, with a mass of $\approx 1.8 \cdot 10^{10}$ \msol, {second to the black hole of Holm 15A, the central galaxy of  galaxy cluster Abell 85, with $\approx 4 \cdot 10^{10}$ \msol} \citep{mehrganetal19}. } It is remarkable that also stellar-mass black holes exhibit relativistic jets, as spectacularly demonstrated by the relatively recent discovery of superluminal motion in a Galactic black hole candidate by \citet{mirabelrodriguez94}.  The self-similarity expressed in Eq. \ref{eq:ss} allows for an invariant jet model and a simple relation between \mbh\ and radio power \citep{heinzsunyaev03}.  The self-similarity notwithstanding, there  is a non-linear relation between BH mass and radio power, with $P_\nu \propto $ \mbh$^{1.3 - 1.4}$, implying that the radio emission normalized to the bolometric luminosity should be much higher for AGN than for microquasars. In the framework of the model of \citet{heinzsunyaev03},  flat spectrum synchrotron jet emission is produced by an inefficient accretion mode. The fundamental plane of black hole activity refers to sources accreting at very low rate {(dimensioneless accretion rate $\dot{m} \lesssim$ 0.01)}, and radiating below a few hundredths of their Eddington luminosity. 

\medskip
\subsection{The quasar Main Sequence}
\label{ms}

\begin{figure}[h!]
	\begin{center}
		\includegraphics[width=9.25cm]{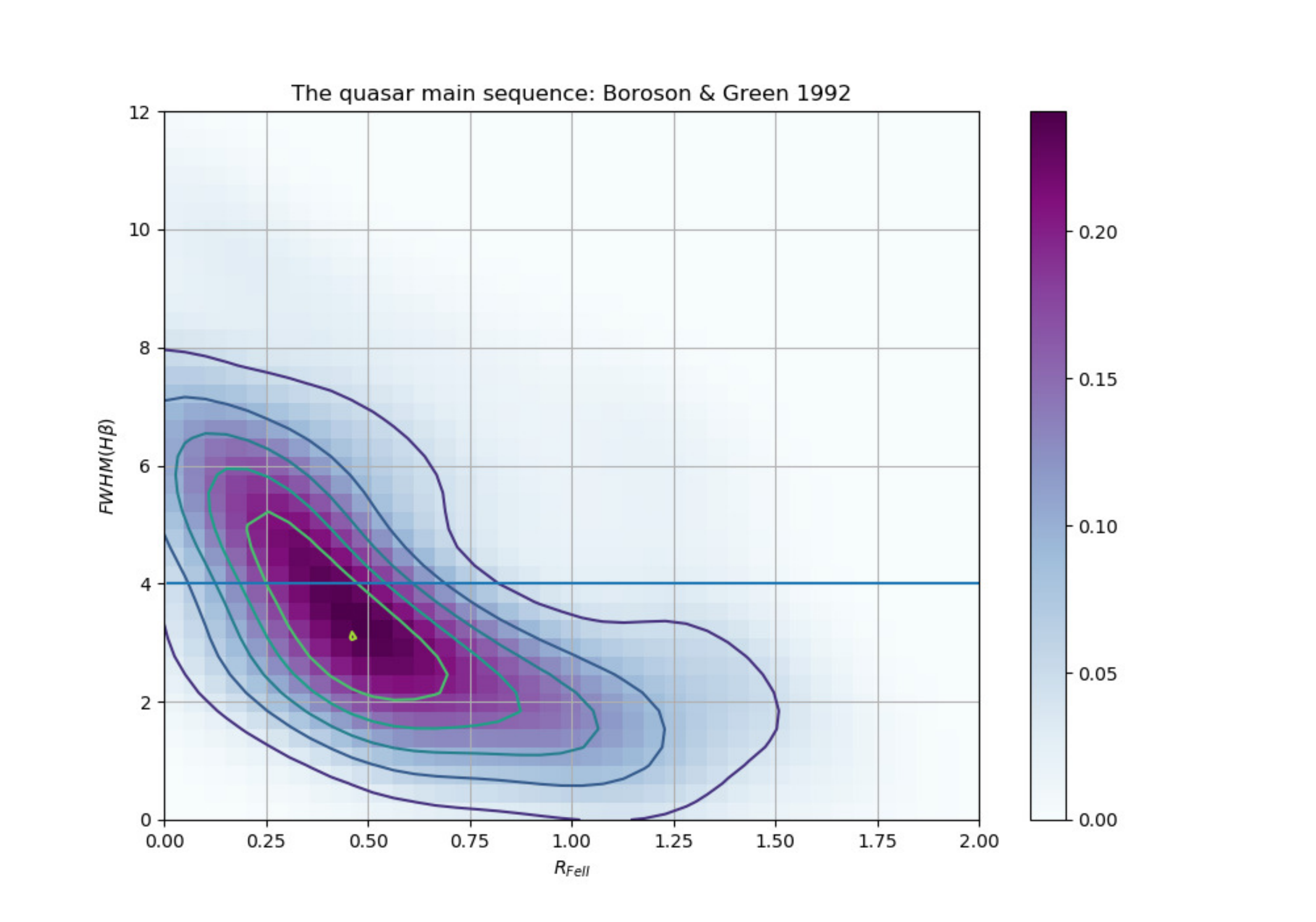}
		\hspace{-1cm}
		\includegraphics[width=9.25cm]{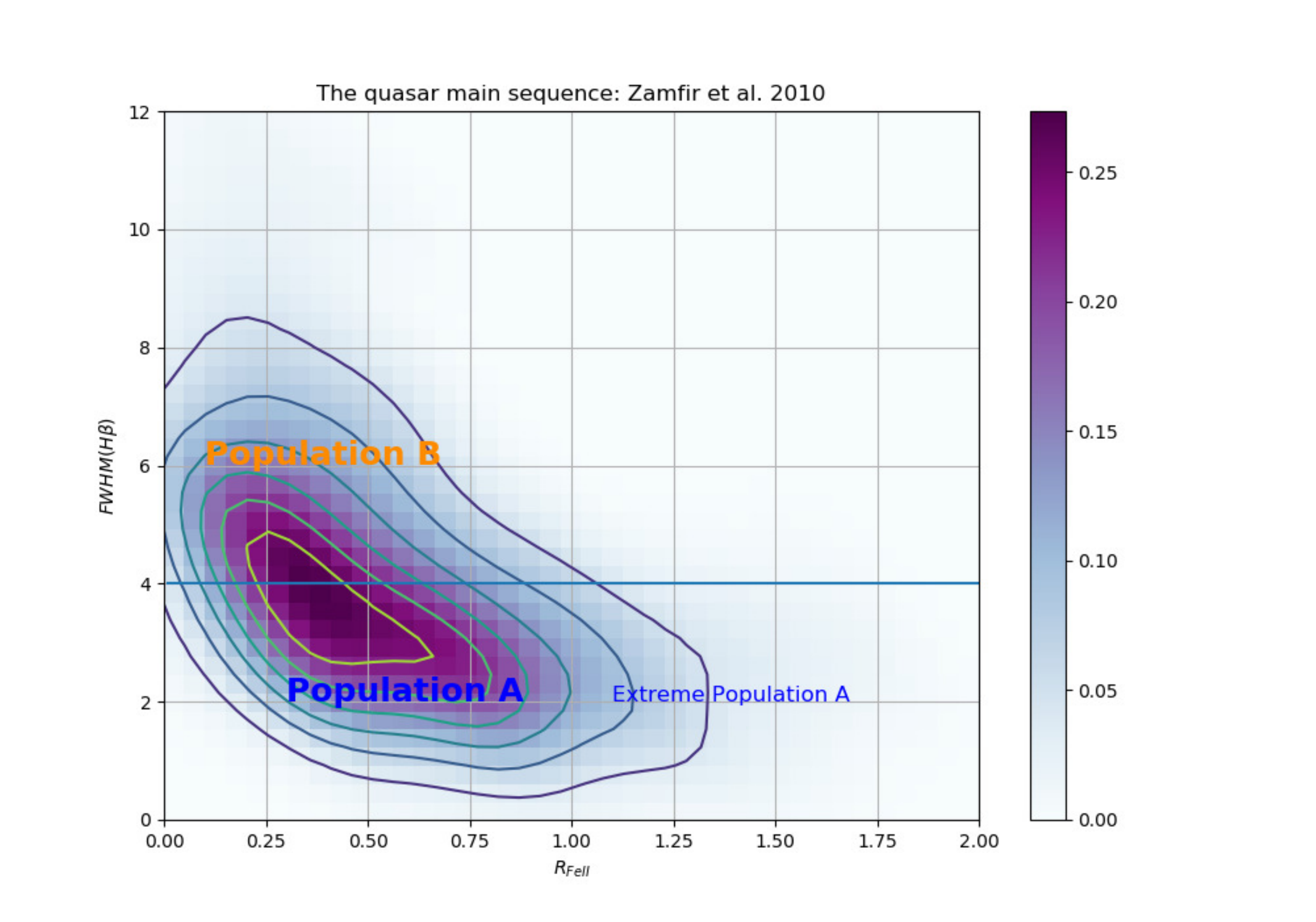}
		\\ 
	\end{center}
	\caption{The quasar main sequence as defined from the original paper by \cite{borosongreen92} based on 88 quasars (left) and the one based on the SDSS sample of 310 low-$z$\ quasars by \citet{zamfiretal10}. {The color shading from cyan to navy blue is proportional to the number density as a function of the \feii\ prominence parameter and of the FWHM of \hb,  and therefore to the source occupation {in} the parameter plane.}    }\label{fig:ms}
\end{figure}

The quasar main sequence is, in many way, analogous to the FP  for black holes in a different accretion mode sustained by higher  { $\dot{m} \sim 0.01 - 1$}. The formulation is rather different, and follows a different discovery path based on the statistical analysis of sources that are predominantly radio-quiet. The quasar main sequence (MS) is defined from the {first Eigenvector (E1) } that was originally identified  by a PCA of about  80 Palomar-Green (PG) quasars and associated with an anti-correlation between {the strength of optical \feii\ emission measured from the prominence of the  emission blend centered at $\lambda$ 4570 \AA\  (\feiiq) with respect to \hb\  (\rfe = I(\feiiq)/I(\hb)) }  and FWHM of \hb\  \citep{borosongreen92}. The E1 MS has withstood the test of time \cite[Fig. \ref{fig:ms},][]{sulenticetal00c,zamfiretal10,popovickovacevic11,kruzceketal11,grupenousek15,shenho14}, and the main optical trend {shown in Fig. \ref{fig:ms}} has been confirmed by samples of more than two order of magnitude larger in size than the original one \cite{shenho14}. 
The importance of \feii\ stems from its extensive  emission from UV to the IR  that can dominate the thermal balance of the low-ionization BLR. The  FWHM(\hb) is associated with  the velocity field in the low-ionization BLR, most likely predominantly  virialized \citep{petersonwandel99}. These two parameters are related to the physical conditions and to the dynamics of the emitting regions, that are in turn influenced by the accretion mode of the central black hole, and its evolutionary stage.  

Trends associated with the MS   have been extended to the radio \cite{sulenticetal03,shenho14,zamfiretal08,gancietal19}, FIR \cite{wangetal06,gancietal19}, IR, \cite{dultzin-hacyanetal99,martinez-aldamaetal15,pandaetal20}, UV \cite{sulenticetal00b,reichardetal03,bachevetal04,sulenticetal06,sniegowskaetal20,baskinlaor05b,richardsetal02,richardsetal05} and X-ray domain \cite{wangetal96,grupeetal01,benschetal15}, and to optical variability as well \cite{maoetal09,bonetal18}.  Table 1 of \citet{fraix-burnetetal17} provides a detailed list of the various parameters that have been measured in the various frequency domains. A summary description of the trends and a justification for the two quasar populations are also provided by  several authors \cite{sulenticetal08,sulenticetal11,sulenticmarziani15}. A nonlinear decay curve  provides a quantitative description of the main sequence in the FWHM — \rfe\ plane \cite{wildyetal19}.

\begin{figure}[h!]
	\begin{center}
	\includegraphics[width=8.cm]{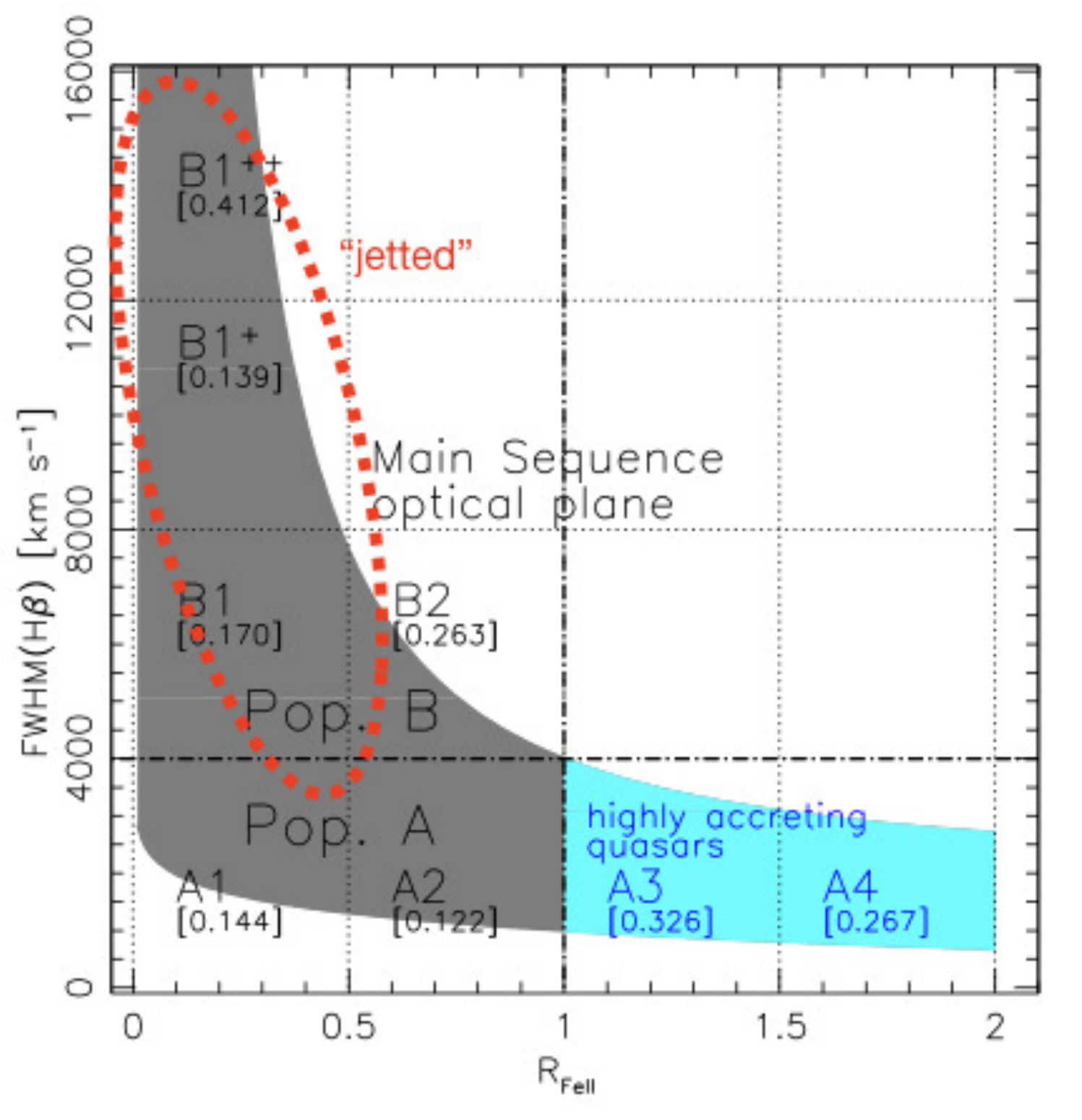}
		\includegraphics[width=8cm]{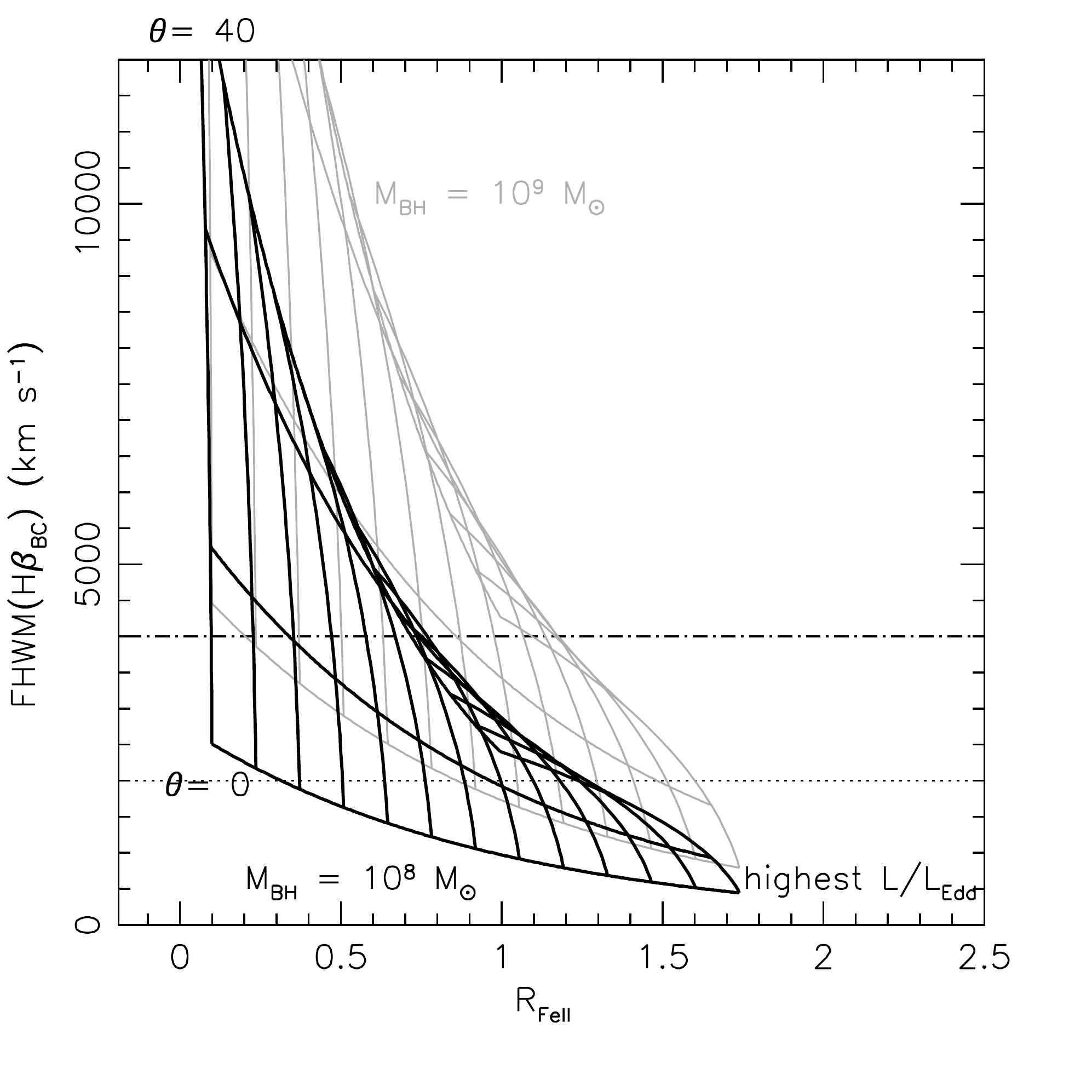}
	\end{center}
	\caption{The optical plane of the quasar main sequence with the occupation accounted for by the combined effect of Eddington ratio and orientation, as claimed by \cite{shenho14}. The two grids were computed for \mbh = $10^8$ \msol\ and   $10^9$ \msol\ (grey), for several values of \lledd\ and for viewing angle $\theta$\ between 0 and 40 degrees, following the toy model described in the text and in more detail in Ref. \citep{marzianietal18a}. In the left panel, the labels identify the areas of Population A and B (respectively below and above the FWHM limit at 4000 \kms), and of extreme Population A (\rfe$\gtrsim 1$).}\label{fig:grid}
\end{figure}

The distribution of the data in the plane \rfe -- FWHM(\hb) makes it expedient to define spectral types \citep[Fig. \ref{fig:grid}, ][]{sulenticetal02,shenho14}. This provides the considerable advantage that a composite spectrum within each bin could be representative of objects in similar physical conditions.  In alternative, a prototype object can be defined for each spectral type and used to analyze systematic changes along the quasar MS.  It is also expedient to distinguish between two populations: Population A made of sources with FWHM(\hb) $\le$ 4,000 \kms and Population B (broader) with FWHM(\hb) $>$ 4,000 \kms. Extreme Population A are quasars with \rfe $\gtrsim 1$\ and extreme Pop. B with undetectable \feii\ emission and the broadest Balmer lines (extreme FWHM \hb\ can reach $\sim 15,000 - 20,000$ \kms\ \cite{eracleoushalpern03,eracleoushalpern04,stratevaetal03}). {Basically, Population B includes sources termed as ``disk dominated", where radiation forces exert a modest influence on the overall dynamics of the gas \citep{richardsetal02}, while Population A is made of quasars radiating at relatively high Eddington ratio \lledd$\gtrsim 0.2$, for which radiation forces are able to maintain a wind that leads to easily identified systematic wavelength displacements toward the blue with respect to the quasar rest frame in the high-ionization emission lines \citep{gaskell82,brothertonetal94,marzianietal96,richardsetal11,coatmanetal16,sulenticetal17}. The extreme of Population A identifies the class of ``strong \feii\ emitters" \citep{liparietal93,grahametal96}.   \feii\ emission overwhelming \hb\ line emission (\rfe$\gtrsim$1) implies extreme Eddington ratio (\lledd $\sim 1$, \citep{marzianisulentic14}) and possibly  super-Eddington accretion rate \citep{wangetal14b,duetal16a,sunshen15,duetal16a,pandaetal18,pandaetal19}.  }


However, along the entire MS, the BLR gas emitting the low-ionization lines belong to predominantly virialized systems \cite{petersonwandel99}. The main asymmetries in the low-ionization line profiles can be explained in the context of a dynamical system whose velocity field is predominantly Keplerian.  The single peaked, symmetric and unshifted profile typical of Population A has been traditionally explained as due to an extended disk \cite{dumontcollinsouffrin90d}, and the same explanation apparently remains valid  in the case of extreme Pop. A AGN that  are   characterised by extreme high-ionisation blueshifts  
\cite{leighlymoore04,sulenticetal07,richards12,marzianietal16a,bischettietal17,sulenticetal17}. The high \civ/\hb\ intensity ratio of the blueshifted emission \cite{marzianietal10} makes it possible that the \hb\ profile remains almost symmetric and can be easily symmetrized by applying a small correction \cite{negreteetal18}.  In general, the distinguish feature of Pop. B sources, a redward asymmetric profile, can be explained by the sum of a disk contribution and emission from a larger distance \cite{bonetal07,bonetal09a}. Reverberation mapping studies of lines from different ionic species has provided evidence of ``ionization stratification"  and velocity-resolved reverberation mapping of sources with asymmetric \hb\  basically confirms the scenario of a Keplerian velocity field \cite{duetal18a,brothertonetal20}. The red-ward asymmetry has been interpreted as due to gravitational and transverse redshift \cite{bonetal15,punslyetal20} or by gas clouds infalling toward the central black hole \citep{wangetal17}.  At the extreme end of Pop. B sources, the profiles are often   very broad and double peaked, accounted for by a bare Keplerian disk model with mild relativistic effects \cite{chenhalpern89,stratevaetal03}. So, all along the quasar MS the low-ionization lines  ({at variance with the high-ionization emission}) appear to be predominantly associated with a bound, Keplerian dynamical system \citep{collinsouffrinetal88,elvis00}. 

Many studies still distinguish between the NLSy1s (FWHM \hb\ $\lesssim 2000$ \kms) and the rest of type-1 AGNs \citep[e.g.,][]{craccoetal16}, and consider NLSy1s an independent class. {There is a general consensus that the limit at 2000 \kms, albeit of historical importance, has no special meaning}. The main reason to extend the limit from 2000 to 4000 \kms\ is that several properties  of NLSy1s are consistent with the ones of ``the rest of Population A'' in the range $2000$ \kms $\lesssim $ FWHM(\hb) $\lesssim 4000$ \kms. The change — in low redshift samples $z \lesssim 1$ — occurs around 4000  \kms, not 2000 \kms\ \citep{craccoetal16}.  On the converse Population A and B can be distinguished on the basis of the Balmer line profiles, and  because of the amplitude of the systematic blueshift of the high-ionization lines with respect to the quasar rest frame. Composite H$\beta$\ profiles of spectral types along the MS are consistent with a Lorentzian for both NLSy1s and the rest of Population A. Other parameters (CIV$\lambda$1549 centroid, \rfe) also span the same ranges in NLSy1s and  the rest of Population A.

The governing accretion parameter accounting for the MS trends is most likely the Eddington ratio, which is related to the mass accretion rate by a monotonic albeit non-linear relation \cite{mineshigeetal00,sadowski11,sadowskietal14}. This explanation —   originally suggested by Boroson \& Green \cite{borosongreen92} — has also withstood the test of time \citep{marzianietal01,boroson02,zamfiretal10,aietal10,xuetal12,shenho14,sunshen15,pandaetal17,pandaetal18}, even if several key pieces needed to connect \lledd\ to the observed parameters remain poorly understood to date. The evidence of a correlation between \rfe\ and \lledd\ is still made murky by the strong effect of orientation on the line broadening, {affecting \mbh\ and \lledd\ computations with both random and systematic errors \citep{marzianietal19}.} \citet{sunshen15} provided evidence of this based on the stellar velocity dispersion of the host spheroid (a proxy for \mbh) anti-correlation  with \rfe, implying that \lledd\ increases with \feii.   Recent approaches include a careful analysis of the role of metallicity and of density and ionization trends \citep{pandaetal18,pandaetal19}, and  confirm \lledd\ as the main physical parameter governing the MS “horizontal branch” along the \rfe\ axis. 

A toy scheme can explain in a qualitative way the occupation of the MS plane under the assumptions that Eddington ratio, mass and an aspect angle $\theta$ (i.e., the angle between the line-of-sight and the accretion disk axis)\ are the parameters setting the location of quasar along the MS  \cite{marzianietal01,marzianietal18a}. If the BLR radius follows a scaling power-law with luminosity ($r \propto L^\mathrm{a}$, \citet{kaspietal00,bentzetal13}), under the standard virial assumption, then 
 \begin{equation}
\mathrm{FWHM}   \propto f_\mathrm{S}(\theta)^{-\frac{1}{2}} \left(\frac{L}{M_\mathrm{BH}}\right)^{-\frac{a}{2}} M_\mathrm{BH}^\frac{1-a}{2} \propto  f_\mathrm{S}^{-\frac{1}{2}}  {L}^\frac{1-a}{2} \left(\frac{L}{M_\mathrm{BH}}\right)^{-\frac{1}{2} }
\label{fwhm}.     
 \end{equation}  
 
 
We can also write  $\mathrm{R_\mathrm{FeII}}$ as a function of $(L/L_\mathrm{Edd} )$ and $\theta$, that needs to be established either empirically or theoretically.  For illustrative purposes, we consider  the “fundamental plane of accreting BHs”  that relates \lledd\ to \rfe\    \citep{duetal16a,bonetal20},  ignoring other relevant factors, such as systematic differences in line shapes and in chemical composition along the MS \cite{pandaetal19,sniegowskaetal20}, and we assume that \rfe\ depends on $\theta$\ following a limb-darkening law \citep{marzianietal01,netzer13}. 

As expected, {the right panel of Fig. \ref{fig:grid} shows that} $\theta$\ predominantly affects FWHM \hb\ and $L/L_\mathrm{Edd}$\ predominantly (but not exclusively) affects  $R_\mathrm{FeII}$. Under the assumptions of the toy scheme the FWHM limit at 4000 \kms should include mainly sources  with $L/L_\mathrm{Edd} \gtrsim 0.1 - 0.2$. Sources at lower \lledd\ are expected to be rare because they should be observed almost pole-on (for example, core-dominated radio-loud quasars whose viewing angle $\theta$\ is relatively small \citep{marzianietal01,zamfiretal08}), and the probability of observing a randomly oriented source at an angle $\theta$\ between the symmetry axis and the line of sight  is $P(\theta) \propto \sin \theta$. Even if such sources are expected to be rare, {  their number increases in flux-limited samples for a Malmquist bias, due to a continuum enhancement via relativistic beaming}.  We can say that separating Pop. A and B at 4000 \kms\  makes sense for low $z$ samples and that, also by a fortunate occurrence,  Pop. A includes mostly relatively high \lledd\ sources. 

\begin{figure}[h!]
	\begin{center}
		\includegraphics[width=12cm]{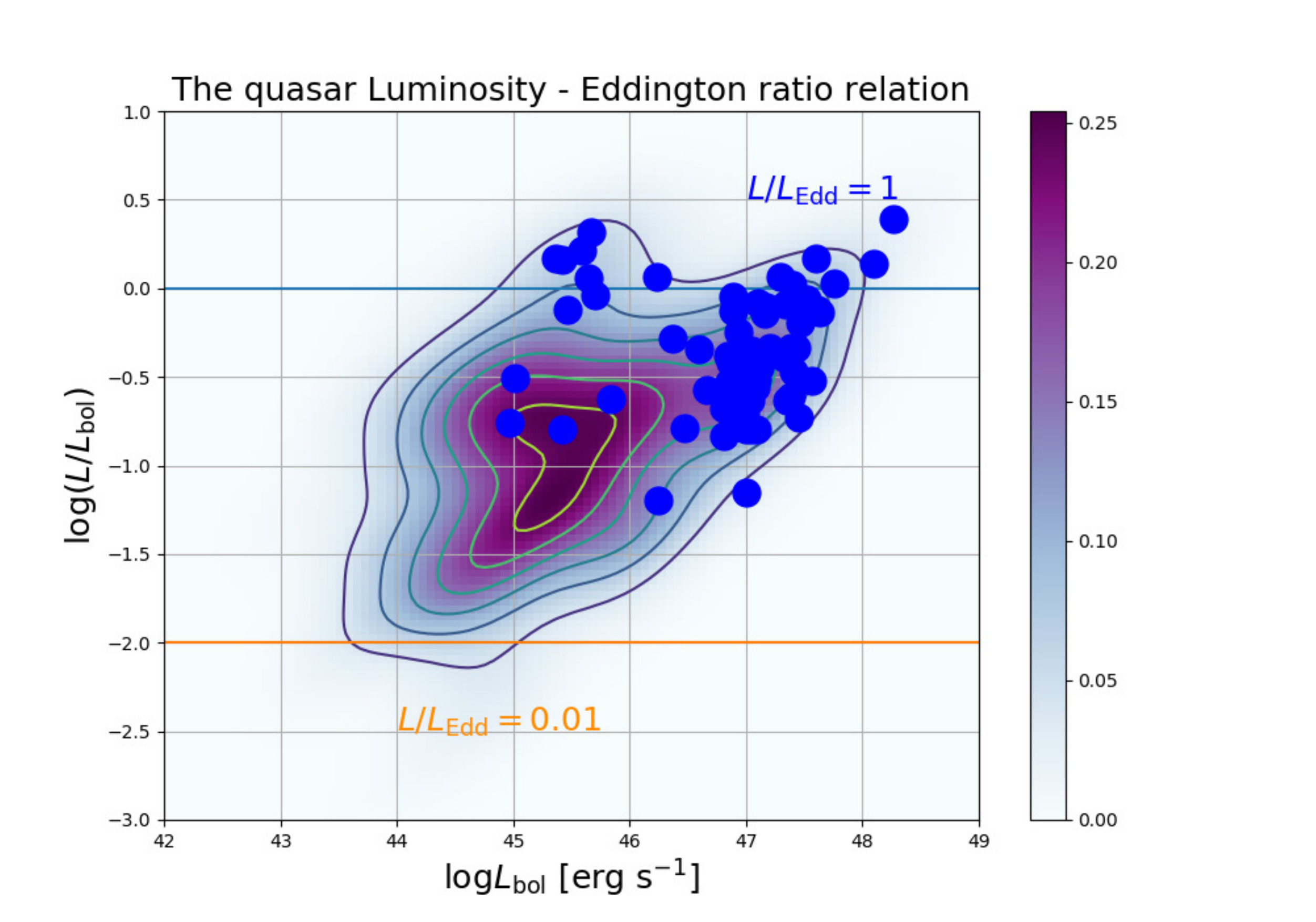}
	\end{center}
	\caption{The relation between Eddington ratio \lledd\ and bolometric luminosity for  the sample described in Fig. \ref{fig:ms}, second panel. Quasars occupy the range 0.01 — 1, but only above Eddington ratio $\approx$ 0.1 large shifts are observed, as shown by the distribution of the blue data points, which represent quasars with the largest \oiiiopt\ blueshift.  }\label{fig:lldd}
\end{figure}

The bolometric luminosity $L$\ can be estimated from optical or UV luminosities.\footnote{There is as yet no consolidated way to compute bolometric corrections, and ideally the bolometric correction should be computed from the spectral energy distribution (SED) of each individual quasar \cite{shangetal11}, or at least for each spectral type along the quasar main sequence \cite{pennelletal17}. Bolometric corrections can be also computed from theoretical considerations on the emission properties of the accretion disk \cite{nemmenbrotherton10,netzer19}. The simplest, and most widely used  approach to compute the bolometric correction is to multiply the monochromatic luminosity by a constant scale factor that is obviously frequency-dependent and roughly $10$ for $\lambda L_\lambda$\ at 5000 \AA, and $\approx 2 - 3$\ for the UV wavelengths where the strongest lines are observed \cite{elvisetal94,woourry02a,richardsetal06}. }      
The diagram  \lledd\ vs. bolometric luminosity (Fig. \ref{fig:lldd}) also provides a strong rationale for the existence of two populations: only above a threshold of \lledd\ $\approx 0.1$\ large shifts are observed.  Data points whose high-ionization lines are strongly blue shifted with respect to the rest frame are superimposed on the distribution of Fig. \ref{fig:lldd}, and are clearly seen   for \lledd$\gtrsim 0.1$\ only. This corresponds to the population A and B of \cite{sulenticetal00a}, of wind and disk-dominated quasars \cite{richardsetal02}, and population 1 and 2 of \citet{collinetal06}.  The data of Fig. \ref{fig:lldd} refer to sources with large blueshift in \oiiiopt, but an equivalent behaviour is observed also for the blueshift of \civ.   At the same time, Fig. \ref{fig:lldd} (and Fig. \ref{fig:mlagn} as well) show the effect of a  strong bias typically affecting   quasar studies over a broad range of redshifts: at high $z$ we detect only the high luminosity sources that corresponds to relatively high \lledd. 



\begin{figure}[h!]
	\begin{center}
		\includegraphics[width=10cm]{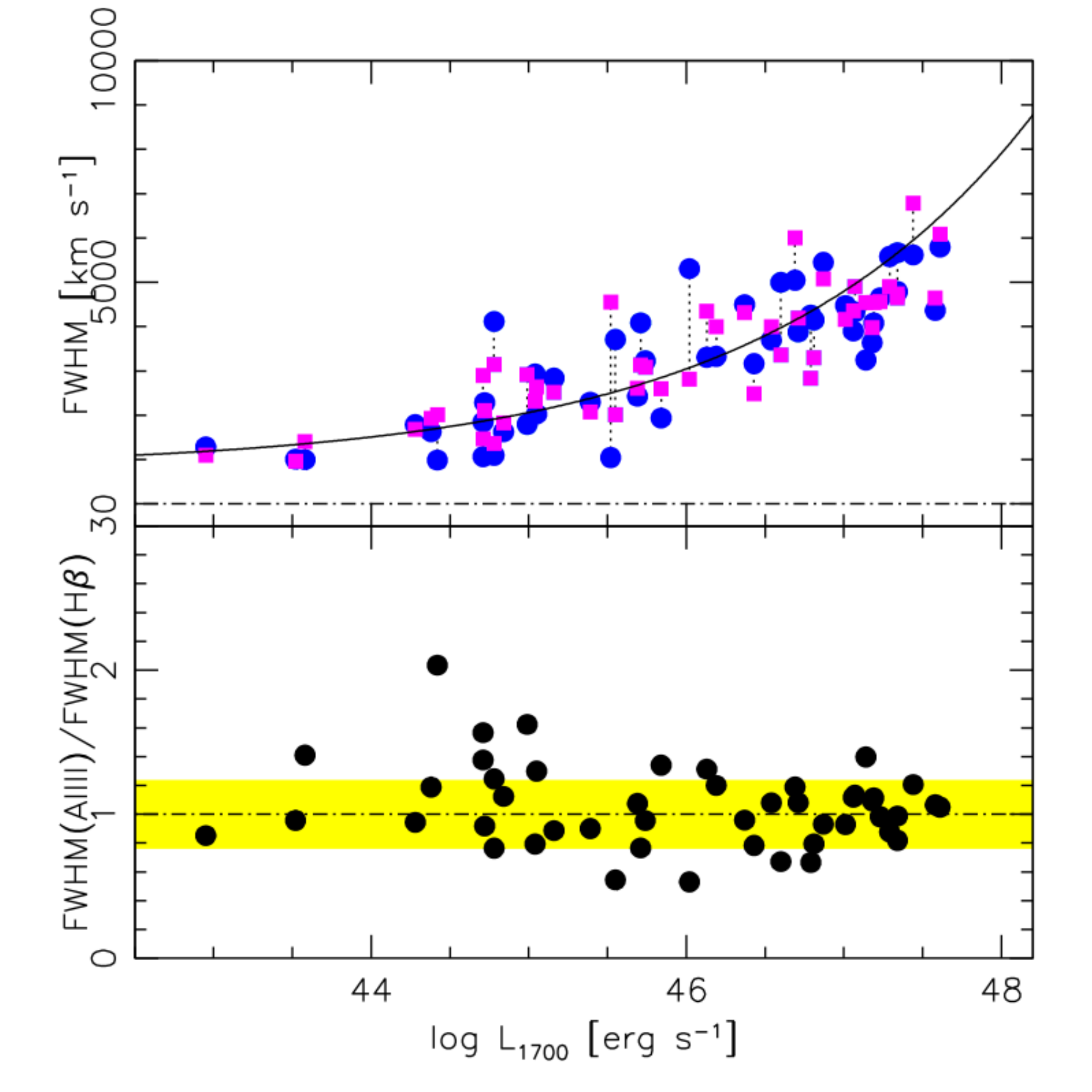}
	\end{center}
	\caption{The relation between FWHM  of \hb\ (blue) and \aliii\ (magenta) and their  FWHM ratio (bottom panel), and bolometric luminosity. The filled line represent the trend FWHM $\propto L^{1/4}$, with arbitrary normalization. The yellow band defines the uncertainty range in the ratio FWHM \aliii / FWHM \hb.  }\label{fig:lfwhm}
\end{figure}

Fig. \ref{fig:ms} refers to low-$z$ ($z \lesssim 1$) samples.  {A complete mapping of the MS at high $L$\ is still missing} (we consider high-luminosity quasars those with bolometric $\log L \gtrsim 47$ [erg/s]): the \hb\ spectral range is therefore accessible only with IR spectrometers to observe the \hb\ spectral regions of high-luminosity quasars that are very rare   at $z \lesssim 1$. A significant progress is expected in the next years, since IR spectral observations covering \hb\ of high-$z$ and high-$L$ quasars are becoming widespread.  A systematic increase in BH mass \mbh\ has a corresponding increase in FWHM. If $a = 0.5$, the FWHM grows with \mbh$^{0.25}$, i.e. a factor of 10 for $\log L$, passing from 44 (relatively low luminosity) to 48 (very luminous quasars). The trend may not be detectable in low-$z$\ flux limited samples, but becomes appreciable if quasars over a wide interval in $L$\  are considered.  At high \mbh, the MS becomes displaced toward higher FWHM values; the  displacement probably accounts for the wedge-shaped appearance of the MS when large samples of quasars are considered \cite{shenho14}. If we consider a limiting Eddington ratio (\lledd $\sim$ 0.1 — 0.2) as a physical criterion for the distinction between Pop. A and B, then the separation based on the FWHM  becomes luminosity dependent. According to the toy scheme, the FWHM of \hb\ (or of any other virialized line) should be $\propto ($\lledd$)^{-\frac{1}{2}} \times L^\frac{1-a}{2}$. Fig. \ref{fig:lfwhm} shows that the  $\propto L^{0.25}$\ for the width of a low- and an intermediate ionisation line.    The maximum \lledd\ should correspond to the minimum FWHM, expected to increase with luminosity as $\propto L^{0.25}$. If the FWHM is plotted against the luminosity, a trend-line nicely envelops the lower FWHM end of the data point distribution \cite{marzianietal09}.

\medskip
\subsection{The BH mass -- luminosity relation}
\label{ml}

\begin{figure}[h!]
	\begin{center}
		\includegraphics[width=12cm]{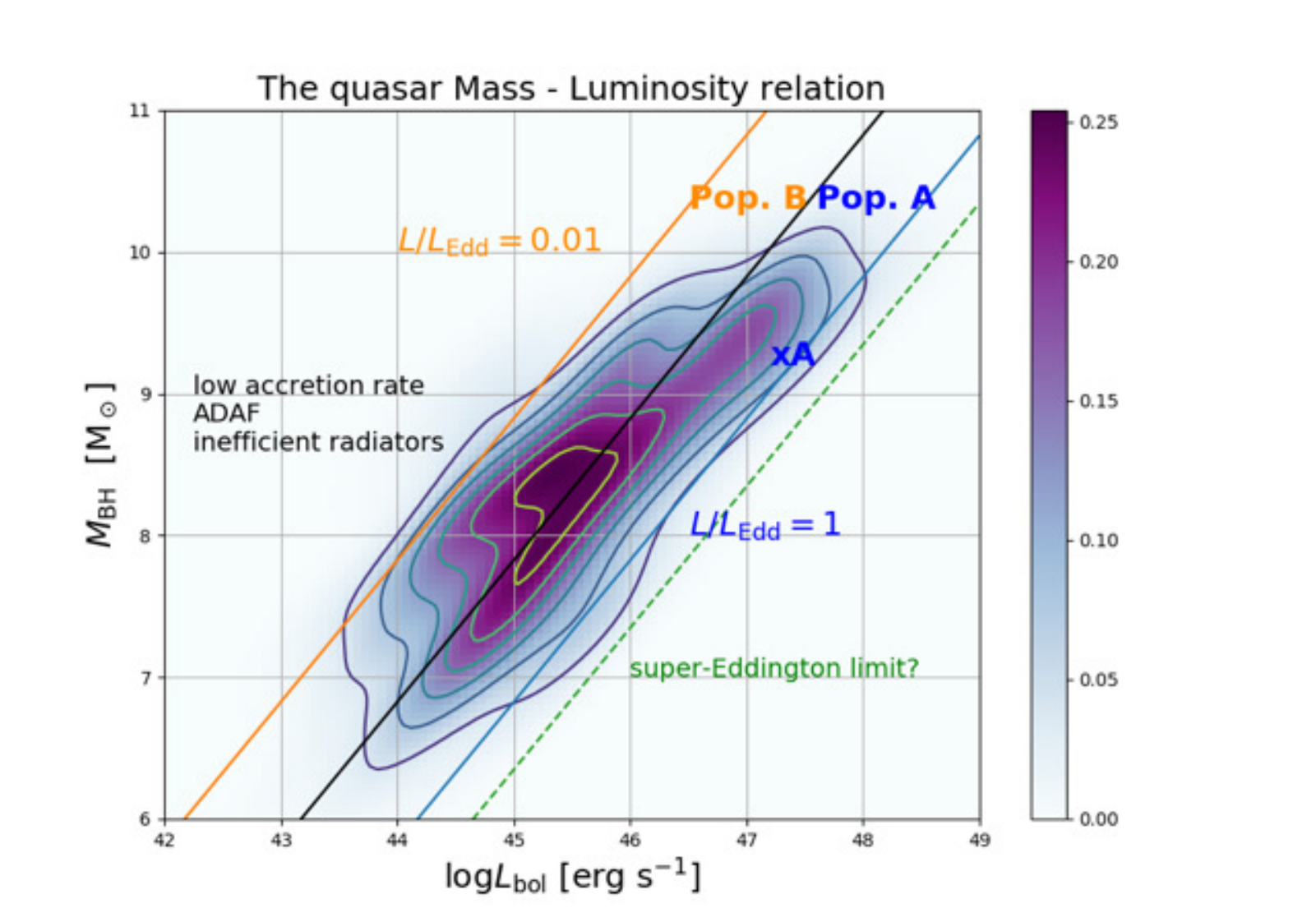}
	\end{center}
	\caption{Mass-luminosity relation for a sample of $\approx$ 330 AGNs, made of 280 low-$z$\ quasars from \cite{marzianietal03b} and high-luminosity 50 HE quasars of the sample described by \citet{sulenticetal04}. The diagonal lines trace the lower $\sim 0.01 \cdot$ \lledd\ and  $\sim 1.00 \cdot $ \lledd. The  wide majority of AGN is included within these limits. }\label{fig:mlagn}
\end{figure}

{Joining the fundamental plane and the main sequence trends for AGN,} four main regimes can be isolated (c.f. \cite{giustiniproga19}) where the physics of the inner accretion and ejection  is expected to change. Observationally, they range from low-luminosity AGN at extremely low accretion rates ($\dot{m} \lesssim 0.01$)\ and Population B quasars radiating at rates $0.01 \lesssim $\lledd$ \lesssim 0.1 - 0.2$, to  Population A sources with \lledd $\gtrsim 0.1 - 0.2$, and extreme Population A sources radiating close or somewhat above the Eddington limit (\lledd $\gtrsim 1$). There is a close formal analogy between the FP of accreting black hole and the MS. Eq. \ref{eq:ss} can be rewritten as an implicit relation between \lledd\ and \mbh. Similarly the MS is a sequence in the plane  FWHM \hb\ — \rfe\ that can be translated into a relation between \lledd\ ($\propto$ \rfe) and \mbh\ ($\propto $ FWHM). The  relations of the MS  are, as in the case of the FP, self-similar over 9 order of magnitude in \mbh\ \citep{zamanovmarziani02}.  Radiation driven winds appear to dominate in the high-ionization line emission in Population A and especially extreme Pop. A, reflecting the importance of the balance between radiation and gravitation forces expressed by \lledd\ in the accretion processes of AGN \cite{ferlandetal09,marzianietal10}, whereas the black hole mass is  the ultimate parameter governing the energetics  \cite{sulenticetal17}. 

The \mbh—luminosity relation can be constructed for large quasars samples once the \mbh\ has been computed (Fig. \ref{fig:mlagn}). Fig. \ref{fig:mlagn} shows that the distribution of quasars in the plane \mbh\ -- $L$\ is constrained within two well-defined diagonal lines, corresponding the \lledd $\approx 0.01$ and \lledd $\approx 1$. The empty area at the top left corner is due to inefficient radiators accreting at very low rate \cite{narayanyi95}, which are most often not type-1 quasars and are difficult to detect; the bottom right area is associated with sources that should be super Eddington radiators. Such  sources are not expected to exist; \lledd $\approx \ a \, few$ could be a physical limit for highly super-Eddington accretion \cite{mineshigeetal00,sadowskietal14}. 

\medskip
\section{A Tully-Fisher law for quasars} 
\label{virial} 


Strong \feii\ emitters have attracted attention since long, but they have been linked to a particular accretion state only recently  \citep{marzianisulentic14,marzianisulentic14a}. The simple selection criterion    \rfe  $>$ 1.0  used for the identification of xA sources from optical data,  corresponds to an equally-simple selection with UV criteria \citep{marzianisulentic14}.  
In addition, the distinguishing features of the UV composite spectrum of \citet{martinez-aldamaetal18} reveals that  the spectrum of xA sources can be recognized by a simple visual inspection.  


Extreme Population  A sources account for $\sim$ 10\% of quasars in low-$z$, optically selected sample FeII in Pop. A.  Lines have low equivalent width: some xAs are weak lined quasars \cite[W(\civ) $\le$ 10 \AA, WLQ][]{diamond-stanicetal09}, whereas WLQs can be considered as the extreme of  Pop. A \citep{marzianietal16a}.  The \ciii\ emission almost disappears. In the plane $\log U - \log n_\mathrm{H}$\ defined by CLOUDY simulations, UV line intensity ratio converge toward extreme values for density (high, $n_\mathrm{H} > 10^{12}-10^{13}$ cm$^{3}$) \cite{negreteetal12,templeetal20}, ionization (low, ionization parameter $U\sim10^{-3} - 10^{-2.5}$).  Extreme values of metallicity are also derived from the intensity ratios CIV/AlIII, CIV/HeII, AlIII/SiIII] \cite{negreteetal12,martinez-aldamaetal18,sniegowskaetal20},  most likely above 10 — 20 times solar or with abundances anomalies that might selectively increase aluminum or silicon, or both. 



xA quasars radiate at extreme \lledd\ along the MS. The \lledd\ dispersion has been found to be small \citep{marzianisulentic14}. {This result} is consistent with accretion disk theory {that  predicts} low radiative efficiency at high accretion rate and that \lledd\ converges toward a limiting value  \citep{mineshigeetal00,abramowiczetal88,sadowskietal14}.  Another important fact is the self similarity of the spectra selected by the \rfe\ criterion: the low-ionization lines become broader with increasing luminosity according to Eq. \ref{fwhm}, but the relative intensity ratios (and so the overall appearance of the spectrum) remains basically unchanged, although some redshift and luminosity effects are expected.   Accretion disk theory predicts that at high accretion rate  a geometrically thick, advection dominated disk should develop \citep{abramowiczetal88,sadowskietal14}.   The innermost part of the disk is puffed up by radiation pressure, while the outermost one remains geometrically thin.  The effect on the BLR structure  can be addressed by two dimensional reverberation mapping and by careful modelling of the coupling between dynamical and physical conditions  \citep{lietal13,pancoastetal14a,lietal18}.  However, this change from the standard thin 
disk provides two key elements for the BLR structure: the existence of a collimated cone-like region, where the high ionization outflows might be produced, and the shadowing of the outer disk where low-ionization emission lines form \citep{wangetal14a}. The low-ionization emitting region may therefore remain shadowed from the intense radiation field that is associated with the continuum observed if the line of sight is { not too} far from the polar axis, and the velocity field { stays unperturbed}. 

\medskip
\subsection{A relation between luminosity and velocity dispersion for quasars}
 
Three conditions are satisfied for xA quasars: (1) constant Eddington ratio \lledd,  close to Eddington limit; (2) The assumption of  virial motions of the low-ionization BLR, so that the black hole mass \mbh\ can be expressed by the virial relation (Eq. \ref{eq:vir}); (3) spectral invariance: for extreme Population A, the ionization parameter $U$ can be written as  $U = {Q(H)}/{4 \pi r_\mathrm{BLR}^2 n_\mathrm{H} c}  \propto  {L}/{r_\mathrm{BLR}^{2}n_\mathrm{H}}$  \citep{netzer13}, where $Q(H)$ is the number of hydrogen ionizing photons. $U$ has to be approximately constant, otherwise we would observe a significant change in the spectral appearance.  The three constraints make it possible to derive a relation between line width (the FWHM of the \hb\ broad component is expressed in units of 1000 \kms) and luminosity:   
\begin{equation}
  L(\rm FWHM)    =   {\mathcal L_{\rm 0}} \cdot  (\rm FWHM)^{4}_{1000}\  {\rm erg  \, s}^{-1} 
  \label{eq:virlum}
  \end{equation}
where ${\mathcal L_{\rm 0}}$\ depends on the square of \lledd, the ionizing range of the spectral energy distribution, and a parameter directly derived from the UV spectra,  the product density times ionization parameter that has been scaled to the typical value $10^{9.6} $cm$^{-3}$ \citep{padovanirafanelli88, matsuokaetal08,negreteetal12}.  Until now,  the FWHM of \hb\ broad component and of \aliii\ have been adopted as VBEs \citep{dultzinetal20,czernyetal20a,marzianietal21}. Equation (\ref{eq:virlum})  implies that  a simple measurement of the FWHM of a low-ionization line yields a $z-$independent estimate of the accretion luminosity  \citep[][c.f. \cite{teerikorpi11}]{marzianisulentic14}. 

The virial luminosity equation is conceptually equivalent to the Tully-Fisher and the early formulation of the Faber Jackson laws for ETGs \citep{faberjackson76,TullyFisher1977}.  { Recent works proposed the ``virial luminosity”  could provide suitable distance indicators because several emission properties appear to be extreme and stable with luminosity scaling with black hole mass at a fixed ratio \citep{wangetal13,lafrancaetal14,wangetal14b}.}  The virial equation has been applied  to xA quasars only (\lledd $\sim$ 1), although in principle could be useful for all quasars with known \lledd, provided a suitable emission line broadened by {virial} motions is used for the luminosity computation. At present, the virial equation can be considered for {\em all xA} quasars distributed  over a wide range of luminosity and redshift, where conventional cosmological distance indicators are not available \cite{czernyetal18,czernyetal20a,marzianietal21}.

\section{Conclusions}
\label{CONCLUSIONS}

In this work we have reviewed only a small part of the big efforts done up to now on the SRs of galaxies and AGN.
We have not addressed for example the correlations that are observed in the X-ray and radio domain, as well as many correlations involving the line emissions visible in the spectra.

It should be now clear that SRs are used continuously in every research area. The aims are different, going from the estimation of masses and distances, or peculiar velocities, or simply to check the output of theories, or to extract from them some useful indications about the physical mechanisms shaping the structure and evolution of galaxies and AGN.

The clear message emerging from this vast panorama of connections between structural, dynamical, gas and stellar population and halo parameters, is that galaxies are very complex objects formed through different channels, that include merging of sub-units, inflows, shocks, collapses, etc., as some of the most influent processes at work. In addition, it is also clear that galaxies vary their properties across the cosmic time, changing their { morphology} and physical characteristics. The simple Hubble morphological classification is therefore only a first naive tentative of summarizing such complexity that today are leading astrophysicists to { adopt} new specific strategies to classify galaxies, describe their properties, and highlight the amount of diversity across the cosmic epochs, but always keeping in mind the necessity of looking at the most important parameters that are able to trace the evolution of galaxies.

This new way of working is now facing the need of sophisticated numerical simulations and new statistical tools able to tackle the big astronomical number of data, exploring different classification schemes and strategies and group galaxies according to their similar evolutionary paths.

The multivariate partitioning analyses appear to be one the most appropriate techniques. The Principal component analysis is one of these tools \cite{Cabanacetal2002, Recio-Blancoetal2006}, but it is not a clustering tool. Many new attempts have used multivariate clustering methods \cite[see e.g.][]{Ellisetal2005, ChattopadhyayChattopadhyay2006, Chattopadhyayetal2007, Chattopadhyayetal2009, Fraix-Burnetetal2009, Sanchez-Almeidaetal2010, Fraix-Burnet2010}. These sophisticated statistical tools are now used in different areas of astrophysics and are giving encouraging results, in particular for the problem of the identification of the galaxy ancestors and the processes more active in the transformation of galaxies \cite{Fraix-Burnetetal2019}.

In conclusion we can say that the world  of SRs is big and complex. A lot of efforts are still necessary to organize such complexity, identify the key relationships having a real physical role for galaxies and AGN  and understand the profound implications behind their intrinsic nature. Possibly the future high-z observations will add new information that will help the clarification of many long-standing open problems.

\newpage
\section*{Abbreviations and acronyms used in the text}

\begin{table}[h] 
\caption{Abbreviations and acronyms used in the text}
\label{abb_acron} 
\begin{tabular}{|c|c|c|c|c|c|} 
\hline
\multicolumn{1}{|c}{N}        &
\multicolumn{1}{|c}{Symbols}  &
\multicolumn{1}{|c}{Meaning}  & 
\multicolumn{1}{|c}{N}        &
\multicolumn{1}{|c}{Symbols}  &
\multicolumn{1}{|c|}{Meaning} \\ 
\hline
1   &     AGN   & Active Galactic Nuclei       &     25  &     JWST  & James Webb Space Telescope  \\           
2   &     ALMA  &Atacama Large Millimeter Array&     26  &           &                             \\           
3   &     BTF   & Baryonic Tully-Fisher        &     27  &     LTGs  & Late Type Galaxies              \\       
4   &     BM    & Baryonic Matter              &     28  &     LZR   & Luminosity-Metallicity Relation \\   
5   &     BH    & Black Hole                   &     29  &     MR    &  Mass -Radius                   \\   
6   &     BLR   & Broad-Line Region            &     30  &     MRR   &  Mass-Radius Relation           \\       
7   &     CoGs   & Clusters of Galaxies        &     31  &     MS    &  Main Sequence                  \\
8   &     CGM   & Circum-Galactic Medium       &     32  &     MZR   & Mass-Metallicity Relation       \\                                                            
9   &     CMR   & Color-Magnitude Diagram      &     33  &     NIR   & Near Infra-Red                  \\
10  &     DEs   & Dwarf Elliptical galaxies    &     34  &     NLSy1 & Narrow Line Seyfert 1           \\       
11  &     DGs   & Dwarf Galaxies               &     35  &     PCA   &  Principal Component Analysis   \\
12  &     DSphs & Dwarf Spheroidal Galaxies    &     36  &     SED   &  Spectral Energy Distribution   \\       
13  &     DM    & Dark Matter                  &     37  &     SF    &  Star Formation                 \\
14  &    EAGLE  &    Evolution and Assembly of &     38  &     SFH   &  Star Formation History         \\
15  &           & Galaxies and their Environments&   39  &     SFR   &  Star Formation Rate            \\ 
16  &     ETGs  & Early Type Galaxies          &     40  &     SKA   &  Square Kilometer Array         \\ 
17  &     FIR   & Far Infra-Red                &     41  &     SMBHs &                                 \\       
18  &     FJ    & Faber - Jackson              &     42  &     SRs   &   Scale Relations               \\
19  &     FP    & Fundamental Plane            &     43  &     TF    &  Tully- Fisher                  \\       
20  &     FOS   & Fiber Optic Switch           &     44  &     VBE   &                                 \\     
21  &     FWHM  & Full Width at Half Maximum   &     45  &     VLBI  &  Very Long Baseline Interferometry \\    
22  &     IGM   & Inter-Galactic Medium        &     46  &     VLTI  &  Very large Telescope Interferometry \\
23  &     IMF   &  Initial Mass Function       &     47  &     WLQ   &  Weak-Lined Quasars              \\       
24  &     ISM   & Inter-Stellar Medium         &     48  &     ZoE   &  Zone of Exclusion              \\
\hline    
\end{tabular}
\end{table}


 \section*{Author Contributions}

  All the authors contributed equally to this work.

  


 
 \section*{Acknowledgments}
 MD want to thank Frontiers for the assistance in the production of the Research Topic.

 \bibliographystyle{frontiersinHLTH&FPHY} 

 \bibliography{test}

 \end{document}